\PassOptionsToPackage{left=0.5cm,right=0.5cm}{geometry}
\documentclass[a4paper,fleqn]{cas-dc}
\usepackage[numbers]{natbib}
\usepackage{float}
\usepackage{graphicx}
\usepackage{algorithm, amsmath}
\usepackage{algpseudocode}
\usepackage{subcaption}
\usepackage{float} 
\usepackage{placeins}
\usepackage{xcolor}
\usepackage[utf8]{inputenc}
\usepackage{array}
\usepackage{mathtools}
\usepackage{amsthm}
\theoremstyle{definitapaion}
\newtheorem{definition}{Definition}[section]
\usepackage{parskip}
\usepackage{tikz}
\usepackage{algorithm}
\usepackage{algpseudocode}
\usetikzlibrary{shapes, arrows.meta, positioning}
\usepackage{booktabs}
\usepackage{graphicx}
\usepackage{multirow}
\usepackage[textsize=tiny]{todonotes}
\usepackage{enumitem}

\begin{document}
\let\WriteBookmarks\relax
\def\floatpagepagefraction{1}
\def\textpagefraction{.001}

% \underline{\textbf{Title of the Paper:}} Portfolio Optimization under Dynamic Rebalancing via Topological Data Analysis and News Sentiments\\

% \underline{\textbf{Details of Authors: }}
% \\
% \underline{\textbf{First and corresponding Author: }}

% \textbf{Name:} Divyanee Garg

% \textbf{Affiliation: }Department of Mathematics, Indian Institute of technology, Delhi, Hauz Khas, New Delhi, 110016, Delhi, India

% \textbf{Email Address:} maz228080@iitd.ac.in
% \\ \\

% \textbf{Acknowledgements} 

% The author sincerely thanks Prof. Aparna Mehra, Department of Mathematics, IIT Delhi, for her valuable guidance, insightful suggestions, and careful review of the manuscript. Puneet Pasricha, Department of Mathematics, IIT Ropar, for his insightful comments. The author sincerely acknowledges the financial support provided through the Prime Minister Research Fellowship by the Ministry of Education, Government of India, under award number PMRF ID-11CS142003.

% \vspace{0.5cm} 

% % \textbf{Funding Information}

% % The author gratefully acknowledges the financial support received under the Prime Minister Research Fellowship (PMRF), Ministry of Education, Government of India.

% % \vspace{0.5cm}

% \textbf{Declaration of Competing Interest}

% The author declares that there are no known competing financial interests or personal relationships that could have appeared to influence the work reported in this paper.

\shorttitle{Portfolio Optimization under Dynamic Rebalancing}
% \shortauthors{D. Garg}
\title [mode = title]{Portfolio Optimization under Dynamic Rebalancing via Topological Data Analysis and News Sentiments}

\author[1]{{Divyanee Garg}}[%auid=000,bioid=1,
                        orcid=0009-0008-1503-1666]
\cormark[1]
 % \cormark[1]
% \fnmark[1]
\ead{maz228080@iitd.ac.in}
% \ead[url]{www.jkkrishnan.in}

% % \credit{Conceptualization, Methodology, Investigation,  Visualization, Programming, Data collection, Empirical analysis, Writing – original draft, Reviewing and editing, and Final form of the manuscript}

% \affiliation[1]{organization={Department of Mathematics, Indian Institute of technology, Delhi},
%                 addressline={Hauz Khas}, 
%                 city={New Delhi},
% %               citysep={}, % Uncomment if no comma needed between city and postcode
%                 postcode={110016}, 
%                 state={Delhi},
%                 country={India}}
% \cortext[cor1]{Corresponding author}

\begin{abstract}
% Understanding the similarity and dissimilarity among financial assets is fundamental to portfolio diversification and risk management. In this paper, we propose a novel asset selection framework based on topological data analysis (TDA) to identify topologically dissimilar assets for portfolio construction. The proposed approach represents each asset in a multidimensional feature space using technical indicators and investor sentiment derived from financial news. Sentiment scores are extracted from daily Refinitiv news headlines using FinBERT and incorporated as an additional feature in the similarity assessment. To capture the intrinsic structure of asset behavior, we compute topological summaries through persistence diagrams and persistence landscapes, and use them to quantify pairwise similarities among assets. These similarity measures are then employed within an agglomerative clustering framework to identify clusters of similar assets and screen a subset of topologically dissimilar assets. The selected assets are subsequently used in a dynamically rebalancing portfolio optimization model that accounts for practical investment constraints such as transaction costs, short selling, and lower and upper bounds.
Understanding similarity among financial assets is essential for portfolio diversification. This paper proposes a novel portfolio framework that addresses three key aspects of asset management: investment-worthy asset selection, optimal weight allocation under practical constraints, including transaction costs, short-selling, and upper and lower bounds, and dynamic rebalancing. The framework integrates Topological Data Analysis (TDA) with multi-dimensional data by combining technical indicators with sentiment scores derived from financial news using FinBERT. A TDA-based distance measure is introduced within agglomerative clustering to select topologically dissimilar assets. The inclusion of sentiment as a feature enhances asset selection, as sentiment signals are highly dynamic and capture rapid changes in market perception and investor behavior that are not reflected in technical indicators. Unlike traditional correlation and Euclidean distances, the proposed approach captures complex, non-linear dependencies through topological summaries such as persistence diagrams and landscapes. 

To capture rapidly evolving market sentiment, the proposed framework adopts a dynamic rolling-window rebalancing strategy with frequent portfolio updates. In addition, a retention mechanism is introduced to preserve high-quality assets across consecutive rebalancing windows, thereby reducing excessive portfolio turnover and transaction costs. Extensive empirical analysis on S\&P500 constituents using a rolling window framework demonstrates that the proposed framework consistently achieves higher returns and better reward-risk ratios than correlation and Euclidean based frameworks, as well as benchmark strategies including Naïve, Index, and portfolios formed from the full asset universe. Notably, the model exhibits robustness by generating positive performance even during periods of heightened market uncertainty, such as the U.S.-Israel-Iran conflict.

% To further examine the robustness of the methodology, we conduct an additional analysis on the S\&P index over 1 January 2026 to 31 March 2026, a period characterized by elevated market stress during the U.S.–Israel–Iran conflict. The findings confirm that the proposed framework remains robust and effective even under adverse market conditions.
\end{abstract}


\maketitle

\section{Introduction}\label{Sec: intro}
Portfolio optimization (PO) encompasses three key processes: asset selection, asset weighting, and asset management. Asset selection involves identifying a subset of promising assets from a wider investment universe by assessing their characteristics, including historical performance and market conditions. Asset weighting allocates capital among these selected assets to construct an efficient portfolio, either by minimizing portfolio risk for a given expected return or maximizing expected return for a specified level of risk \cite{Markowitz1952PortfolioSelection}. Finally, asset management entails the ongoing monitoring and adjustment of the portfolio over time, particularly through periodic rebalancing in response to evolving market conditions and new information. Collectively, these processes form the backbone of modern portfolio construction, essential to both financial economics and practical investment management. 

Asset management \cite{al2018outperformance} can be categorized into passive and active strategies. Passive management \cite{dhingra2024comprehensive}, often involving index funds, aims to replicate a benchmark index's performance with minimal trading and lower risk, appealing to risk-averse investors. Conversely, active management \cite{garg2025enhanced, yu2022dynamic} frequently adjusts portfolios to outperform benchmarks, attracting more risk-tolerant investors willing to accept higher transaction costs and risk for potentially higher returns. This study emphasizes on actively managed portfolios. 

Despite the foundational role of asset selection \cite{yu2024dynamic}, much existing literature focuses on the asset-weighting stage without selecting investment-worthy assets, treating portfolio design as primarily an optimization problem. However, the choice of assets significantly impacts portfolio performance, highlighting the critical nature of the asset selection step. Markowitz's principle of diversification \cite{Markowitz1952PortfolioSelection, migliavacca2023bibliometric} reinforces that spreading investments across diverse asset classes reduces risk. Conversely, excessive diversification can incur higher transaction costs, complicate portfolio management, and dilute the potential benefits of active asset selection. Keynes et al. \cite{keynes1983keynes} advocated for concentrated investment in a limited number of strategically chosen assets, thus favoring focused strategies. This dynamic between diversification and concentration has sparked interest in sparse portfolio selection \cite{brodie2009sparse, dai2018some}, aiming to construct portfolios with relatively few assets. Methods to induce sparsity typically fall into three categories: regularization-based approaches, cardinality-constrained optimization models, and clustering via unsupervised learning. Regularization methods introduce penalty terms, typically based on the $\ell_1$-norm of portfolio weights, to shrink insignificant asset positions toward zero \cite{goel2025risk, kremer2022sparse}. Alternatively, cardinality constraints directly restrict the number of selected assets \cite{kalayci2020efficient, xu2024efficient}. However, regularization approaches often require careful hyperparameter tuning and may lead to risk concentration in high-dimensional settings \cite{kremer2022sparse, wu2024sparse}. Similarly, cardinality-constrained portfolio models are non-convex and NP-hard, resulting in significant computational complexity and reliance on specialized heuristic algorithms for practical implementation \cite{wang2026exact}. In contrast, clustering-based asset selection provides a computationally efficient alternative by grouping assets into homogeneous clusters, simplifying the identification of representative assets. Clustering-based methods have shown strong potential for sparse portfolio construction by reducing dimensionality. Studies have shown the effectiveness of agglomerative clustering for portfolio construction using correlation-based similarity \cite{leon2017clustering} and its applicability and superiority in pair trading frameworks \cite{han2023pairs}. Consequently, this work employs agglomerative clustering (also known as hierarchical clustering) to identify investment-worthy assets for portfolio construction. For recent advancements in agglomerative clustering methods, see Ran et al. \cite{ran2023comprehensive}.

Financial markets generate both structured data, such as prices and technical indicators, and unstructured data, including financial news headlines. While structured data can easily integrate into quantitative models, extracting insights from textual data remains complex due to financial terminology and contextual dependencies. Recent advances in domain-specific language models have led to architectures such as FinBERT \cite{araci2019finbert} and FinLLaMA \cite{konstantinidis2024finllama}, which extend BERT \cite{devlin2019bert} and LLaMA \cite{touvron2023llama} using financial-domain training corpora. In particular, FinBERT employs transformer-based contextual learning to better capture financial semantics and sentiment compared to traditional lexicon-based and general-purpose language models \cite{araci2019finbert, deng2017adapting, renault2017intraday}. FinBERT is pre-trained and fine-tuned on financial documents, enabling better understanding of financial language and sentiment, and outperforming standard BERT and dictionary-based methods in financial sentiment analysis \cite{mahendran2025comparative}.

Recent studies highlight the importance of investor sentiment and multi-source data in financial decision-making. Investor sentiment has been linked to asset volatility and market behavior \cite{siganos2017divergence, wei2017informativeness}, while integrating financial news, social media sentiment, and technical indicators has been shown to improve asset prediction and portfolio performance \cite{weng2017stock, yu2022dynamic}. Similarly, Zhou et al. \cite{zhou2021big} propose an asset selection framework based on data envelopment analysis that incorporates multiple data sources, including technical and social media data. A comprehensive review of textual data-driven asset market prediction is provided in Nardo et al. \cite{nardo2016walking}.

The effective integration of these heterogeneous data sources can provide a more comprehensive representation of asset behavior and improve investment decision-making. In particular, financial news sentiment contains valuable information about market perception and investor reaction that may not be fully reflected in historical price-based variables and indicators alone. Motivated by this, we employ FinBERT to extract sentiment scores from financial news headlines and incorporate them into the multi-dimensional feature representation used for agglomerative clustering. Consequently, the portfolio construction process comprises asset selection based on multi-dimensional data, including news sentiment, followed by investment weight determination and dynamic rebalancing.

The effectiveness of any clustering approach largely depends on the choice of distance measure, with correlation-based \cite{leon2017clustering} and Euclidean distances \cite{han2023pairs} being among the most commonly used. However, Pearson correlation can be misleading when returns do not follow Gaussian or elliptical distributions, possibly leading to erroneous similarity assessments between financial variables \cite{embrechts2003using}. As a result, highly correlated assets may still exhibit different return patterns, leading to misleading similarity assessments and unintended portfolio exposures. Similarly, Euclidean distance captures only pointwise differences and may overlook complex nonlinear relationships in financial data. These limitations become more pronounced when asset trading data are represented using multidimensional features rather than univariate return series.

These limitations motivate the use of Topological Data Analysis (TDA), which is effective in extracting structural and non-linear information from multidimensional data. Unlike correlation and Euclidean measures, TDA captures qualitative features such as connected components, loops, and void providing insights into the underlying patterns in financial datasets, particularly in non-linear contexts \cite{aromi2021topological, goel2020topological}. These features provide insights into the intrinsic organization of data, particularly in settings characterized by nonlinearity and complex dependence patterns. For instance, Goel et al. \cite{goel2020topological} proposed a TDA norm-based asset filtering strategy that outperformed correlation-based approaches, while Aromi et al. \cite{aromi2021topological} demonstrated the effectiveness of persistence landscapes (PLs) in capturing structural properties of financial data.

TDA is typically applied to multidimensional point-cloud data, whereas univariate time series require an additional embedding step to reconstruct a higher-dimensional space. In contrast, our framework naturally forms a four-dimensional point-cloud representation for each asset using relative strength index (RSI), stochastic oscillator (SO), moving average convergence divergence (MACD), and sentiment scores, enabling a direct application of TDA. Motivated by this advantage, the present study integrates multi-dimensional asset representations with TDA to construct more informative distance measures for capturing asset similarity. Specifically, we compute distance matrices based on persistence diagrams (PDs) and persistence landscapes (PLs), and subsequently employ these within an agglomerative clustering framework for effective asset selection. In particular, persistence homology (PH), a key tool in TDA, summarizes these topological features across multiple scales through PDs and PLs, offering a robust representation that is less sensitive to noise and small perturbations.

TDA has been successfully applied in various domains, including finance, such as in pair trading \cite{majumdar2023pairs}, classification \cite{karan2021time}, pattern recognition \cite{carlsson2014topological}, and biological data analysis \cite{lum2013extracting, papamarkou2024position}. In finance, TDA has shown promise in detecting structural market changes and financial crises \cite{gidea2018topological}, though its application to portfolio construction remains relatively limited \cite{goel2025sparse, rivera2019topological, sokerin2024portfolio}. Topologically distinct assets exhibit fundamentally different structural and temporal characteristics. Since TDA captures higher-order structural relationships beyond linear dependence, topologically distinct assets are less likely to exhibit similar market behavior, thereby improving diversification, which is desirable for portfolio construction. Motivated by this, we employ TDA-based similarity measures for asset selection.

We employ two TDA-based distance measures, namely the Average Wasserstein Distance (AWD) on PDs and the Average Persistence Landscape (APL) distance on PLs \cite{goel2025sparse}. These distances are computed on multidimensional asset point clouds constructed using technical indicators and sentiment scores. To analyze the impact of sentiment information, we consider both 3-dimensional representations based only on technical indicators and 4-dimensional representations that additionally include sentiment scores, and compare their effects on clustering and subsequent portfolio construction.

For benchmarking, we also consider correlation and Euclidean distances under both feature settings, with and without sentiment, resulting in eight distinct distance measures. The selected assets are then incorporated into a dynamic rebalancing mean--variance (DRMV) framework. Following \cite{yu2017incorporating, yu2011portfolio}, the model incorporates practical constraints such as transaction costs, short-selling decisions, and lower and upper bounds on holdings, thereby improving the realism and implementability of the portfolio strategy. Incorporating these real-world constraints, particularly transaction costs, is essential, as they are often neglected in existing portfolio construction studies despite their significant impact on portfolio turnover, implementability, and realized performance.

The proposed framework is implemented using a rolling-window rebalancing mechanism. At each rebalancing date, assets are first selected through the proposed clustering-based filtering approach and then allocated using the (DRMV) model. The allocation from the previous period is incorporated into the current optimization problem to dynamically account for transaction costs and reduce unnecessary turnover, thereby improving portfolio stability and practical applicability. We conduct extensive empirical analysis using eight different distance measures and compare the proposed framework against several benchmark strategies, including the full-universe (DRMV) model without clustering, the equally weighted portfolio, the market index, and the buy-and-hold version of the proposed framework. This comparison enables a systematic evaluation of the effects of clustering, sentiment integration, TDA-based similarity measures, and dynamic rebalancing on portfolio performance.

The proposed framework is evaluated using daily returns of S\&P500 constituents from January 2025 to December 2025, and further tested during the recent U.S.--Israel--Iran conflict period from October 2025 to March 2026 to assess robustness under market uncertainty. The empirical results show that TDA-based portfolios, particularly those incorporating sentiments with indicators, consistently outperform correlation and Euclidean based portfolios as well as benchmark strategies. %These findings highlight the effectiveness of combining topological features with news-driven sentiment for developing sparse, adaptive, and data-driven portfolio strategies.

\subsection{Motivation}

Portfolio diversification relies on identifying sufficiently distinct assets to reduce portfolio risk without sacrificing return. Most existing portfolio construction and asset screening methods use similarity measures such as Pearson correlation \cite{jung2016clustering, michis2022multiscale, millington2021construction, puerto2020clustering} and Euclidean distance \cite{mattera2025time}. However, these measures often fail to capture the nonlinear, noisy, and time-varying structure of financial markets \cite{goel2025sparse}, which may limit diversification benefits.

Meanwhile, modern financial markets are influenced by heterogeneous data sources, including historical prices, technical indicators, and investor sentiment from financial news and media \cite{han2023pairs, yu2022dynamic, zhou2021big}. Although these sources contain valuable information, they are often analyzed separately or used only during portfolio allocation \cite{yu2022dynamic}. This motivates the need for a unified framework that integrates multidimensional asset information for asset selection.

In this context, TDA provides a promising alternative by capturing structural and persistent geometric patterns in complex and high-dimensional data through tools such as PDs and PLs. Unlike traditional similarity measures, TDA can effectively identify nonlinear dependence structures and hidden patterns, making it well-suited for financial applications. Furthermore, financial markets are frequently affected by geopolitical events, economic uncertainty, and sentiment-driven reactions, motivating the development of robust and adaptive portfolio strategies. These considerations motivate the proposed framework, which integrates TDA, multidimensional asset representations, sentiment information, and dynamic portfolio optimization (PO) into a unified investment framework.

\subsection{Objectives}

The primary objective of this study is to develop an asset screening and portfolio construction framework that improves portfolio performance by identifying topologically dissimilar assets using multidimensional financial information. Each asset is represented using technical indicators (RSI, SO, and MACD) along with sentiment scores extracted from financial news headlines using FinBERT, thereby incorporating both quantitative and textual information into the asset representation.

The second objective is to measure asset similarity using topological summaries rather than conventional linear measures. Specifically, TDA is employed to construct PD and PL, from which pairwise similarity measures are computed and integrated into an agglomerative clustering framework for asset selection.

The third objective is to examine the effectiveness of the proposed TDA-based asset screening within the (DRMV) framework that incorporates practical constraints such as transaction costs, short selling, and rolling rebalancing. Finally, the study evaluates the robustness of the proposed framework under volatile and stressed market conditions. Overall, the study investigates whether TDA-based multidimensional representations and sentiment integration can improve asset selection and portfolio performance compared to conventional approaches.

\subsection{Main findings}

The empirical results demonstrate that the proposed framework effectively identifies informative asset subsets and improves portfolio performance. TDA-based similarity measures derived from PDs and PLs provide more informative asset screening than conventional correlation and Euclidean distance measures. The inclusion of investor sentiment extracted from financial news further enhances asset representation and clustering quality, leading to improved asset selection.

Portfolios constructed using the proposed TDA-based filtering and dynamic rebalancing framework consistently outperform traditional filtering methods and benchmark strategies, including the na\"ive portfolio, market index, optimization without filtering, and buy-and-hold strategies. The proposed framework achieves superior return and reward-to-risk ratios while maintaining sparse and implementable portfolios under practical constraints.

The robustness analysis further shows that the proposed methodology remains effective during periods of market stress and geopolitical uncertainty. Overall, the findings highlight the potential of combining TDA, multidimensional asset representations, and sentiment information for robust and practical PO.

\subsection{Contributions}

This paper makes several methodological and empirical contributions to the literature on the application of TDA, asset selection, and PO. First, the study develops a multidimensional asset representation that integrates technical indicators and investor sentiment extracted from financial news using FinBERT, enabling a richer characterization of asset behavior and incorporating alternative textual information directly into the asset similarity structure. Second, the paper introduces a TDA-driven asset screening framework based on topological dissimilarity. By employing topological distance within an agglomerative clustering framework, the proposed approach identifies structurally distinct assets to enhance portfolio diversification and sparse asset selection.

Third, the study provides a comprehensive empirical comparison against benchmark strategies, including correlation- and Euclidean-based filtering methods, the na\"ive portfolio, market index, and optimization over the full asset universe. The results demonstrate the economic value of integrating TDA, multidimensional representations, and sentiment information in dynamic portfolio construction. Overall, the paper contributes to the intersection of financial optimization, machine learning, and TDA by proposing an interpretable, data-driven, and practically implementable framework for asset selection and PO.

\section{The description of data sources}\label{Sect: data sources}

Traditional asset selection methods mainly rely on price-based information. However, financial markets are influenced by multiple heterogeneous data sources, including technical indicators and investor sentiment reflected in news headlines. While technical indicators capture historical market dynamics, sentiment data provide behavioral and forward-looking information.

\subsection{Technical indicator}\label{Sec: Technical indicator}

The proposed framework utilizes daily open, high, low, and close asset prices to compute three widely used technical indicators: RSI, SO, and MACD \cite{appel2005technical, basak2019predicting, weng2017stock}. These indicators are employed as input features for the asset selection process. A brief description of each indicator is provided in Appendix A.

\subsection{News-based sentiment indicators}\label{Sec: sentiment}

Financial news reflects investor perception and market expectations that may not be fully captured by historical prices \cite{baker2007investor, de1990noise}. Prior studies show that sentiment-driven textual information significantly influences asset prices and trading behavior \cite{checkley2017hasty}. Motivated by this, we incorporate firm-level financial news headlines obtained from the \emph{Refinitiv Eikon} platform through the LSEG workspace API.

\textbf{Framework for measuring sentiment score:}

Since TDA requires quantitative inputs, textual financial news must first be transformed into numerical representations. To achieve this, sentiment extraction techniques are employed to quantify the polarity and intensity of sentiment expressed in news headlines.

Traditional dictionary-based approaches, such as the financial lexicon proposed by Loughran and McDonald \cite{loughran2011liability}, may not adequately capture contextual information, nuanced tone, or sarcasm in financial text. To address these limitations, this study employs FinBERT \cite{araci2019finbert}, a transformer-based model developed by Prosus AI for financial sentiment analysis. FinBERT has shown strong performance in financial NLP applications and has been widely utilized in recent studies \cite{jun2024predicting, lee2025large, sidogi2021stock}. In addition, its computational efficiency makes it suitable for high-frequency and real-time financial data processing. Validation studies by Mantshimuli \cite{mantshimuli2025sentiment} further demonstrate its superior classification accuracy compared with alternative benchmark models.

The sentiment extraction procedure consists of four steps:  
(i) collecting firm-level financial news headlines from the Refinitiv platform,  
(ii) translating non-English headlines into English,  
(iii) classifying each headline using FinBERT to obtain posterior probabilities for positive, negative, and neutral sentiments, denoted by $P_{\text{pos}}$, $P_{\text{neg}}$, and $P_{\text{neu}}$, respectively, and  
(iv) constructing the sentiment score as
\[
S_t = P_{\text{pos}} - P_{\text{neg}}, \quad S_t \in [-1,1],
\]
where positive and negative values of $S_t$ indicate optimistic and pessimistic sentiment, respectively, and larger absolute values represent stronger sentiment intensity.

\begin{figure}
    \centering
    \includegraphics[width=0.6\linewidth]{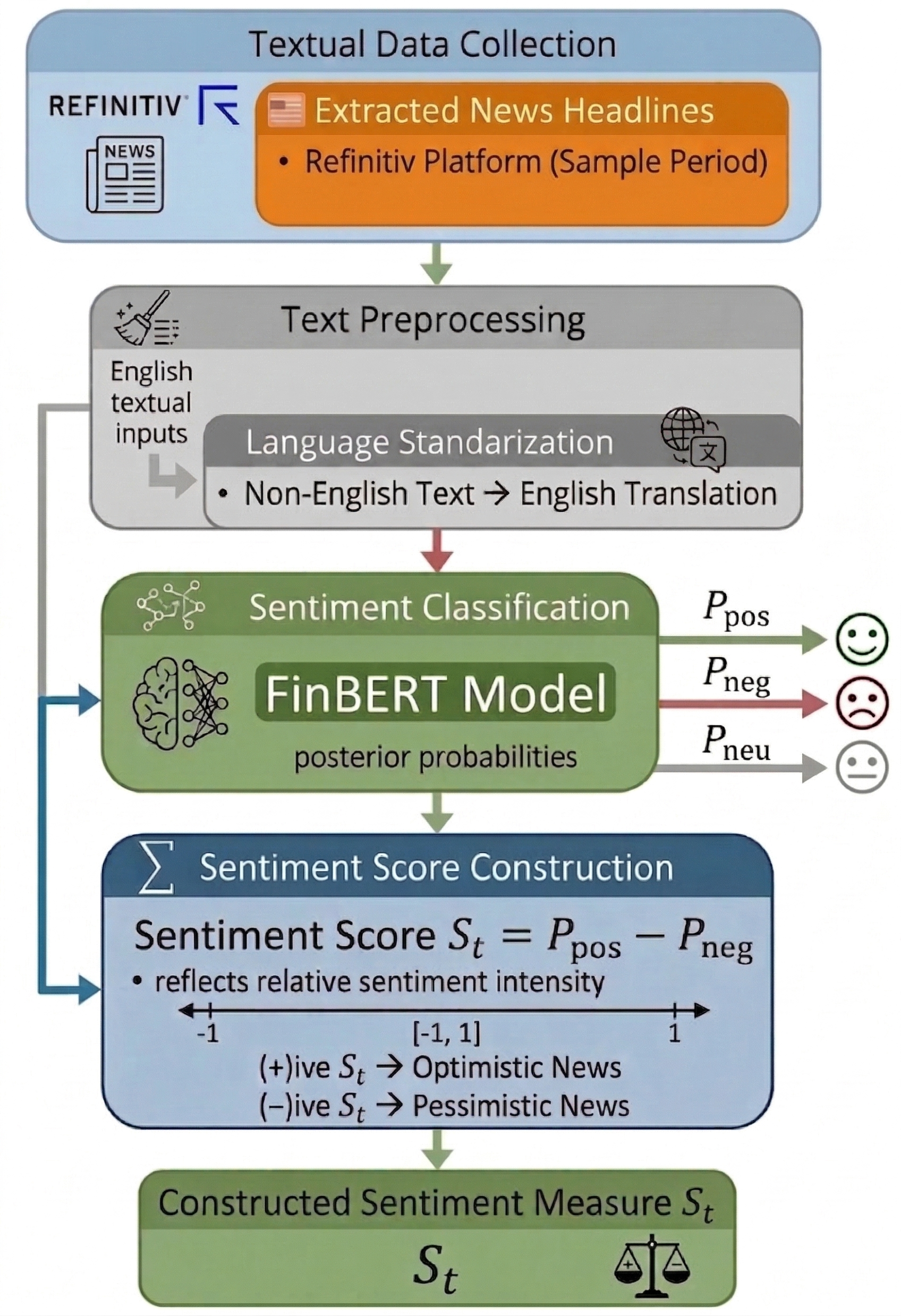}
    \caption{Flowchart for sentiment score computation from news headlines. Google Gemini was used to refine and enhance the visual presentation of a flowchart developed by the author. The resulting figure was subsequently reviewed, modified, and finalized by the author.
    }
    \label{fig: Sentiment score}
\end{figure}

The overall sentiment extraction framework is illustrated in Figure \ref{fig: Sentiment score}. For each asset, the daily sentiment score is computed as the average sentiment score of all associated news headlines that day. To ensure consistency with trading data, sentiment values from non-trading days are reassigned to the immediately following trading day. By integrating sentiment scores with RSI, SO, and MACD, a four-dimensional representation is constructed for each asset.

% \paragraph{Rationale for using FinBERT:}
% FinBERT is selected due to its domain-specific financial training, ability to capture contextual semantics through transformer-based embeddings, and strong empirical performance in financial sentiment analysis.

\textbf{Data preprocessing:}
Before clustering and topological analysis, all input variables are standardized to ensure equal contribution in distance computations. Since RSI, SO, MACD, and sentiment scores are measured on different scales, standardization transforms them to a comparable scale and prevents the clustering process from being biased toward features with larger numerical magnitudes.

\section{Topological data analysis}
A key requirement of TDA is that the data must be represented as a point cloud in a multidimensional space. In financial applications, however, portfolio construction is often based only on return data, which forms a one-dimensional time series and does not naturally provide such a representation. Therefore, existing studies \cite{goel2025sparse, goel2025risk, goel2020topological} commonly use Takens’ embedding \cite{takens2006detecting} to reconstruct a higher-dimensional space from a single time series. Although theoretically sound, this approach generates an artificial point cloud and depends heavily on heuristic choices of embedding parameters such as delay and embedding dimension.  The choice of these parameters may influence the extracted topological features. In contrast, our approach constructs the point cloud directly from multiple economically meaningful time-series features. Each observation consists of three technical indicators (RSI, SO, and MACD) along with a sentiment indicator derived from FinBERT, resulting in a natural four-dimensional representation of a point cloud. This enables the extraction of topological structures directly from multidimensional market and behavioral information, avoiding the need for artificial embedding.

After constructing the point cloud representation, the next step is to build an appropriate simplicial complex. A \(m\)-simplex\footnote{A \(m\)-simplex $[u_0, u_1, \ldots, u_m]$ is a convex hull $\left\{ \sum_{i=0}^{m} \eta_i u_i \;\middle|\; \sum_{i=0}^{m} \eta_i = 1 \text{ and } \eta_i \geq 0 \text{ for all } i \right\}$ of \(m + 1\) geometrically independent points \(\{u_0, u_1, \ldots, u_m\}\).} is defined as the convex hull of \(m+1\) affinely independent points \(\{u_0, u_1, \ldots, u_m\}\), where each \(u_i \in \mathbb{R}^q\). A simplicial complex \(C\) is a collection of simplices satisfying two conditions: (i) Every face\footnote{The faces of a \(k\)-simplex \([u_0, \ldots, u_k]\) are the \((k-1)\)-simplices spanned by subsets of \(\{u_0, \ldots, u_k\}\). } of a simplex in \(C\) is also contained in \(C\); that is, if \(\sigma \in C\) and \(\tau \subseteq \sigma\), then \(\tau \in C\); (ii) The intersection of any two simplices in \(C\) is either empty or a common face of both. Equivalently, a geometric simplicial complex in $\mathbb{R}^q$ is a collection of simplices for which every face of a simplex is also included and the intersection of any two simplices is either empty or a common face.

Basically, the simplicial complex of a point cloud $Z=\{Z_{t_1}\}_{{t_1}=1}^{T_1}, Z_{t_1} \in \mathbb{R}^q$ is the space comprising points, edges, triangles, tetrahedra, and higher-dimensional polytopes. Note that 0-dimensional simplices represent data points (vertices), 1-dimensional simplices are connected pairs of vertices (edges), and 2-dimensional simplices are filled triangles determined by their three vertices. Several constructions are available to build a simplicial complex, including the Čech complex \cite{bubenik2015statistical} and the Vietoris–Rips complex \cite{ghrist2008barcodes}. In this study, we adopt the Vietoris–Rips (Rips) complex due to its computational tractability and widespread use in applied TDA. The fundamental idea of the Rips complex is to connect sufficiently close points. This leads to the question of how “sufficient closeness” is determined. This notion is governed by a scale parameter $\epsilon > 0$, which determines whether pairs of points are connected. The choice of $\epsilon$ directly influences the resulting topological structure. If $\epsilon$ is too small, few or no edges are formed, and the complex consists primarily of isolated vertices, yielding limited structural information. Conversely, if $\epsilon$ is excessively large, all points become mutually connected, resulting in a single connected component and the disappearance of meaningful topological features.

Rather than selecting a single arbitrary value of $\epsilon$, TDA addresses this issue by examining the evolution of the simplicial complex across a continuum of scales. Specifically, we construct a sequence of Vietoris–Rips complexes indexed by an increasing set of scale parameters, forming a filtration $\{R(Z, \epsilon_n)\}_{n \in \mathbb{N}}$ where $Z \subset \mathbb{R}^q$ denotes the point cloud and ${\epsilon_n}$ is a non-decreasing sequence with $\epsilon_0 = 0$. This nested sequence satisfies \( R(Z, \epsilon_1) \subset R(Z, \epsilon_2) \) whenever \( \epsilon_1 < \epsilon_2 \), allowing the topological structure of the data to be studied as a function of scale. 

Formally, for a threshold $\epsilon>0$, the Vietoris--Rips complex $R(Z,\epsilon)$ contains all $m$-simplices whose vertices are pairwise within distance $2\epsilon$. In particular, an edge between two points $z_a,z_b\in Z$ is included if $d(z_a,z_b)<2\epsilon$. A triangle is formed when all pairwise distances among three vertices satisfy this condition. More generally, an $m$-simplex $\{z_0,\dots,z_m\}$ is included in $R(Z,\epsilon)$ whenever every pair of vertices lies within distance $2\epsilon$.

As $\epsilon$ increases, the collection $\{R(Z,\epsilon)\}$ forms a filtration, i.e., a nested sequence of simplicial complexes. Across this filtration, topological features such as connected components, loops, and higher-dimensional voids appear (birth) and disappear (death) as the scale parameter changes. Their disappearance occurs due to the addition of simplices that merge components or fill existing holes. Features that persist over a wide range of $\epsilon$ are considered robust and geometrically meaningful, as they reflect intrinsic structures of the point cloud, whereas short-lived features are generally treated as noise.

Given the filtration $\{R(Z,\epsilon)\}_{\epsilon>0}$, the homology groups $H_r(R(Z,\epsilon))$, denoted simply by $H_r$, for $r=0,1,2,\ldots$, are computed to capture different types of topological structure of the simplicial complex. In particular, $H_0$ represents connected components, describing how data points cluster as $\epsilon$ increases, while $H_1$ characterizes one-dimensional loops formed by closed cycles that are not filled by higher-dimensional simplices. Higher-order groups such as $H_2$ correspond to two-dimensional voids or cavities enclosed by simplices. As the scale parameter changes, these topological features may appear, merge, or disappear due to the addition of simplices in the filtration. This evolution of features across scales is known as \emph{persistent homology}, where each feature is identified by its birth and death scales, providing a multi-scale representation of the underlying topological structure of the data.

For a fixed homological dimension $r$, each topological feature is characterized by a birth time $b$ and a death time $d$, representing the values of $\epsilon$ at which the feature appears and disappears, respectively. The set of all such birth--death pairs forms the PD $\mathcal{D}_Z$, defined as a multiset of points in $W =\{(b,d)\in\mathbb{R}^2 \mid 0\leq b\leq d\}$. Each point $(b,d)$ in the diagram corresponds to a topological feature that is created at scale $b$ and vanishes at scale $d$ during the filtration.

The primary aim of applying TDA is to uncover the intrinsic geometric structure of the four-dimensional point cloud formed from technical and sentiment features through PH over multiple scales of the Vietoris--Rips filtration. %For each homological dimension $r\geq0$, the Betti number $\beta_r$ counts the number of independent $r$-dimensional topological features, where $\beta_0$, $\beta_1$, and $\beta_2$ represent connected components, loops, and voids, respectively. These topological measures capture nonlinear geometric patterns in the data that may not be identified using conventional distance-based methods such as correlation and Euclidean distances. 
By constructing the filtration and analyzing its PH, a multiscale representation %of the Betti numbers 
is obtained. In this study, we mainly consider $r=0$ and $r=1$, corresponding to connected components and loops, respectively. To compare the topological structures of different assets, a distance measure between PDs is required. For this purpose, we employ the $p$-Wasserstein distance, which quantifies the optimal matching cost between the birth--death pairs of two PDs.

To analyze PD quantitatively, a suitable distance metric is required. One of the most commonly used measures for comparing PD is the $p$-Wasserstein distance.

\begin{definition}
The $p$-Wasserstein distance \cite{cohen2010lipschitz} between two PDs $\mathcal{D}_{Z_i}$ and $\mathcal{D}_{Z_j}$ is defined as
\begin{equation}
W_p(\mathcal{D}_{Z_i},\mathcal{D}_{Z_j})
=
\left(
\inf_{\gamma}
\sum_{z\in\mathcal{D}_{Z_i}}
\|z-\gamma(z)\|_\infty^p
\right)^{1/p},
\end{equation}
where $p\geq1$, $\gamma$ is a bijection between the diagrams, and $\|\cdot\|_\infty$ denotes the $L^\infty$ norm.
\end{definition}

Although PD captures rich topological information, its metric space under the Wasserstein distance is incomplete, limiting direct statistical and machine learning analysis \cite{bubenik2015statistical}. To overcome this, PLs were introduced as a functional representation of PDs. A PLs map birth--death pairs into continuous piecewise-linear functions, where significant topological features appear as prominent peaks.

\begin{definition}
Given birth--death pairs $(b_i,d_i)$, $i=1,\ldots,m$, the PLs are defined as
\[
\zeta(k,t)=\zeta_k(t), \quad  \zeta:\mathbb{N}\times\mathbb{R}\rightarrow[0,\infty),
\]
where $\zeta_k(t)$ is the $k$-th largest value among $\{\Lambda_{(b_i,d_i)}(t)\}_{i=1}^m$, and $\zeta_k(t)=0$ for $k>m$. The function $\Lambda_{(b,d)}(t)$ is defined as
\begin{equation}
\Lambda_{(b,d)}(t)=
\begin{cases}
t-b, & t\in\left[b,\frac{b+d}{2}\right],\\
d-t, & t\in\left[\frac{b+d}{2},d\right],\\
0, & \text{otherwise}.
\end{cases}
\end{equation}
\end{definition}

To construct a PL, each birth--death pair $(b,d)$ in the PD is transformed into a triangular function with center $(b+d)/2$ and height $(d-b)/2$. These functions are stacked to form a sequence of piecewise-linear functions, where the $k$-th landscape function $\zeta_k(t)$ is defined as the $k$-th largest value among all triangular functions at $t$.

PLs provide a functional representation of PDs, enabling the use of tools from functional analysis and statistical learning. In particular, PLs belong to the Banach space $L^p(\mathbb{N}\times\mathbb{R})$, consisting of sequences of functions $\zeta=(\zeta_k)_{k\in\mathbb{N}}$, where each $\zeta_k:\mathbb{R}\rightarrow\mathbb{R}$ denotes the $k$-th PL. To quantify the magnitude of a PL and to compare landscapes obtained from different datasets, the $L^p$ norm is commonly used:
\begin{equation}
\|\zeta\|_p
=
\left(
\sum_{k=1}^{\infty}\|\zeta_k\|_p^p
\right)^{1/p},
\end{equation}
where $\|\cdot\|_p$ is the standard $L^p$ norm. Equipped with this norm, the space of PLs becomes a Banach space. The norm provides a numerical summary of the PLs and measures the overall prominence of topological features in the data. In practice, the cases $p=1$ and $p=2$ are most commonly used, allowing PLs to be treated as functional data objects for statistical analysis such as averaging and hypothesis testing. However, the mapping from PD to PL is generally not one-to-one, meaning that different PD may yield the same PL.

\section{Asset selection procedure}\label{sec: TDA + clustering}
\subsection{Similarity measure between assets}\label{Sec : distance}
Since TDA operates on point clouds in multi-dimensional spaces, consider a $d$-dimensional time series $Z=\{z_t\in\mathbb{R}^d\}_{t=1}^{T_1}$, where $d=4$ in our case and $T_1$ denotes the in-sample period. The corresponding point cloud representation is
\begin{equation}
Z=
\begin{bmatrix}
z_1\\
z_2\\
\vdots\\
z_{T_1}
\end{bmatrix}
=
\begin{bmatrix}
z_1^{(1)} & z_1^{(2)} & z_1^{(3)} & z_1^{(4)}\\
z_2^{(1)} & z_2^{(2)} & z_2^{(3)} & z_2^{(4)}\\
\vdots & \vdots & \vdots & \vdots\\
z_{T_1}^{(1)} & z_{T_1}^{(2)} & z_{T_1}^{(3)} & z_{T_1}^{(4)}
\end{bmatrix},
\end{equation}
where each row corresponds to a $d$-dimensional observation at time $t$.

Treating the entire series as a single point cloud ignores temporal ordering, since persistent homology depends only on the geometric structure of the point cloud. Consequently, two series with similar geometric structures but different temporal dynamics may produce identical PD (for example, one being the time-reversal of the other). To preserve temporal information, we employ a sliding-window framework.

For a fixed window length $L\in\{1,\ldots,T_1\}$, the time series is divided into $K$ sub-series:
\begin{equation}\label{Eq: subseries}
\begin{aligned}
\mathrm{sub}(Z)&=\Big\{
(z_1,\ldots,z_L)_1,\;
(z_{1+s},\ldots,z_{L+s})_2,\;
\ldots,\\ & (z_{T_1-L+1},\ldots,z_{T_1})_K
\Big\} \\
&= \Big\{ Z_1, Z_2, \ldots, Z_K \Big\}.
\end{aligned}
\end{equation}
where the step size $s\in\{1,\ldots,L\}$ satisfies $T_1-L=s(K-1)$. Here, $s=L$ corresponds to non-overlapping windows, while $s=1$ gives maximal overlap. For each sub-series $\{Z_k\}_{k=1}^K$, a PD $\mathcal{D}_{Z_k}$ is computed.

Given two $d$-dimensional time series $Z_i=\{z_{it}\}_{t=1}^{T_1}$ and $Z_j=\{z_{jt}\}_{t=1}^{T_1}$ corresponding to asset $i$ and $j$, their similarity is measured by aggregating the $p$-Wasserstein distances between the PDs of corresponding sub-series.

\begin{definition}
For $1\leq p<\infty$, the average $p$-Wasserstein distance is defined as
\begin{equation}\label{eq: AWD}
d_{\mathrm{AWD}}(Z_i,Z_j)
=
\sum_{k=1}^{K}
w_k\,W_p(\mathcal{D}_{Z_{ik}},\mathcal{D}_{Z_{jk}}),
\end{equation}
where $W_p(\mathcal{D}_{Z_{ik}},\mathcal{D}_{Z_{jk}})$ denotes the $p$-Wasserstein distance between the PDs of the $k$-th sub-series of $Z_i$ and $Z_j$, respectively. The weights satisfy $w_k>0$ and $\sum_{k=1}^{K}w_k=1$.
\end{definition}

This framework preserves both the geometric information captured by persistent homology and the temporal evolution of the series. A similar distance can be defined using PL. Let $\zeta_{Z_{ik}}$ and $\zeta_{Z_{jk}}$ denote the PL corresponding to the PD $\mathcal{D}_{Z_{ik}}$ and $\mathcal{D}_{Z_{jk}}$ of the $k$-th sub-series of $Z_i$ and $Z_j$, respectively.

\begin{definition}
For $1\leq p<\infty$, the average $p$-landscape distance between $Z_i$ and $Z_j$ is defined as
\begin{equation}\label{eq: ALD}
d_{\mathrm{ALD}}(Z_i,Z_j)
=
\sum_{k=1}^{K}
w_k
\left\|
\zeta_{Z_{ik}}-\zeta_{Z_{jk}}
\right\|_{p},
\end{equation}
where $\|\cdot\|_p$ denotes the $L^p$ norm, and $w_k>0$ with $\sum_{k=1}^{K}w_k=1$.
\end{definition}

In this study, we set $w_k=\frac{1}{K}$. The step-by-step procedures for computing $d_{\mathrm{AWD}}$ and $d_{\mathrm{ALD}}$ are summarized in Algorithms \ref{alg:pointcloud}, \ref{alg:rips}, and \ref{alg:persistence}. As benchmark measures, we also consider correlation-based and Euclidean-based distances for multivariate time series, denoted by $d_{\mathrm{AC}}$ and $d_{\mathrm{AE}}$, respectively\footnote{The definitions of these distances are provided in Appendix B.}. These benchmark measures are compared with the proposed TDA-based distances $d_{\mathrm{AWD}}$ and $d_{\mathrm{ALD}}$\footnote{Distances computed using the four-dimensional feature space (technical indicators and sentiment score) are denoted by AWDS, APLS, ACS, and AES, whereas those computed using only technical indicators are denoted by AWDI, APLI, ACI, and AEI.}. 

\begin{figure}
    \centering
    \includegraphics[width=0.9\linewidth]{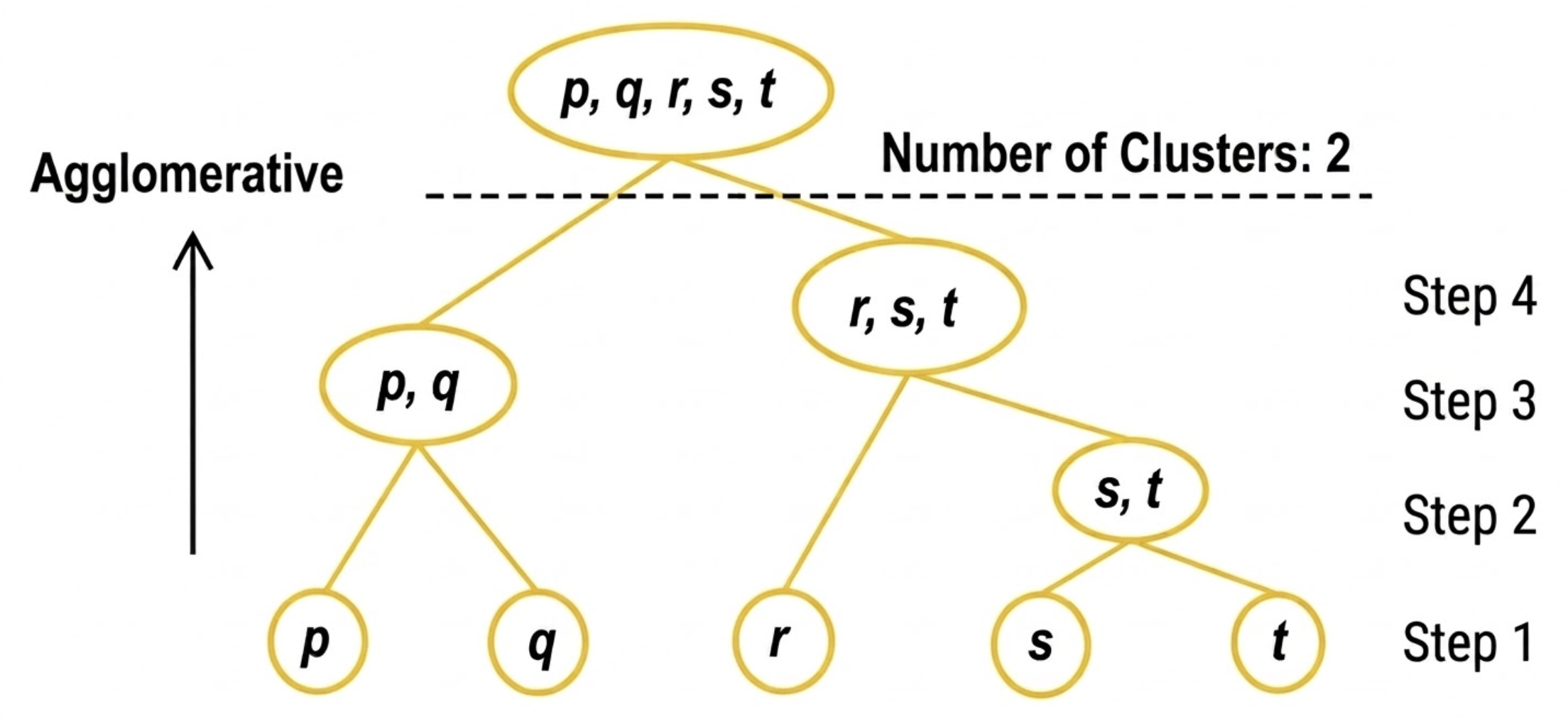}
    \caption{Illustration of agglomerative clustering.}
    \label{fig: Hirarchical}
\end{figure}

To group assets based on similarity, we employ agglomerative clustering (see Figure \ref{fig: Hirarchical}) using these eight pairwise distance measures. The detailed clustering procedure is provided in Appendix C.

\subsection{Asset selection}\label{Sec: asset selection}

Let $\mathcal{N}=\{1,2,\ldots,n\}$ denote the set of available assets. At each rebalancing period $t$, assets are clustered using historical data, producing a selected asset set $\mathcal{S}_t\subseteq\mathcal{N}$ with cardinality $k_t=|\mathcal{S}_t|$. Since clustering results vary across rolling windows, $k_t$ is time-varying. The portfolio strategy is implemented using a rolling-window framework with in-sample period $T_1$ and out-of-sample period $T_2$. At each rebalancing step:

\begin{itemize}

\item Clustering is performed via agglomerative clustering described in Appendix C.

\item Since assets within the same cluster are expected to exhibit similar behavior and risk-return characteristics, selecting assets from different clusters helps improve portfolio diversification. Therefore, representative assets are selected from each cluster based on their Sharpe ratio (SR) computed over the in-sample period. Let $n_c$ denote the number of assets in cluster $c$. Clusters with $n_c=1$ are treated as outliers and excluded. For clusters with $n_c\geq2$, assets are ranked in descending order of SR, and the top $k_c=\left\lceil \rho n_c \right\rceil$ assets are selected, where $\rho$ denotes the selection proportion and $\lceil\cdot\rceil$ is the ceiling operator. In this study, $\rho=0.10$, implying that approximately the top $10\%$ of assets are selected from each cluster. The final selected set $\mathcal{S}_t$ is obtained by taking the union of selected assets across all clusters, promoting diversification while maintaining a sparse portfolio.

\item Since clustering outcomes vary across windows, the selected asset set may change substantially over time, leading to high portfolio turnover and transaction costs. To address this, we incorporate a retention mechanism using the previously selected asset set $\mathcal{S}_{t-1}$. Assets in $\mathcal{S}_{t-1}$ but not in $\mathcal{S}_t$ are re-evaluated based on their SR, and assets ranked within the next top 10\% are retained in the portfolio. This mechanism reduces unnecessary trading while preserving assets with stable historical performance.

\end{itemize}

\begin{figure}
    \centering
    \includegraphics[width=0.6\linewidth]{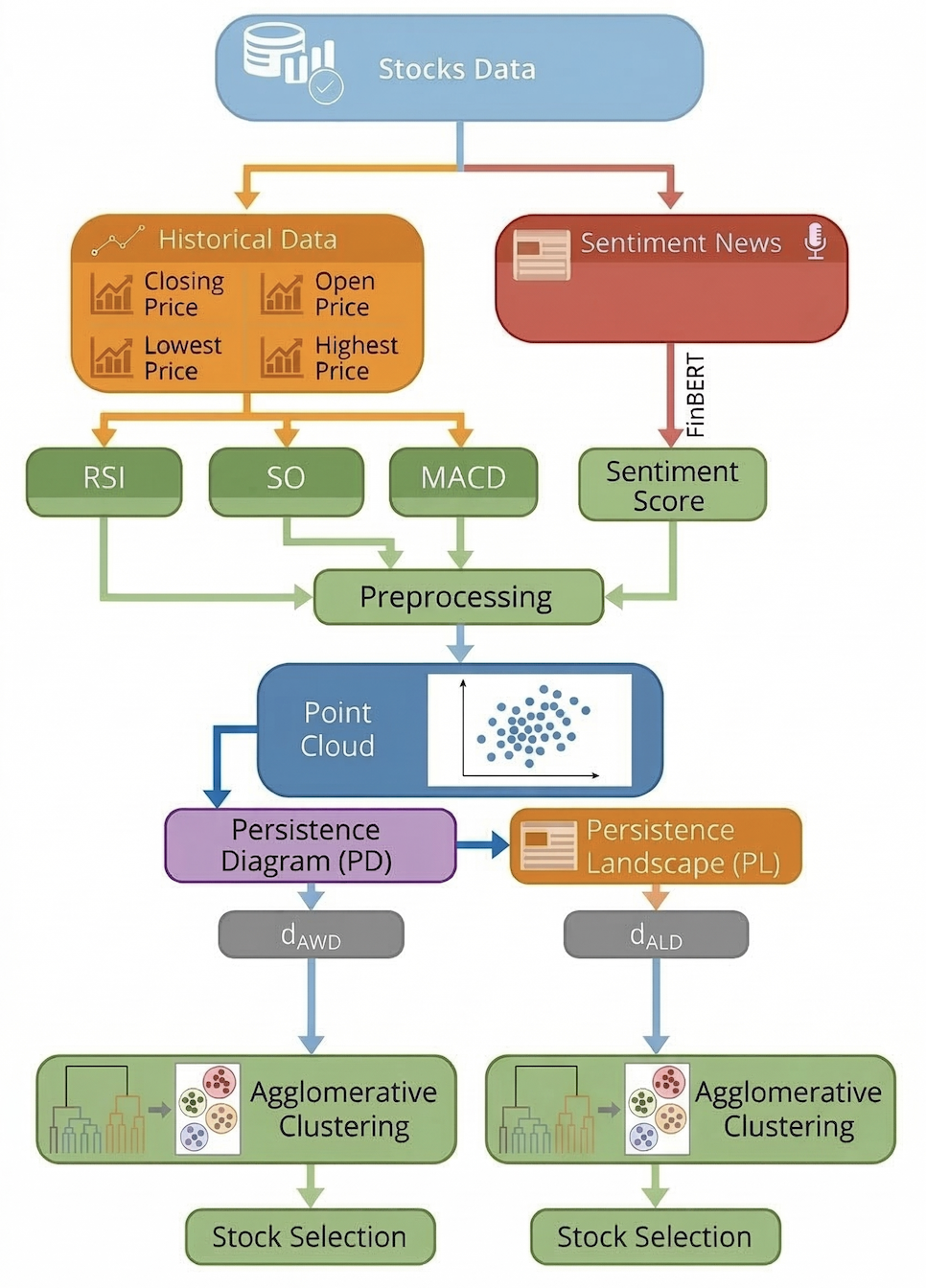}
    \caption{Flowchart of the proposed asset selection framework integrating data preprocessing, agglomerative clustering, and final asset selection. The initial flowchart concept was developed by the author and subsequently enhanced using Google Gemini to improve its visual presentation. The final figure was carefully reviewed and edited by the author.
    }
    \label{fig: Point cloud/asset selection}
\end{figure}

Figure \ref{fig: Point cloud/asset selection} illustrates the overall asset selection framework, from data collection and preprocessing to clustering and final asset selection.

\section{Portfolio formation and trading strategy}\label{Sec: PO model}
Assets are selected at the end of each in-sample period using the agglomerative clustering framework and distance measures described in Section \ref{sec: TDA + clustering}. This selection process reduces the dimensionality of the investment universe and identifies representative assets for portfolio construction. The PO problem is then formulated over the selected asset set $\mathcal{S}_t$.

In this study, we employ a mean--variance framework to determine the optimal portfolio weights. Let $\bar{r}_t\in\mathbb{R}^{k_t}$ and $C_t\in\mathbb{R}^{k_t\times k_t}$ denote the expected return vector and covariance matrix of the selected assets, respectively, where $k_t=|\mathcal{S}_t|$. Based on these quantities, we propose a dynamic rebalancing PO framework that combines clustering-based asset selection with mean--variance optimization, while incorporating short-selling constraints and transaction costs within the budget constraint. The model is solved repeatedly in a rolling-window setting to adapt the portfolio to changing market conditions. Given the selected asset set $\mathcal{S}_t$, the portfolio rebalancing model following \cite{yu2011portfolio,yu2017incorporating} is formulated as follows:

\begin{figure}[h]
\begin{align}
\max \quad
&   \bar{r}_t^\top (w_t^{+}-w_t^{-}) - \lambda (w_t^{+}-w_t^{-})^\top C_t (w_t^{+}-w_t^{-}) \notag \\
& - \mathbf{1}^\top (p_1 l_t^{+}
+ p_2 l_t^{-}
+ p_3 s_t^{+}
+ p_4 s_t^{-}) \tag{DRMV}\\
\text{s.t.} \quad
& w_t^{+} = w_{t-1}^{+} + l_t^{+} - l_t^{-}
\label{eq:longupdate}
\\
& w_t^{-} = w_{t-1}^{-} + s_t^{+} - s_t^{-}
\label{eq:shortupdate}
\\
& m^+ u_t \le w_t^{+} \le M^+ u_t
\label{eq:longbound}
\\
& m^- v_t \le w_t^{-} \le M^- v_t
\label{eq:shortbound}
\\
& u_t + v_t \le \mathbf{1}
\label{eq:exclusive}
\\
& \mathbf{1}^\top
\left(
w_t^{+} + k w_t^{-}
+ p_1 l_t^{+}
+ p_2 l_t^{-}
+ p_3 s_t^{+}
+ p_4 s_t^{-}
\right) = 1
\label{eq:budget}
\\
& u_t,v_t \in \{0,1\}^{k_t}
\\
& w_t^{+},w_t^{-},l_t^{+},l_t^{-},s_t^{+},s_t^{-}\in \mathbb{R}_+^{k_t},\\
& t = 1,\ldots, M. \notag
\end{align}
\end{figure}

Here, $\lambda\in[0,1]$ is the trade-off parameter balancing expected return and portfolio risk, and its optimal value is determined through a grid search over $[0,1]$. Further, $M$ denotes the total number of rolling windows (rebalancing periods) in the out-of-sample horizon.

The parameters $m^{+}$ and $M^{+}$ denote the lower and upper bounds on long positions, while $m^{-}$ and $M^{-}$ represent the corresponding bounds on short positions. These bounds control the minimum and maximum exposure to individual assets, preventing extremely small allocations and promoting diversification. The binary variables $u_t$ and $v_t$ indicate whether an asset is included in the long or short portfolio, respectively. Constraint (\ref{eq:exclusive}) ensures that an asset can be held either long or short, but not both simultaneously. If $u_{i,t}=v_{i,t}=0$, then asset $i$ is excluded from the portfolio.

\textbf{Decision variables}

The vectors $w_t^{+},w_t^{-}\in\mathbb{R}^{k_t}$ denote the long and short portfolio weights after rebalancing at time $t$, with net portfolio position $x_t=w_t^{+}-w_t^{-}$. Initially, the portfolio is assumed to be empty, i.e., $w_0^{+}=0$ and $w_0^{-}=0$. The vectors $l_t^{+}$ and $l_t^{-}$ represent purchases and sales from existing long positions, whereas $s_t^{+}$ and $s_t^{-}$ denote newly initiated short positions and repurchases of previously shorted assets. The binary vectors $u_t$ and $v_t$ determine asset inclusion in long and short portfolios, respectively.

\textbf{Transaction costs}

Transaction costs are incorporated using parameters $p_1,p_2,p_3,$ and $p_4$ within the budget constraint, which represent the proportional costs associated with buying assets, selling long positions, initiating short positions, and repurchasing assets to close short positions, respectively. Since the set of selected assets $\mathcal{S}_t$ obtained from clustering may vary across rolling windows, the number of selected assets $k_t = |\mathcal{S}_t|$ can differ from that in the previous period $k_{t-1}$. As a result, the portfolio composition needs to be adjusted at each rebalancing step. However, small variations in clustering outcomes across consecutive windows may lead to frequent changes in the selected asset set, resulting in excessive portfolio turnover and increased transaction costs. To address this issue, we incorporate a continuity mechanism consistent with the asset selection procedure described in Section~\ref{Sec: asset selection}. Portfolio adjustments are categorized into three cases:

\begin{enumerate}

\item \textbf{Common assets:}  
If an asset $i$ belongs to both $\mathcal{S}_{t-1}$ and $\mathcal{S}_t$, its previous long and short weights are carried forward and used as the initial positions for the current optimization problem. These weights correspond to $w_{t-1}^{+}$ and $w_{t-1}^{-}$ in the rebalancing equations.

% \item \textbf{Dropped assets:}  
% If $i\in\mathcal{S}_{t-1}$ but $i\notin\mathcal{S}_t$, existing positions must be liquidated. Let $\mathcal{D}_t=\mathcal{S}_{t-1}\setminus\mathcal{S}_t$ denote the set of dropped assets at the end of the rebalancing period $t$. The associated liquidation cost is
% \[
% TC_t^{\mathrm{drop}}
% =
% \sum_{i\in\mathcal{D}_t}
% \left(
% p_2 w_{i,t-1}^{+}
% +
% p_4 w_{i,t-1}^{-}
% \right),
% \]
% where $w_{i,t-1}^{+}$ and $w_{i,t-1}^{-}$ denote the long and short portfolio weights of asset $i$ held at the previous rebalancing period, respectively. These costs are deducted from the portfolio value and added in total transaction cost during performance evaluation.

\item \textbf{Dropped assets:} If $i\in\mathcal{S}_{t-1}$ but $i\notin\mathcal{S}_t$, the asset is removed from the investable universe. To ensure proper liquidation, the asset is temporarily retained in the optimization at rebalancing period $t$ with its portfolio weights fixed at
\[
w_{i,t}^{+}=0,\qquad
w_{i,t}^{-}=0.
\]
Consequently, no new long or short positions are initiated,
\[
l_{i,t}^{+}=0,\qquad
s_{i,t}^{+}=0,
\]
while the existing positions are fully liquidated,
\[
l_{i,t}^{-}=w_{i,t-1}^{+},\qquad
s_{i,t}^{-}=w_{i,t-1}^{-}.
\]
Accordingly, the associated transaction costs are automatically accounted for through the (DRMV) model.

% and any existing position must be fully liquidated. Consequently, no new long or short positions are initiated, 
% \[ 
% l_{i,t}^{+}=0,\qquad s_{i,t}^{+}=0, 
% \] 
% while the existing positions are completely closed, 
% \[ 
% l_{i,t}^{-}=w_{i,t-1}^{+},\qquad s_{i,t}^{-}=w_{i,t-1}^{-}, 
% \] 
% resulting in 
% \[ 
% w_{i,t}^{+}=0,\qquad w_{i,t}^{-}=0. 
% \] 
% Let \[ \mathcal{D}_t=\mathcal{S}_{t-1}\setminus\mathcal{S}_t \] denote the set of dropped assets at rebalancing period $t$. The corresponding liquidation transaction cost is \[ TC_t^{\mathrm{drop}} = \sum_{i\in\mathcal{D}_t} \left( p_2 w_{i,t-1}^{+} + p_4 w_{i,t-1}^{-} \right), \] where $p_2$ and $p_4$ denote the proportional transaction costs for selling long positions and repurchasing short positions, respectively. This liquidation cost is deducted from the portfolio value and included in the total transaction cost during the out-of-sample performance evaluation.

\item \textbf{Newly selected assets:}  
If $i\in\mathcal{S}_t$ but $i\notin\mathcal{S}_{t-1}$, it has no existing position carried over from the previous rebalancing period. Therefore, its long and short holdings for period $t-1$ are set to zero, i.e., $w_{i,t-1}^{+}=0$ and $w_{i,t-1}^{-}=0$. The optimal portfolio allocation for such assets is then determined entirely by the optimization model at time $t$.

\end{enumerate}

This procedure aligns portfolio weights from consecutive windows and ensures consistent rebalancing under a dynamically changing asset universe.

\section{Experimental setup}\label{Sec: Experimental Setup}

\subsection{Sample data}

Firm-level financial news data are obtained from the LSEG Refinitiv Workspace news database, which provides high-frequency textual information on a daily basis. %Specifically, up to five news headlines per day are collected for each constituent stock of the S\&P500 Index over the period from January 1, 2024 to January 31, 2025 (approximately 250 trading days). 
Specifically, news headlines per day are collected for each constituent stock of the S\&P500 over the period from January 1, 2025 to December 31, 2025 (approximately 250 trading days). 
Corresponding market data, including daily opening, closing, highest, and lowest prices, are also collected for the same set of stocks.

Stocks with missing observations during the sample period are excluded to maintain data consistency, resulting in a final sample of $493$ assets. Sentiment scores derived from the news data are combined with technical indicators to construct a multidimensional feature representation for each asset. These features are subsequently used to generate point cloud representations, compute similarity measures between assets, and implement the clustering and asset selection framework described in Section~\ref{sec: TDA + clustering}.

\subsection{Rolling window with cross-validation framework}

\begin{figure}
    \centering
    \includegraphics[width=0.9\linewidth]{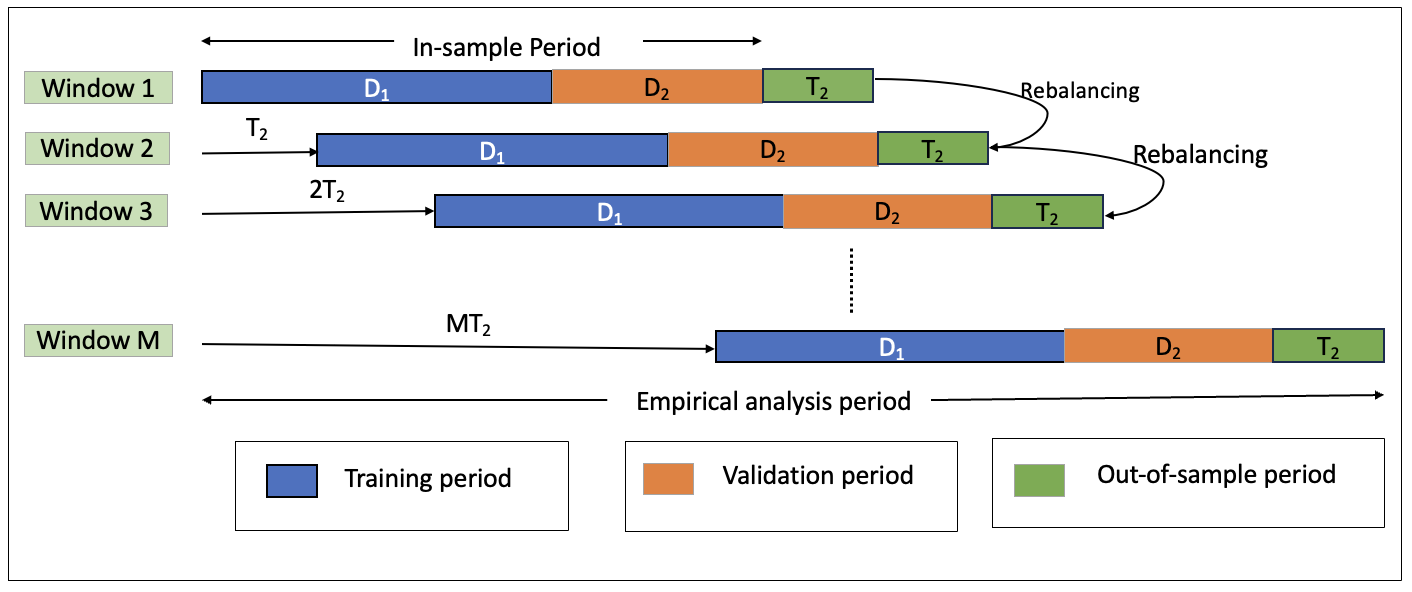}
    \caption{Rolling-window framework with cross-validation. The in-sample period is divided into a training period $D_1$ and a validation period $D_2$ for selecting the parameter $\lambda$ in the (DRMV) model. The optimized portfolio is then evaluated over the out-of-sample period with rebalancing every $T_2$ days.}
    \label{fig: rolling window}
\end{figure}

To simulate a realistic investment setting and avoid look-ahead bias, we employ a rolling-window framework (see Figure \ref{fig: rolling window}). The procedure is summarized as follows:

\begin{enumerate}

\item \textbf{Window construction:}  
At each step, a window\footnote{Throughout the paper, we consider \(T_1 = D_1 + D_2\), where \(D_1\) denotes the training period, \(D_2\) denotes the validation period, and \(T_1\) represents the overall in-sample period.} of length $D_1+D_2+T_2$ is considered, where the first $T_1=D_1+D_2$ observations form the in-sample period and the remaining $T_2$ observations are used for out-of-sample evaluation.

\item \textbf{Stock selection:}  
Representative stocks are selected from the in-sample period using the clustering framework described in Section~\ref{sec: TDA + clustering}.

\item \textbf{In-sample partitioning and parameter selection:}  
After stock selection, the in-sample period is divided into two consecutive segments for selecting the trade-off parameter $\lambda$:
\begin{itemize}

\item \textbf{Training period ($D_1$ days):}  
The first $D_1$ observations are used to estimate the (DRMV) model for the selected subset of stocks. A grid search is performed over a predefined set of candidate values $\lambda \in \{0.2, 0.4, 0.6, 0.8, 1\}$.

\item \textbf{Validation period ($D_2$ days):}  
The remaining $D_2$ observations are used to evaluate the portfolios corresponding to each candidate $\lambda$. The value maximizing the SR is selected for portfolio construction.

\end{itemize}

\item \textbf{Portfolio optimization:}  
Using the selected assets and optimal $\lambda$, the (DRMV) model is solved over the entire in-sample period to obtain the optimal portfolio weights.

\item \textbf{Out-of-sample evaluation:}  
The portfolio is held fixed over the next $T_2$ trading days, and its performance is evaluated on the out-of-sample period.

\item \textbf{Rolling update:}  
The window is shifted forward by $T_2$ days, and the entire procedure, including clustering, parameter selection, and PO, is repeated.

\item \textbf{Performance aggregation:}  
The process continues until the dataset is exhausted, and the overall strategy performance is obtained by aggregating the out-of-sample results across all rolling windows.

\end{enumerate}

The daily return of asset $i$ on day $t_1$ is computed as
\begin{equation*}
r_{it_1}
=
\frac{C_{i(t_1)}-C_{i(t_1-1)}}{C_{i(t_1-1)}},
\end{equation*}
where $C_{i(t_1)}$ denotes the closing price of asset $i$ on day $t_1$.

\subsection{Sample period and parameters}
\begin{table*}[h!]
\centering
\small
\caption{Number of rolling windows ($M$) obtained for different choices of in-sample period ($T_1$) and out-of-sample period ($T_2$).}
\label{tab:windows}
\begin{tabular}{c|cc}
\hline
\textbf{Out-of-sample Period ($T_2$)} 
& \multicolumn{2}{c}{\textbf{Number of Rolling Windows ($M$)}} \\
\cline{2-3}
& {$T_1 = 63$ (3 months)} & {$T_1 = 84$ (4 months)} \\
\hline
3  & 62 & 55 \\
5  & 37 & 33 \\
10 & 18 & 16 \\
\hline
\end{tabular}
\end{table*}

In the first rolling window, an in-sample period of $T_1=63$\footnote{For \(T_1=63\), the in-sample period consists of a training period \(D_1=42\) trading days and a validation period \(D_2=21\) trading days.} trading days (approximately three months) is used for portfolio construction, followed by an out-of-sample period of $T_2$ trading days for performance evaluation. To examine the effect of rebalancing frequency, we consider $T_2\in\{3,5,10\}$. The choice of relatively short out-of-sample horizons is motivated by the inclusion of sentiment-based features in our framework. Financial sentiment is inherently short-lived and tends to lose its predictive power over longer horizons. The number of rolling windows $M$ depends on the choice of $T_1$ and $T_2$.

Although historical price data of assets are readily available, collecting firm-level news headlines from the LSEG Refinitiv Workspace is computationally intensive due to API rate limits and large-scale data extraction requirements. Consequently, the empirical analysis is restricted to a one-year sample period, which still provides a sufficient number of rolling windows for evaluation. For robustness, we additionally consider a longer in-sample period of $T_1=84$\footnote{For \(T_1=84\), the in-sample period consists of a training period \(D_1=63\) trading days and a validation period \(D_2=21\) trading days.} trading days. Table~\ref{tab:windows} reports the number of rolling windows corresponding to different choices of $T_1$ and $T_2$.

Transaction costs are incorporated consistent with those commonly observed in the U.S. market \cite{yu2017incorporating} by setting all transaction cost parameters \((p_j,\; j=1,2,3,4)\) at 25 basis points of the trading value, with trading margin $k=100\%$. To ensure realistic portfolio allocations, lower and upper bounds are imposed on long and short positions. Specifically, for long positions, we set $m^{+}=5\%$ and $M^{+}=20\%$, while for short positions, we set $m^{-}=1\%$ and $M^{-}=5\%$. These settings, consistent with \cite{yu2011portfolio}, are intended to maintain diversification and limit excessive short exposure.

\subsection{Performance measures}\label{sub: Performance measures}

The portfolio performance is evaluated using out-of-sample data after calibrating the model on the corresponding in-sample period. Let $M$ denote the total number of rolling windows and $T_2$ the out-of-sample horizon. The daily out-of-sample portfolio return is computed as $r_{t_1}^{*} = \sum_{i=1}^{n} r_{it_1}x_i^{*},  t_1=1,\ldots,MT_2,$ where $x_i^{*}=w_i^{+*}-w_i^{-*}$ denotes the optimal portfolio weight of asset $i$, and $r_{it_1}$ is the realized return of asset $i$ on day $t_1$.

The out-of-sample performance is evaluated using the following measures:

\begin{enumerate}

\item \textbf{Mean:} The average out-of-sample portfolio return is computed as $\hat{\mu}_p
=
\frac{1}{MT_2}
\sum_{t_1=1}^{MT_2} r_{t_1}^{*}$.

\item \textbf{Standard deviation (SD):} The out-of-sample portfolio risk is measured using the SD of returns as $\hat{\sigma}
=
\sqrt{
\frac{
\sum_{t_1=1}^{MT_2}
(\hat{\mu}_p-r_{t_1}^{*})^2
}{MT_2}
}$.

\item \textbf{VaR:}  
VaR at the 95\% confidence level \cite{linsmeier1996risk}.

\item \textbf{CVaR:}  
CVaR at the 95\% confidence level \cite{rockafellar2002deviation}.

\item \textbf{Min:} The minimum out-of-sample portfolio return is given by $\min_{t_1=1,\ldots,MT_2} r_{t_1}^{*}.$

\item \textbf{Max:} The maximum out-of-sample portfolio return is given by $\max_{t_1=1,\ldots,MT_2} r_{t_1}^{*}$.

\item \textbf{Maximum drawdown (MDD):}  
Largest peak-to-trough decline in cumulative portfolio value.

\item \textbf{Average drawdown (ADD):}  
Average decline from previously attained peaks during the out-of-sample period.

\item \textbf{Sharpe ratio (SR):} The SR \cite{sharpe1998sharpe} measures excess return per unit of portfolio risk by 
\[
\mathrm{SR}
=
\frac{\hat{\mu}_p-r_f}{\hat{\sigma}},
\qquad \hat{\mu}_p \ge r_f,
\]
where $r_f$ denotes the risk-free rate.

\item \textbf{Stable tail-adjusted return ratio (STARR):} The STARR ratio measures excess return relative to downside tail risk:
\[
\mathrm{STARR}
=
\frac{\hat{\mu}_p-r_f}{\mathrm{CVaR}},
\qquad
\hat{\mu}_p>r_f.
\]

\item \textbf{Sterling ratio:} It measures excess return per unit of average drawdown risk:
\[
\mathrm{Sterling\ ratio}
=
\frac{\hat{\mu}_p-r_f}{\mathrm{ADD}},
\qquad
\hat{\mu}_p>r_f.
\]

\item \textbf{Cumulative returns:} 
Total portfolio growth over the out-of-sample horizon.

\end{enumerate}

We also conduct statistical significance tests to examine whether the out-of-sample performance of one strategy differs significantly from the other. In particular, we employ two statistical tests corresponding to mean return and SR, namely the t-test and Sharpe-test \cite{ledoit2008robust}.\footnote{Detailed formulations of these statistical tests are provided in Appendix D.}

\section{Empirical results}\label{Sec: Epirical results}

This section presents the empirical results of the proposed framework. We first analyze the clustering structures obtained from agglomerative clustering using different distance measures, namely AWD, ALD, AC, and AE distances, with and without the sentiment feature. These measures capture different notions of similarity among assets, ranging from topological structures to conventional statistical and geometric relationships. We then evaluate the out-of-sample portfolio performance of the proposed framework using the performance measures defined in Section \ref{sub: Performance measures}.

\subsection{Clustering characteristics}

To analyze the clustering structure, we employ agglomerative clustering with a percentile-based threshold determined from nearest-neighbor distances in the asset distance matrix. We report results for $\alpha=0.95$ and $0.99$, as these values produce a sufficient number of clusters for portfolio construction. Table \ref{tab: clustering results} presents the clustering results for four distance measures: AWD, APL, AC, and AE.

To examine the impact of sentiment information, we compare clustering results obtained using all four feature representation ($\mathrm{C_{AWDS}}$, $\mathrm{C_{APLS}}$, $\mathrm{C_{ACS}}$, and $\mathrm{C_{AES}}$) with those based only on technical indicators ($\mathrm{C_{AWDI}}$, $\mathrm{C_{APLI}}$, $\mathrm{C_{ACI}}$, and $\mathrm{C_{AEI}}$). The reported results correspond to an in-sample period of 63 days and an out-of-sample period of 3 days, yielding 62 rolling windows. All clustering statistics are averaged across these windows and rounded to the nearest integer.

Unlike k-means clustering, the proposed agglomerative clustering framework is able to identify outliers. Among the 493 stocks, the number of clustered stocks ranges from 452 to 480, while the number of outliers varies between 13 (2.63\%) and 41 (8.32\%). As the threshold level increases from $\alpha=95\%$ to $\alpha=99\%$, the number of outliers decreases, consistent with the fact that more clusters merge at higher threshold levels. The resulting cluster structure is well distributed and not overly concentrated. Across all specifications, the largest cluster contains between 62 and 92 stocks, corresponding to approximately 13\%--19\% of the total stock universe. The second and third largest clusters contain 25--69 and 19--42 stocks, respectively, indicating the presence of several meaningful clusters beyond the dominant group. Overall, the agglomerative clustering framework produces a balanced cluster structure across all distance measures.

\begin{figure*}[htbp]
    \centering
    \begin{minipage}{0.32\textwidth}
        \centering
        \includegraphics[width=\linewidth, height=3.5cm]{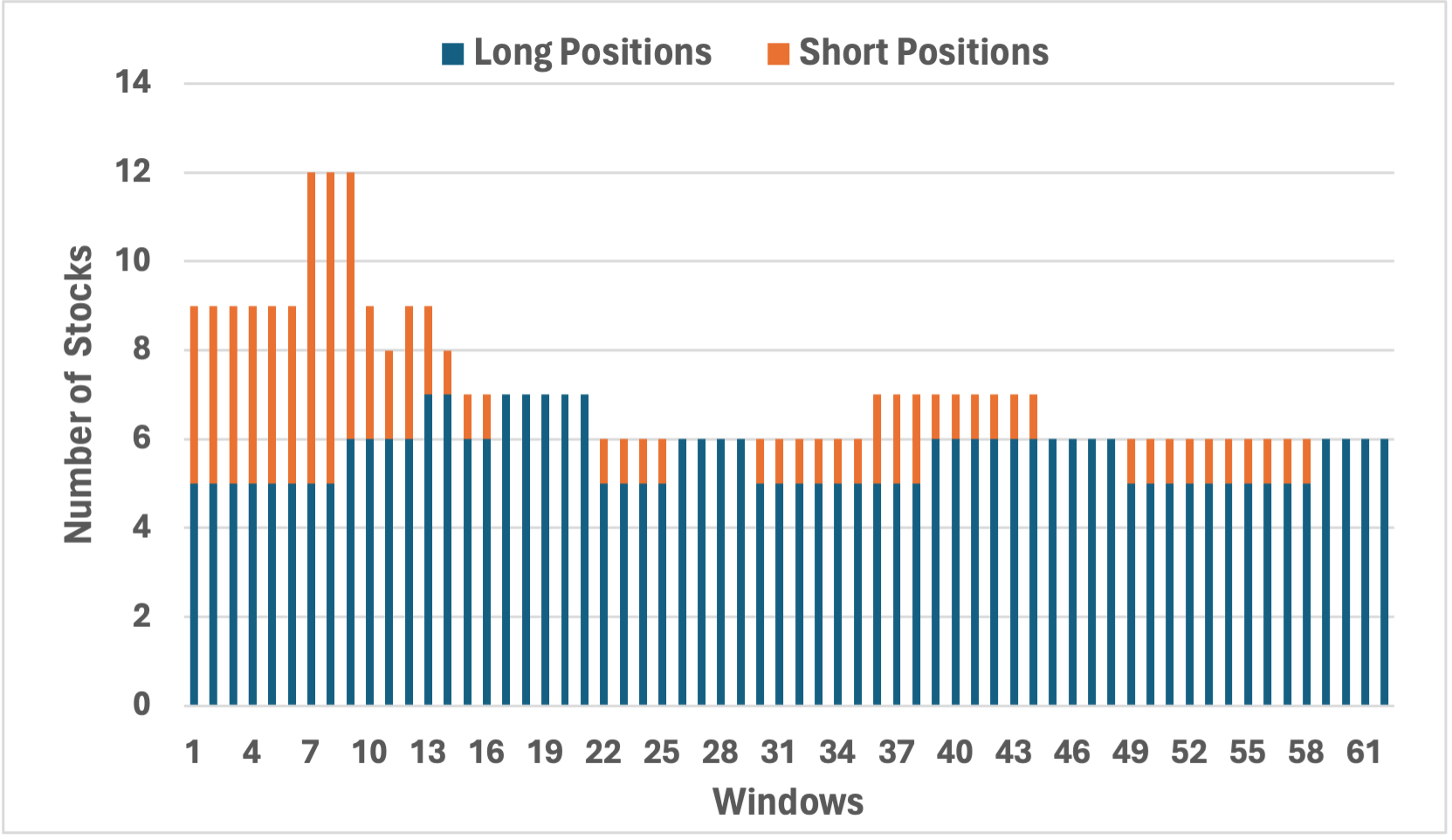}
    \end{minipage}
    \hfill
    \begin{minipage}{0.32\textwidth}
        \centering
        \includegraphics[width=\linewidth, height=3.5cm]{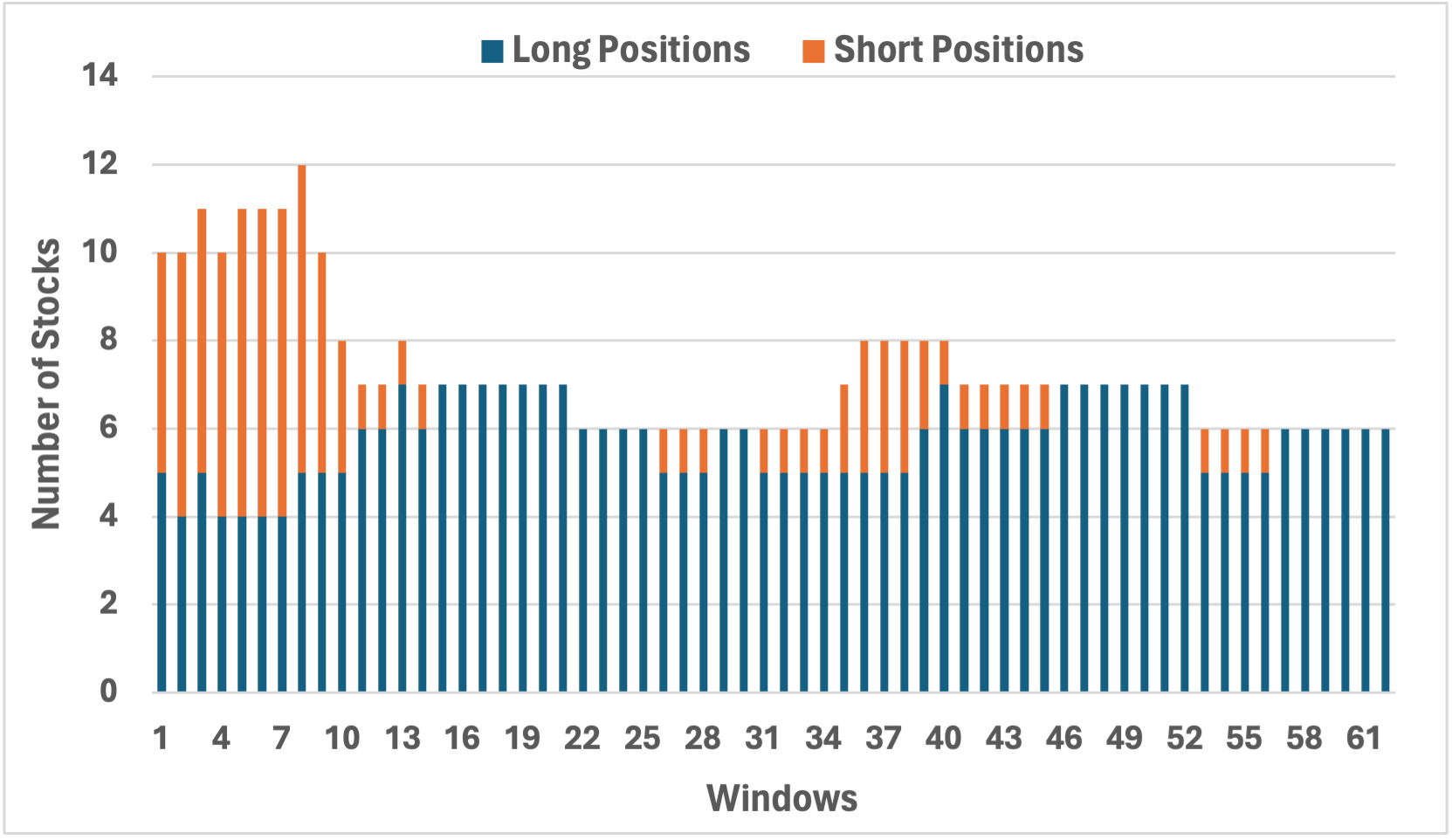}
    \end{minipage}
    \hfill
    \begin{minipage}{0.32\textwidth}
        \centering
        \includegraphics[width=\linewidth, height=3.5cm]{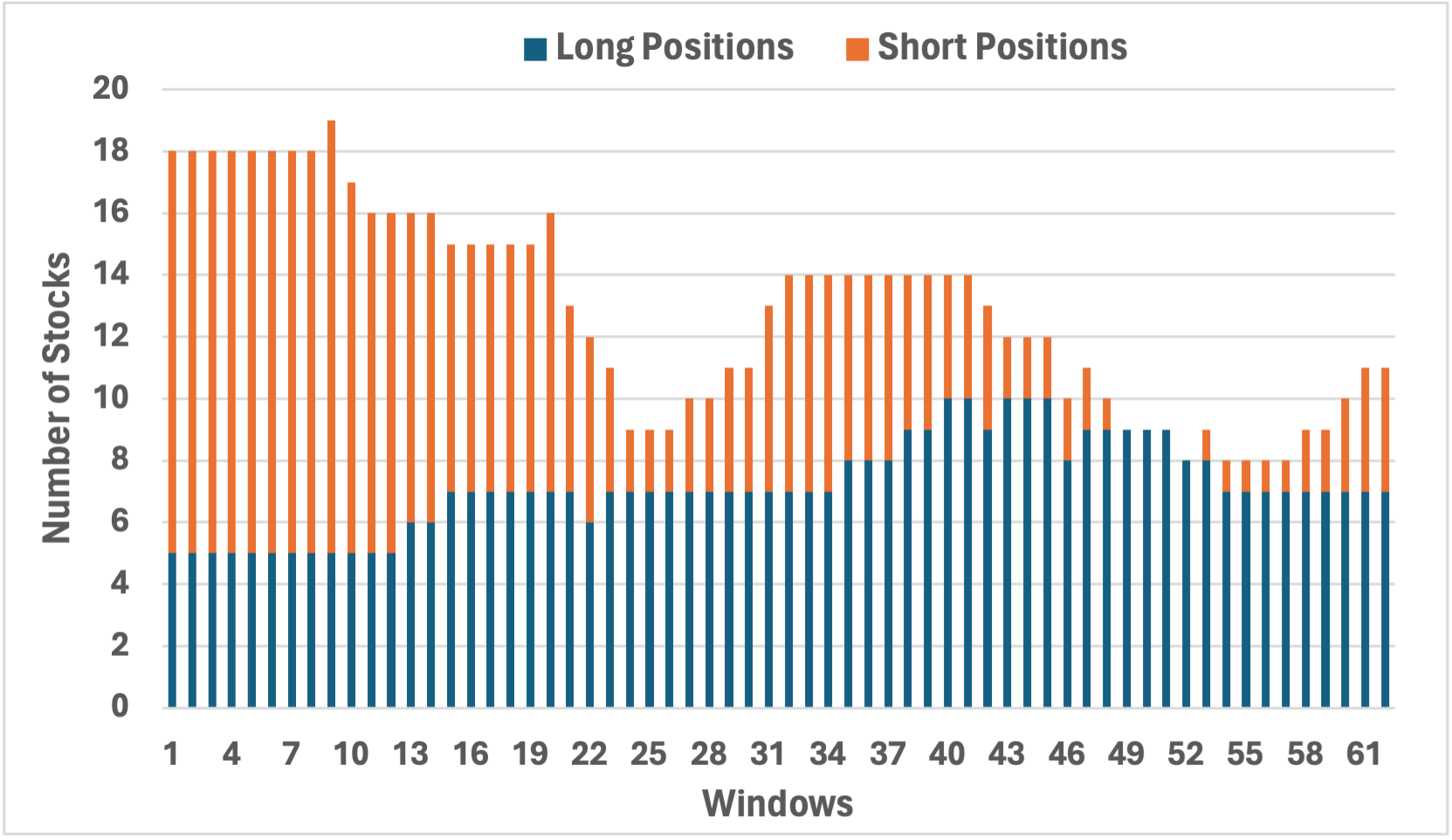}
    \end{minipage}
    \caption{Comparison of total number of long and short positions across rolling windows for an in-sample period of 63 trading days and an out-of-sample period of 3 trading days under the \(\mathrm{(DRMV)_{AWDS}}\) (left), the \(\mathrm{(DRMV)_{AWDI}}\) (center), and the \(\mathrm{(DRMV)_{A}}\) (right) models.}
    \label{fig: no. of stocks}
\end{figure*}

We also report the average numbers of traded, long, and short positions across all rolling windows for the eight clustering-based models and for $\mathrm{(DRMV)_A}$, where no clustering is performed. Since these averages do not fully capture the temporal variation in portfolio composition, Figure~\ref{fig: no. of stocks} illustrates the numbers of long and short positions across rolling windows for $\mathrm{(DRMV)_{AWDS}}$, $\mathrm{(DRMV)_{AWDI}}$, and $\mathrm{(DRMV)_A}$ under an in-sample period of 63 days and an out-of-sample period of 3 days.

It can be observed that, compared with the clustering-based models $\mathrm{(DRMV)_{AWDS}}$ and $\mathrm{(DRMV)_{AWDI}}$\footnote{Same trend is consistently observed across all other distance-based clustering methodologies.}, the model $\mathrm{(DRMV)_A}$ tends to trade a larger number of assets across rolling windows. Moreover, the number of traded assets under $\mathrm{(DRMV)_A}$ exhibits greater variation over time, indicating a less stable portfolio composition. This higher and more unstable trading activity likely contributes to the larger transaction costs incurred by $\mathrm{(DRMV)_A}$. The result is intuitive, as optimization over the full stock universe provides a larger set of candidate assets, leading to more frequent portfolio adjustments across rebalancing periods.

% \begin{figure*}[htbp]
%     \centering
%     \begin{subfigure}{0.45\textwidth}
%         \centering
%         \includegraphics[width=\linewidth]{3d_number of stocks_AWDS.png}
%         % \caption{$\mathrm{(DRMV)_{AWDS}}$}
%         \label{fig:graph1}
%     \end{subfigure}
%     \hfill
%     \begin{subfigure}{0.45\textwidth}
%         \centering
%         \includegraphics[width=\linewidth]{3d_number of stocks_AWDI.png}
%         % \caption{($\mathrm{(DRMV)_{AWDI}}$}
%         \label{fig:graph2}
%     \end{subfigure}
%     \hfill
%     \begin{subfigure}{0.45\textwidth}
%         \centering
%         \includegraphics[width=\linewidth]{3d_number of stocks_(DRMV)_A.png}
%     \end{subfigure}
%     \caption{Comparison of total number of long and short positions across rolling windows for an in-sample period of 63 trading days and an out-of-sample period of 3 trading days under the \(\mathrm{(DRMV)_{AWDS}}\) (top-left), the \(\mathrm{(DRMV)_{AWDI}}\) (top-right), and the \(\mathrm{(DRMV)_{A}}\) (bottom) models.}
%     \label{fig: no. of stocks}
% \end{figure*}

Table~\ref{tab:ari_results} reports the Average Adjusted Rand Index (AARI) \cite{zhang2012generalized} and Average Silhouette Score (ASS) \cite{rousseeuw1987silhouettes} for all eight clustering specifications under an in-sample period of 3 months and an out-of-sample period of 3 days, resulting in 62 rolling windows\footnote{The reported AARI and ASS values are averaged across all 62 rolling windows.}. Since the AARI measures the similarity of cluster assignments across consecutive windows, higher values indicate greater temporal stability. Across all specifications, the AARI values range from 0.70964 to 0.80948, indicating reasonably stable clustering structures over time.

Among the different distance measures, the correlation and Euclidean based methods generally yield higher AARI values, particularly for $\mathrm{C_{ACI}}$ and $\mathrm{C_{AEI}}$, suggesting more stable cluster assignments. In contrast, the TDA-based methods, especially $\mathrm{C_{AWDS}}$, produce relatively lower AARI values, indicating greater sensitivity to evolving and nonlinear market dynamics. Furthermore, incorporating sentiment information generally reduces the AARI values relative to the corresponding indicator-only specifications, suggesting that sentiment introduces additional time-varying information into the clustering process and produces more adaptive clustering structures that are responsive to changing market perceptions and short-term fluctuations.

The ASS values, which measure intra-cluster cohesion and inter-cluster separation, range from 0.22621 to 0.27710 across all specifications. Although moderate, these values still indicate meaningful clustering structures, which is expected in financial datasets \cite{leon2017clustering, cen2022financial} due to noisy, overlapping, and time-varying asset behavior. Among the different methods, the TDA-based specifications, particularly $\mathrm{C_{AWDS}}$ and $\mathrm{C_{APLS}}$, generally achieve higher ASS values, indicating better-separated clusters. In contrast, the correlation and Euclidean based methods, despite exhibiting greater temporal stability, tend to produce slightly lower ASS values. This is because these measures mainly capture linear dependencies or global geometric similarities, resulting in smoother but less distinct cluster boundaries. Overall, the results suggest a trade-off between clustering stability and clustering quality across different distance specifications.

\begin{figure*}[htbp]
\centering

% Row 1
\begin{subfigure}{0.23\textwidth}
    \centering
    \includegraphics[width=\linewidth]{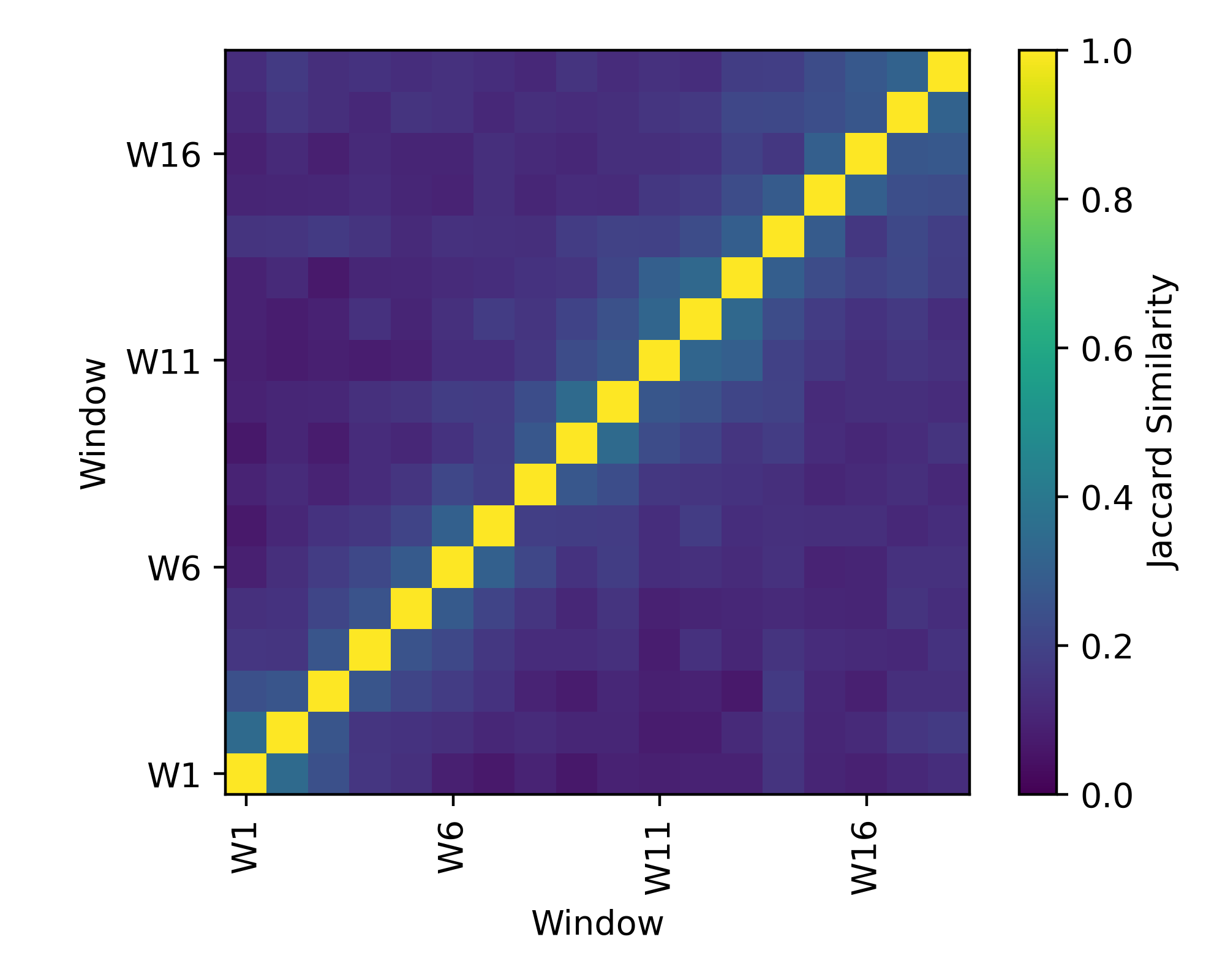}
    \caption{AWDS}
\end{subfigure}
\hfill
\begin{subfigure}{0.23\textwidth}
    \centering
    \includegraphics[width=\linewidth]{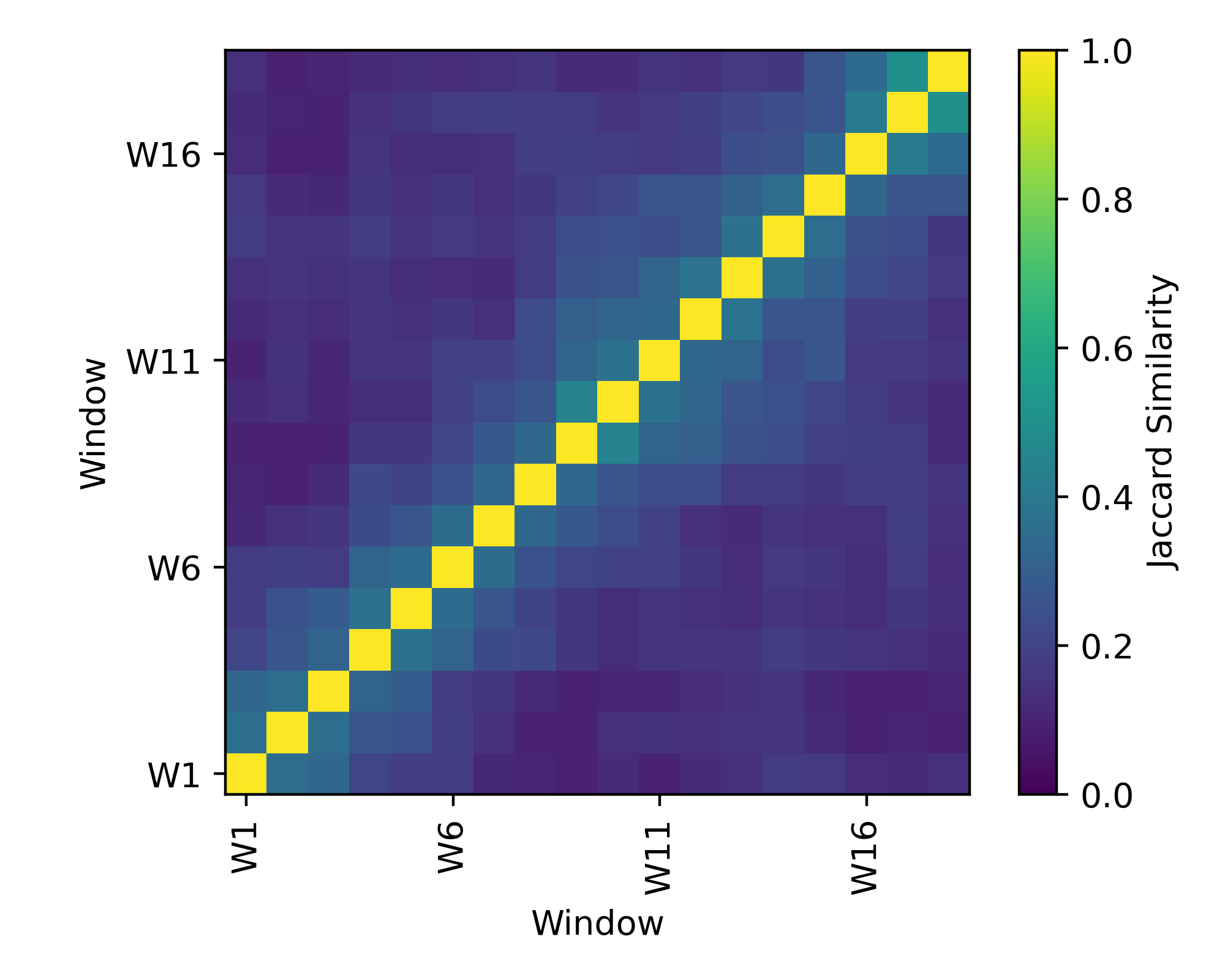}
    \caption{AWDI}
\end{subfigure}
\hfill
\begin{subfigure}{0.23\textwidth}
    \centering
    \includegraphics[width=\linewidth]{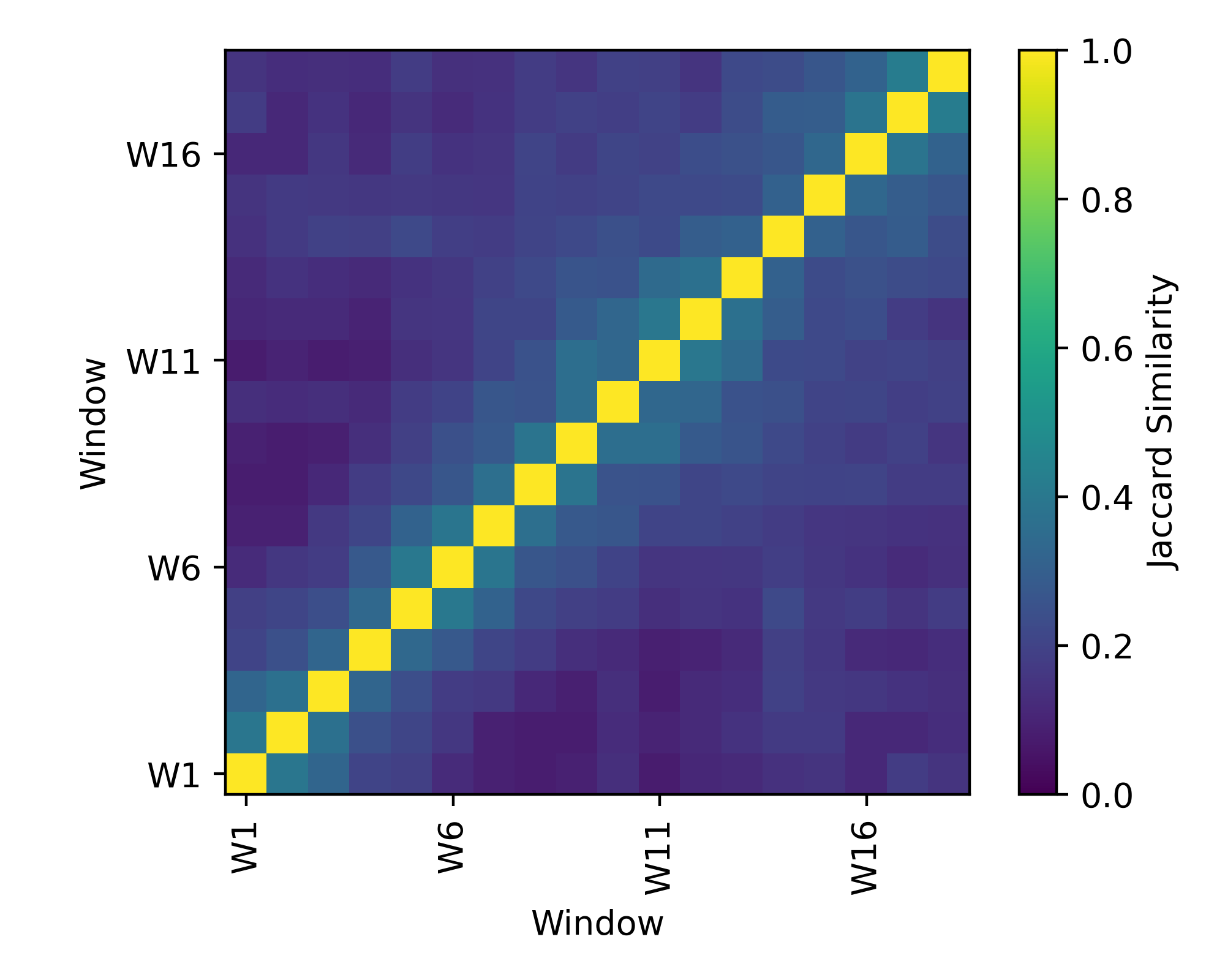}
    \caption{APLS}
\end{subfigure}
\hfill
\begin{subfigure}{0.23\textwidth}
    \centering
    \includegraphics[width=\linewidth]{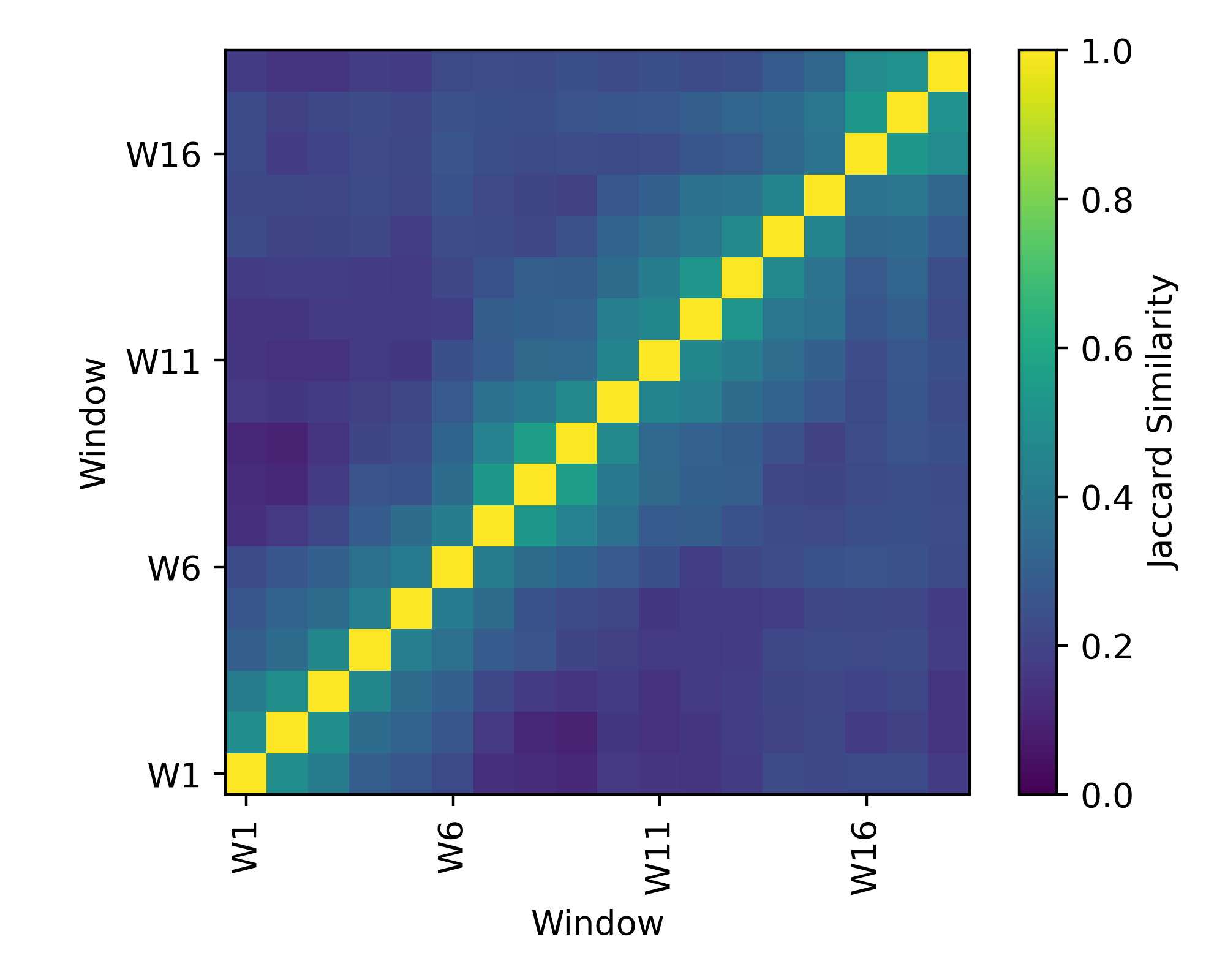}
    \caption{APLI}
\end{subfigure}

\vspace{0.3cm}

% Row 2
\begin{subfigure}{0.23\textwidth}
    \centering
    \includegraphics[width=\linewidth]{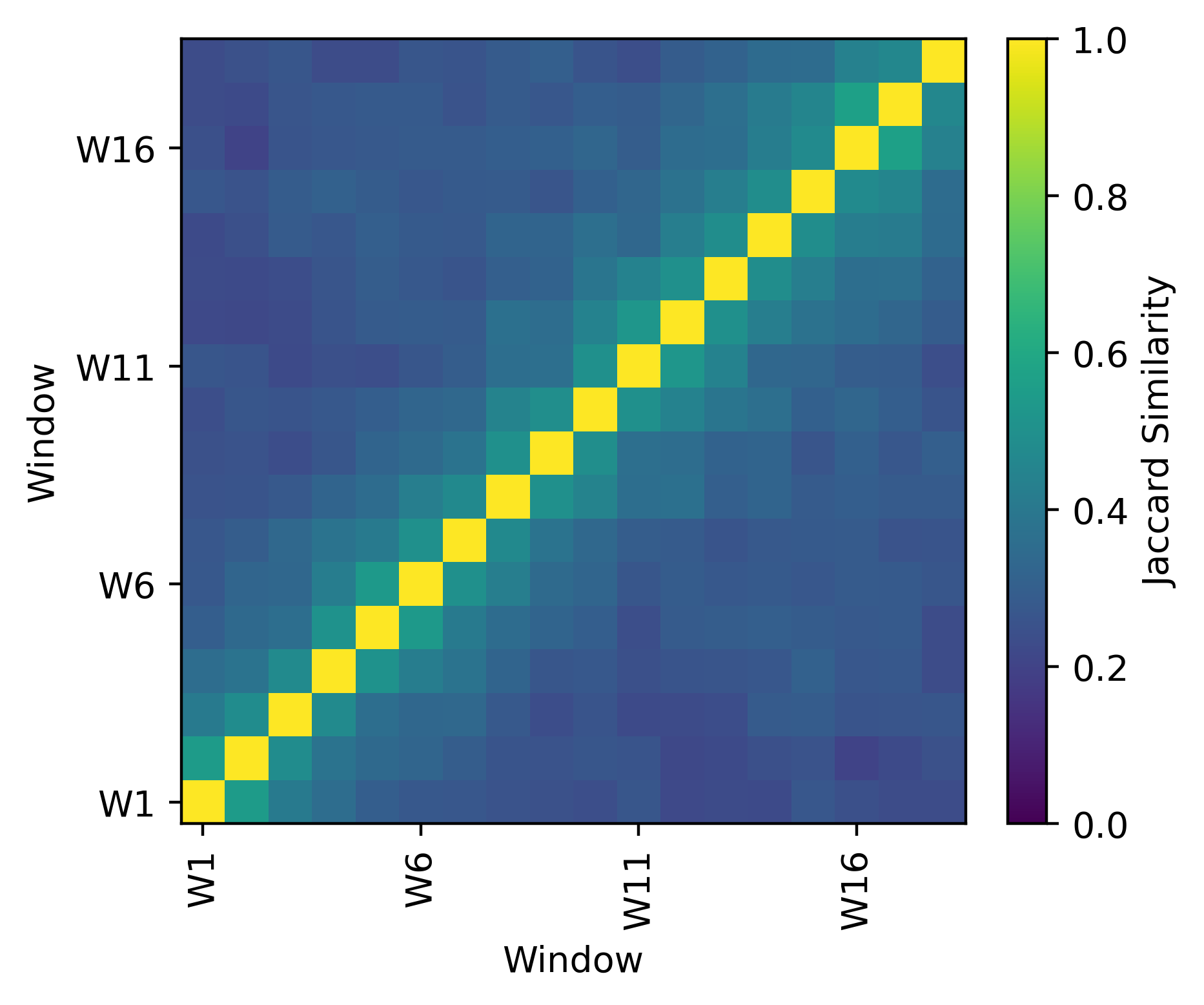}
    \caption{ACS}
\end{subfigure}
\hfill
\begin{subfigure}{0.23\textwidth}
    \centering
    \includegraphics[width=\linewidth]{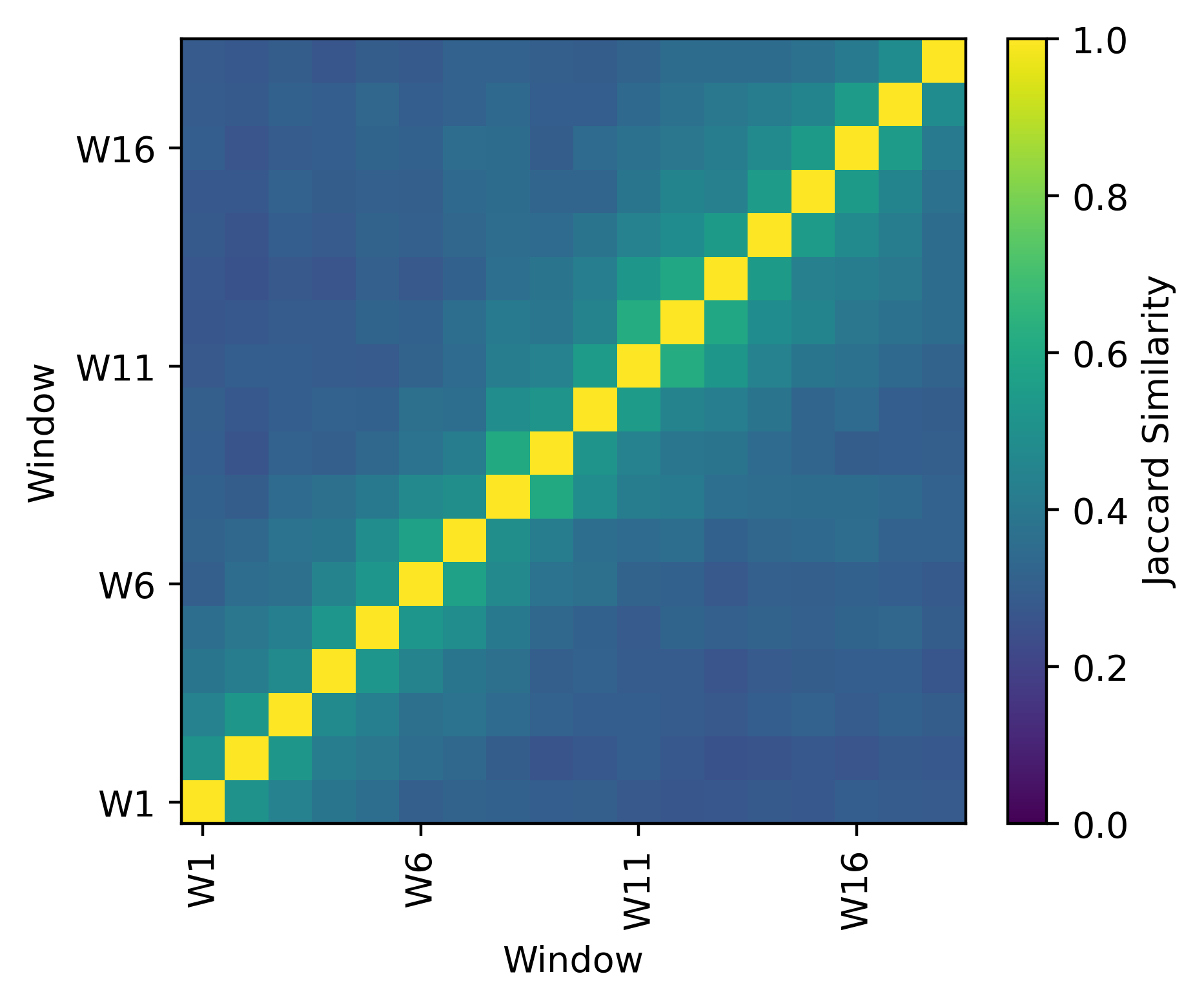}
    \caption{ACI}
\end{subfigure}
\hfill
\begin{subfigure}{0.23\textwidth}
    \centering
    \includegraphics[width=\linewidth]{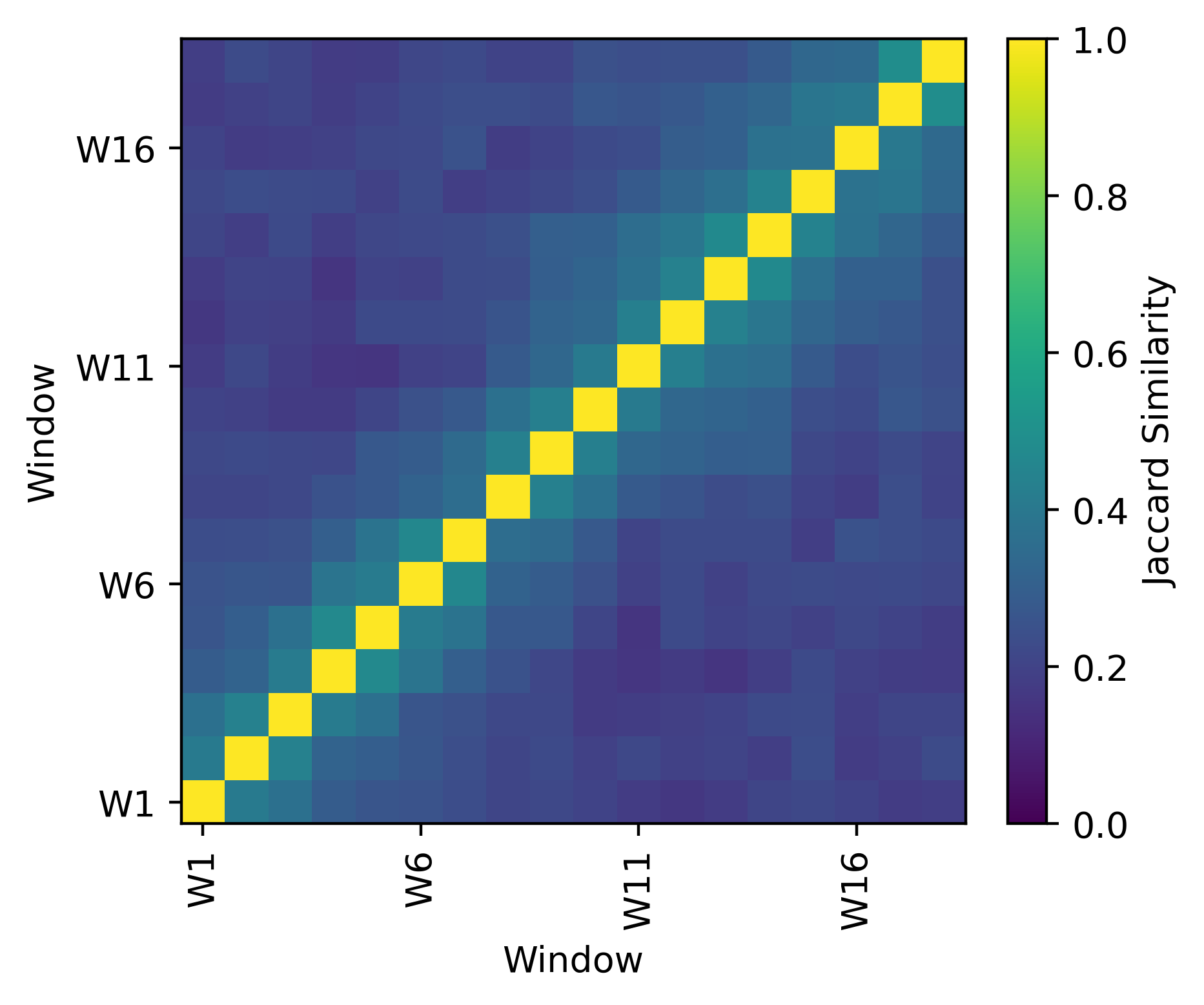}
    \caption{AES}
\end{subfigure}
\hfill
\begin{subfigure}{0.23\textwidth}
    \centering
    \includegraphics[width=\linewidth]{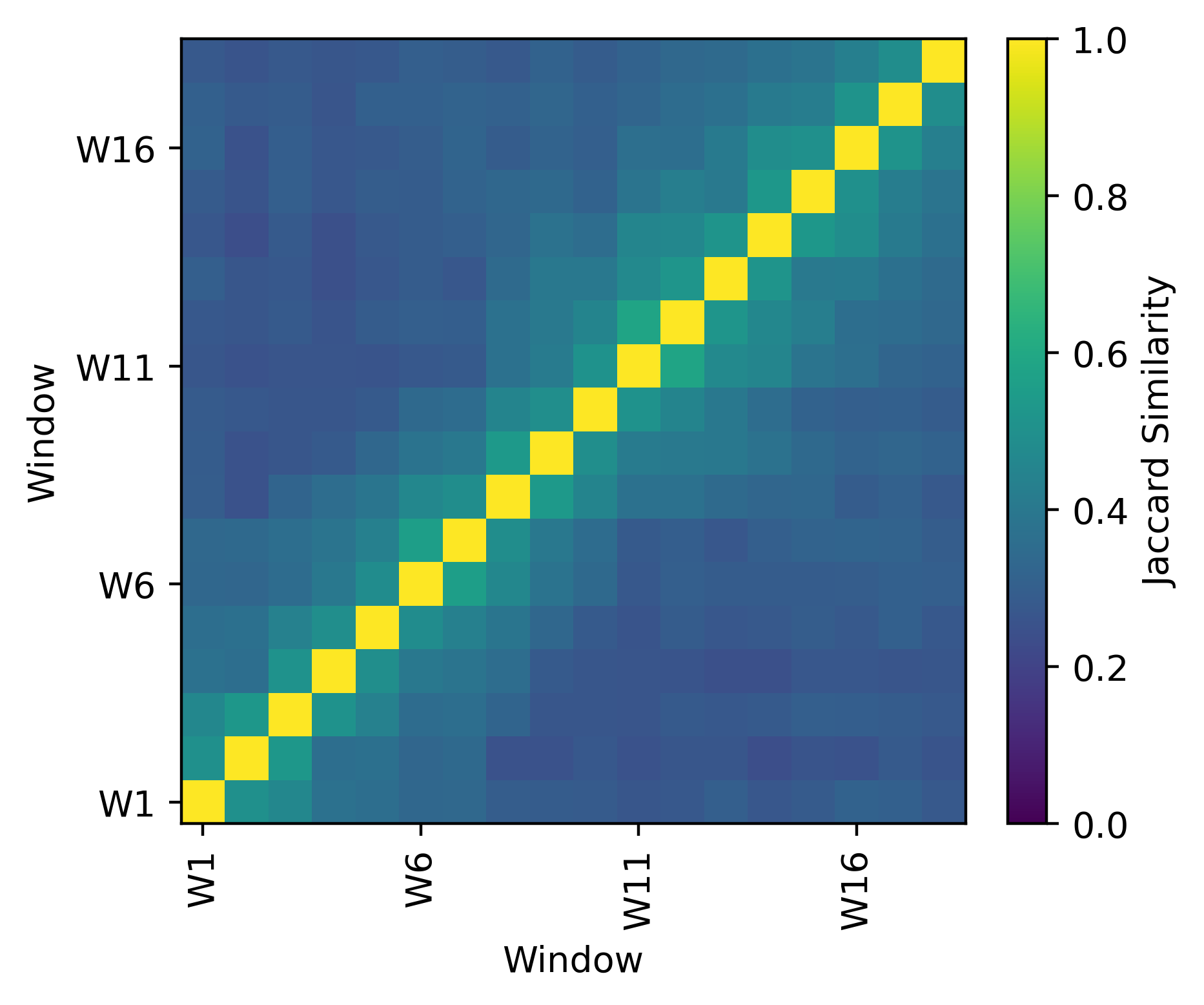}
    \caption{AEI}
\end{subfigure}
\caption{Average Jaccard similarity heatmaps of selected stocks across the eight filtering strategies.}
\label{fig:jaccard_heatmaps}
\end{figure*}

Next, we analyze the composition of selected stocks across rolling windows using the Jaccard similarity (JS) between the corresponding selected stock sets\footnote{Figure \ref{fig:jaccard_heatmaps} reports pairwise Jaccard similarity values for the proposed clustering framework under different distance measures with an in-sample period of 3 months and an out-of-sample period of 10 days, resulting in 18 rolling windows.}. The results are presented in Figure \ref{fig:jaccard_heatmaps}. For all eight clustering specifications, the highest JS values are concentrated near the main diagonal, indicating that nearby rolling windows produce similar selected stock sets and that portfolio composition evolves gradually over time. As expected, the similarity generally declines as one moves farther away from the diagonal. For instance, the selected stock universe in Window 1 is more similar to that in Window 2 than to that in Window 18.

Second, the heatmaps corresponding to the AWDS, AWDI, APLS, and APLI exhibit relatively greater variability in the off-diagonal regions, with more pronounced fluctuations in similarity across distant windows. This suggests that the selected stocks under these TDA-based filtering strategies are more responsive to changes in the underlying market structure and may capture evolving non-linear relationships among assets more effectively. In contrast, the correlation and Euclidean distance-based clustering, namely ACS, ACI, AES, and AEI, display comparatively smoother heatmaps. In particular, ACI and AEI show broader bands of moderate similarity away from the diagonal, indicating greater temporal consistency in stock selection, consistent with the higher AARI values. Comparisons between sentiment-based and non-sentiment-based variants further show that incorporating sentiment information generally increases variation in the similarity structure across windows, suggesting that sentiment-based filtering leads to a more adaptive and responsive stock selection process.

\begin{figure*}[htbp]
    \centering
    \begin{subfigure}{0.32\textwidth}
        \centering
        \includegraphics[width=\linewidth]{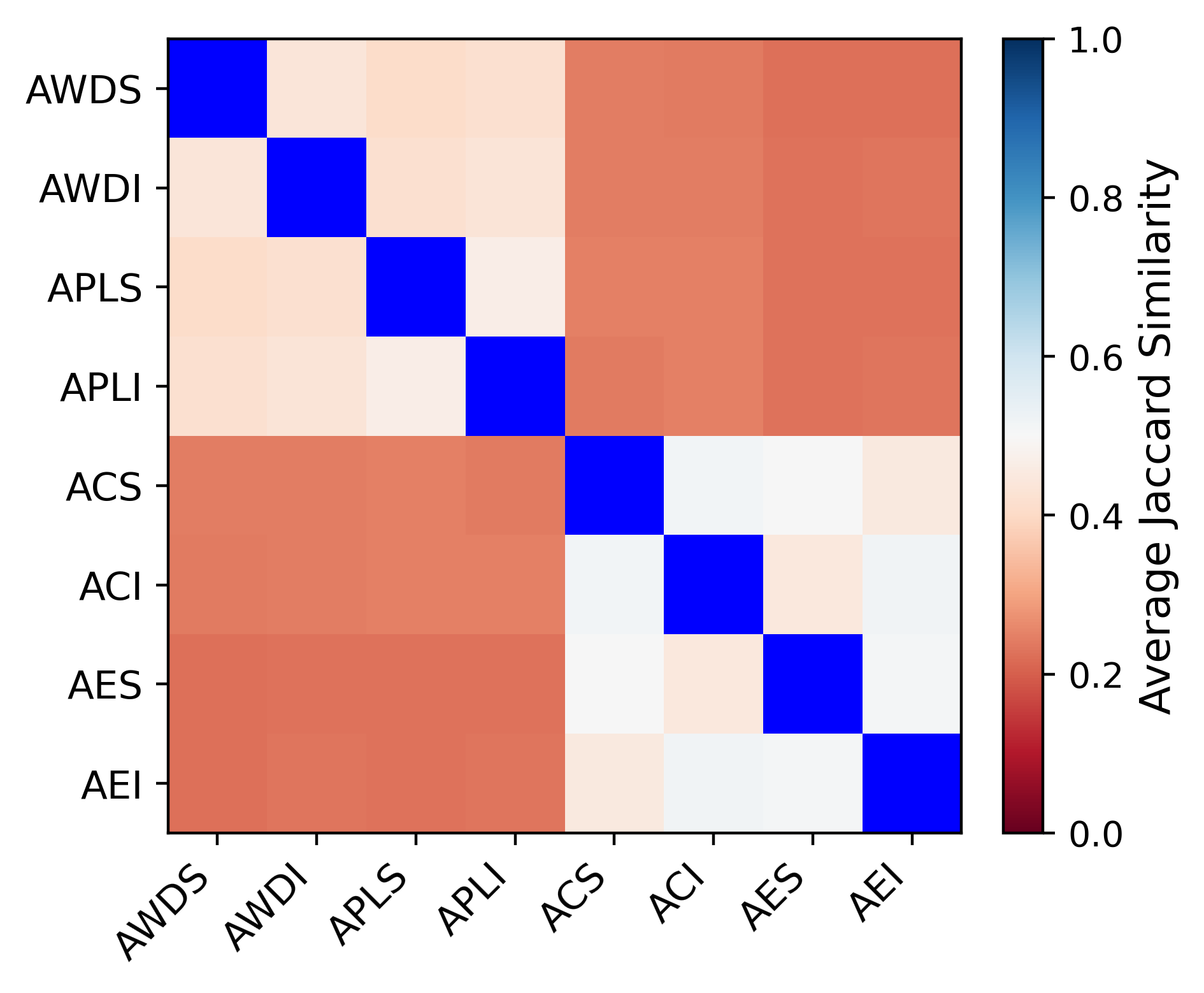}
    \end{subfigure}
    \hfill
    \begin{subfigure}{0.32\textwidth}
        \centering
        \includegraphics[width=\linewidth]{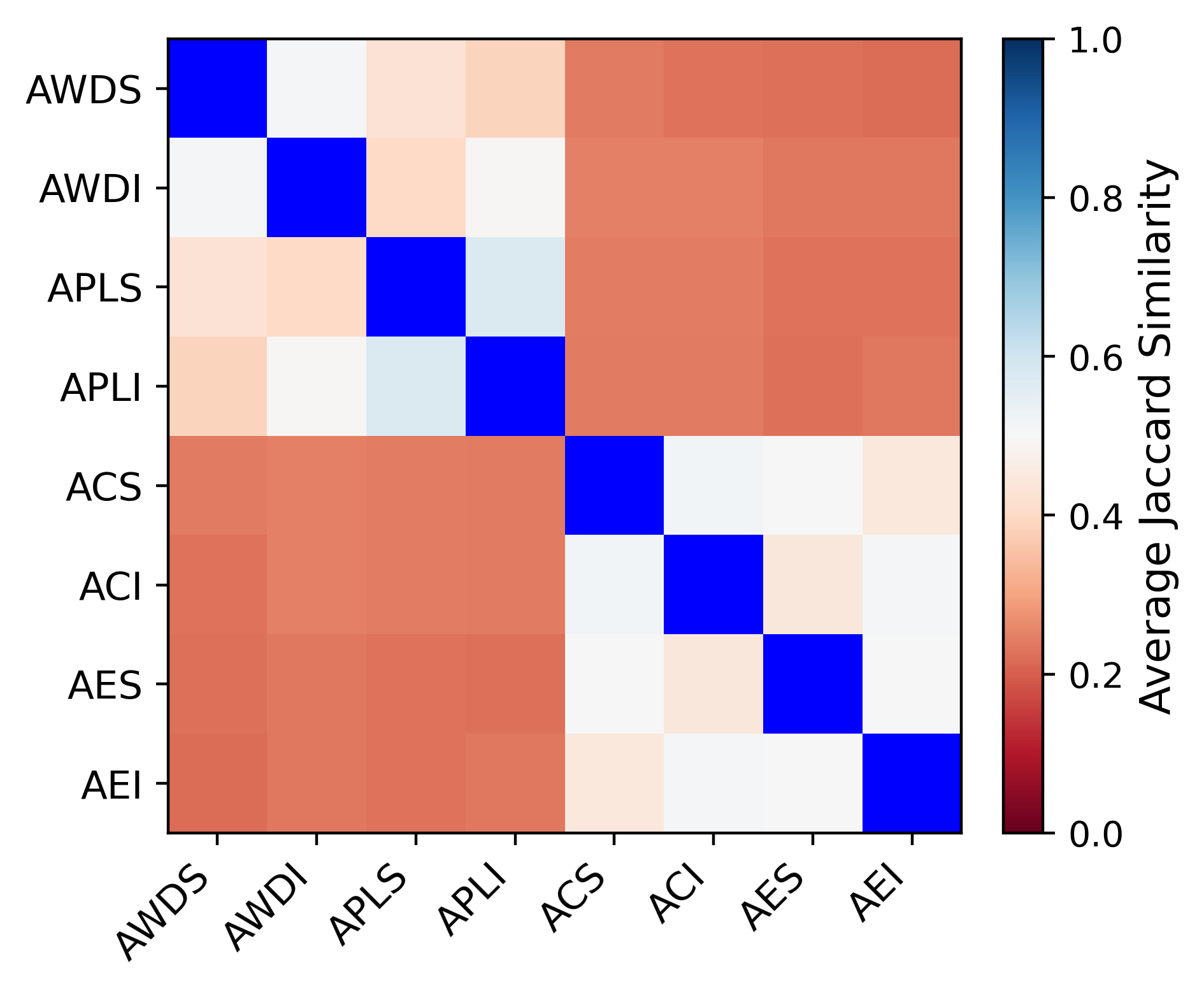}
    \end{subfigure}
    \hfill
    \begin{subfigure}{0.32\textwidth}
        \centering
        \includegraphics[width=\linewidth]{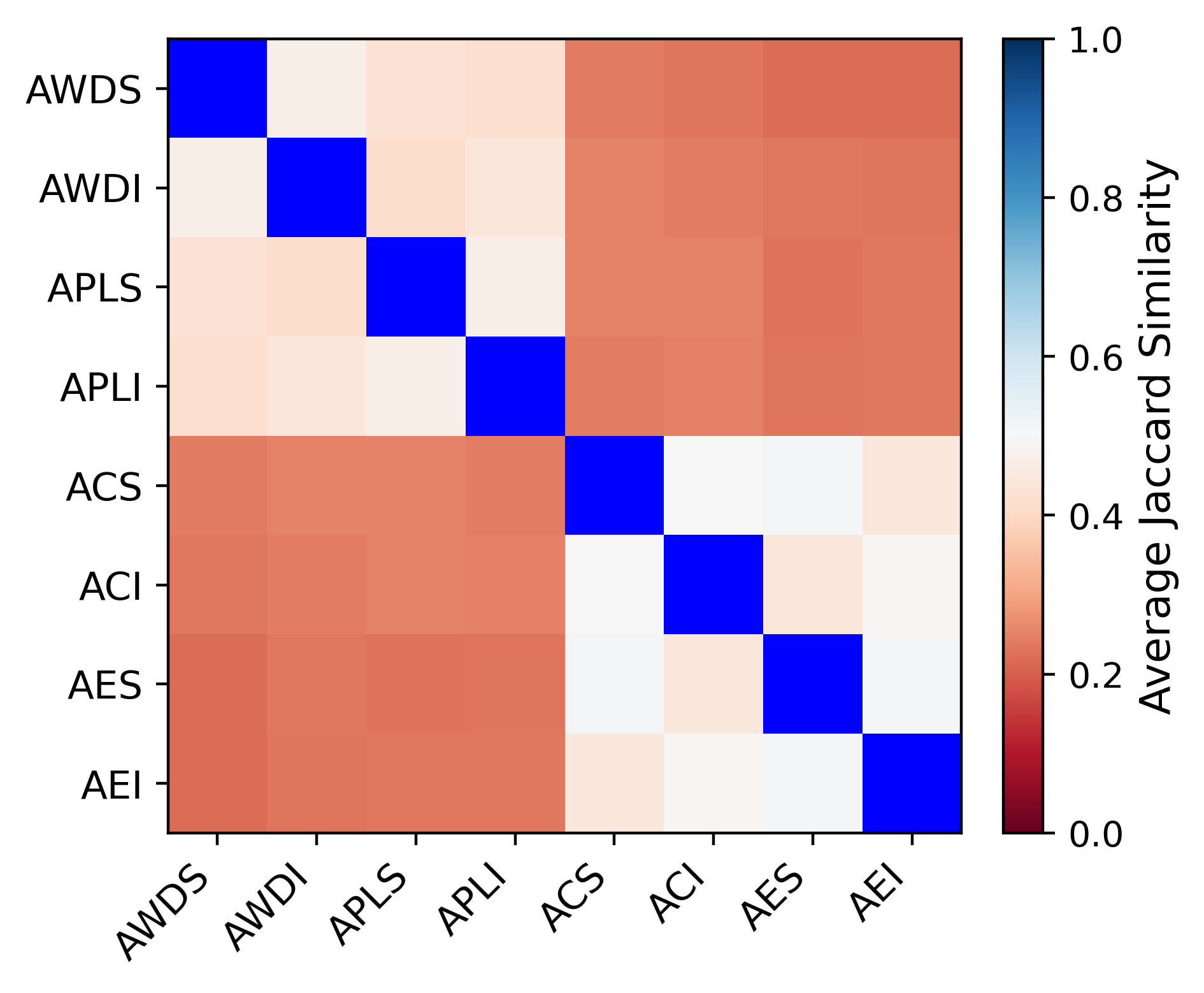}
    \end{subfigure}
    \caption{Comparison of average Jaccard similarity on selected stocks across the eight filtering strategies corresponding to different rebalancing frequencies, namely 3-day (left), 5-day (center), and 10-day (right), with 3 month in-sample period and $\alpha=95\%$.}
    \label{fig: Jaccard across 8 distance}
\end{figure*}

Finally, we compare the composition of selected stocks across the eight distance measures using JS. For each rolling window, the JS is computed between the selected stock sets obtained from every pair of distance measures, and the resulting values are averaged across all windows. Figure \ref{fig: Jaccard across 8 distance} presents the average pairwise JS for the 3-day, 5-day, and 10-day rebalancing horizons under $\alpha=95\%$. A clear block structure is observed across all rebalancing frequencies. The TDA-based methods (AWDS, AWDI, APLS, and APLI) exhibit moderate similarity among themselves, indicating that they tend to select relatively similar stock subsets. Similarly, the correlation and Euclidean methods (ACS, ACI, AES, and AEI) also show high within-group similarity. In contrast, the similarity between the TDA-based and conventional distance-based methods is lower, suggesting that the two groups identify materially different stock subsets for portfolio construction. Nevertheless, the presence of moderate cross-group similarity between TDA and correlation or Euclidean, indicates that TDA-based methods still capture some common market structure while additionally identifying nonlinear asset relationships beyond conventional correlation and geometric dependencies. This overall pattern remains consistent across all rebalancing horizons.

\begin{table*}[h!]
\centering
\scriptsize
\caption{Clustering Results for in-sample period $T_1 = 63$ and out-of-sample period $T_2 = 3$ for eight different distance based agglomerative clustering under $\alpha=95\%$ and $\alpha=99\%$, including the $\mathrm{(DRMV)_A}$ model. The reported values represent averages over all 62 windows and are rounded up to the nearest integer.}
\label{tab: clustering results}
\begin{tabular}{lccccccccc}
\hline
\textbf{Metric} 
& \multicolumn{2}{c}{$\mathrm{C_{AWDS}}$} 
& \multicolumn{2}{c}{$\mathrm{C_{APLS}}$} 
& \multicolumn{2}{c}{$\mathrm{C_{ACS}}$} 
& \multicolumn{2}{c}{$\mathrm{C_{AES}}$} &  $\mathrm{(DRMV)_{A}}$ \\
\cline{2-10}
& $\alpha=95\%$ & $\alpha=99\%$ 
& $\alpha=95\%$ & $\alpha=99\%$ 
& $\alpha=95\%$ & $\alpha=99\%$ 
& $\alpha=95\%$ & $\alpha=99\%$ \\
\hline

Number of clusters & 64 & 34 & 50 & 23 & 69 & 44 & 70 & 53 & -\\
Number of stocks in total & 493 & 493 & 493 & 493 & 493 & 493 & 493 & 493 & 493\\
Number of stocks in clusters & 455 & 474 & 458 & 477 & 456 & 472 & 452 & 473 & -\\
Number of outliers & 38 & 19 & 35 & 16 & 37 & 21 & 41 & 20 & -\\
Largest cluster & 75 & 86 & 70 & 81 & 78 & 89 & 77 & 84 & -\\
2nd largest cluster & 35 & 57 & 48 & 58 & 25 & 36 & 27 & 43 & -\\
3rd largest cluster & 21 & 30 & 36 & 35 & 19 & 27 & 20 & 26 & -\\
Total selected stocks & 72 & 66 & 75 & 59 & 72 & 69 & 84 & 78 & -\\
Total traded stocks & 8 & 8 & 9 & 8 & 9  & 8 & 8 & 9 & 13\\
Stocks in long position & 6 & 7 & 6 & 6 & 6 & 5 & 6 & 6 & 8 \\
Stock in short position & 2 & 2 & 2 & 3 & 2 & 3 & 2 & 3 & 6 \\
\hline

\textbf{Metric} 
& \multicolumn{2}{c}{$\mathrm{C_{AWDI}}$} 
& \multicolumn{2}{c}{$\mathrm{C_{APLI}}$} 
& \multicolumn{2}{c}{$\mathrm{C_{ACI}}$} 
& \multicolumn{2}{c}{$\mathrm{C_{AEI}}$} &  $\mathrm{(DRMV)_{A}}$\\
\cline{2-10}
& $\alpha=95\%$ & $\alpha=99\%$ 
& $\alpha=95\%$ & $\alpha=99\%$ 
& $\alpha=95\%$ & $\alpha=99\%$ 
& $\alpha=95\%$ & $\alpha=99\%$ \\
\cline{2-10}

Number of clusters & 55 & 26 & 45 & 21 & 76 & 50 & 72 & 48 &-\\
Number of stocks in total & 493 & 493 & 493 & 493 & 493 & 493 & 493 & 493 & 493\\
Number of stocks in clusters & 461 & 478 & 470 & 480 & 461 & 476 & 458 & 474 &-\\
Number of outliers & 32 & 15 & 23 & 13 & 32 & 17 & 35 & 19 &-\\
Largest cluster & 62 & 83 & 74 & 87 & 83 & 90 & 82 & 92 &-\\
2nd largest cluster & 38 & 45 & 46 & 69 & 26 & 35 & 26 & 37 &- \\
3rd largest cluster & 21 & 33 & 31 & 42 & 23 & 24 & 21 & 28 &-\\
Total selected stocks & 68 & 63 & 73 & 60 & 83 & 76 & 80 & 75 &-\\
Total traded stocks & 8 & 8 & 8 & 9 & 9 & 9 & 8 & 8 &13\\
Stocks in long position & 6 & 6 & 6 & 5 & 6 & 5 & 6 & 5 & 8\\
Stock in short position & 2 & 2 & 3 & 3 & 3 & 3 & 2 & 3 & 6\\
\hline
\end{tabular}
\end{table*}

\begin{table*}[h!]
\centering
\small
\caption{Average Adjusted Rand Index (AARI) and Average Silhouette Score (ASS) for in-sample period $T_1 = 63$ and out-of-sample period $T_2 = 3$ for eight different distance-based agglomerative clustering under $\alpha = 95\%$ and $\alpha = 99\%$}
\label{tab:ari_results}
\resizebox{0.70\textwidth}{!}{
\begin{tabular}{lcccccccc}
\hline
\textbf{Metric} 
& \multicolumn{2}{c}{{$\mathrm{C_{AWDS}}$}} 
& \multicolumn{2}{c}{{$\mathrm{C_{APLS}}$}} 
& \multicolumn{2}{c}{{$\mathrm{C_{ACS}}$}} 
& \multicolumn{2}{c}{{$\mathrm{C_{AES}}$}} \\
\cline{2-9}
& $\alpha=95\%$ & $\alpha=99\%$ 
& $\alpha=95\%$ & $\alpha=99\%$ 
& $\alpha=95\%$ & $\alpha=99\%$ 
& $\alpha=95\%$ & $\alpha=99\%$ \\
\hline
AARI & 0.73418 &  0.70964 & 0.75762 & 0.76384 & 0.78024  & 0.77284 & 0.76698 & 0.79681\\
ASS & 0.27710 & 0.25831 & 0.26406 & 0.24679 & 0.25722 & 0.25039 & 0.24276 & 0.22621\\
\hline
\textbf{Metric} 
& \multicolumn{2}{c}{{$\mathrm{C_{AWDI}}$}} 
& \multicolumn{2}{c}{{$\mathrm{C_{APLI}}$}} 
& \multicolumn{2}{c}{{$\mathrm{C_{ACI}}$}} 
& \multicolumn{2}{c}{{$\mathrm{C_{AEI}}$}} \\
\cline{2-9}
& $\alpha=95\%$ & $\alpha=99\%$ 
& $\alpha=95\%$ & $\alpha=99\%$ 
& $\alpha=95\%$ & $\alpha=99\%$ 
& $\alpha=95\%$ & $\alpha=99\%$ \\
\hline
AARI 
& 0.756811 & 0.73021 & 0.76946  & 0.77567 & 0.77127 & 0.80948 & 0.78677 &  0.77011 \\
ASS & 0.25388 & 0.24840 & 0.25153 & 0.24937 & 0.24445 & 0.23232 & 0.24282 & 0.22861 \\
\hline
\end{tabular}}
\end{table*}

% \begin{figure*}[htbp]
%     \centering
%     \begin{subfigure}{0.45\textwidth}
%         \centering
%         \includegraphics[width=\linewidth]{Avg_Jaccard_3M3D.png}
%         % \caption{$\mathrm{(DRMV)_{AWDS}}$}
%     \end{subfigure}
%     \hfill
%     \begin{subfigure}{0.45\textwidth}
%         \centering
%         \includegraphics[width=\linewidth]{Avg_Jaccard_3M5D.png}
%         % \caption{($\mathrm{(DRMV)_{AWDI}}$}
%     \end{subfigure}
%     \hfill
%     \begin{subfigure}{0.45\textwidth}
%         \centering
%         \includegraphics[width=\linewidth]{Avg_Jaccard_3M10D.png}
%     \end{subfigure}
%     \caption{Comparison of average Jaccard similarity on selected stocks across the eight filtering strategies corresponding to different rebalancing frequencies, namely 3-day (top left), 5-day (top right), and 10-day (bottom center) with 3 month in-sample period and $\alpha=95\%$.}
%     \label{fig: Jaccard across 8 distance}
% \end{figure*}

\subsection{Performance analysis}

Following the asset selection phase, the optimal portfolio is constructed by solving the (DRMV) model over the filtered asset universe for each of the eight distance measures. The resulting portfolios are denoted by $\mathrm{(DRMV)_{AWDS}}$, $\mathrm{(DRMV)_{APLS}}$, $\mathrm{(DRMV)_{ACS}}$, $\mathrm{(DRMV)_{AES}}$, $\mathrm{(DRMV)_{AWDI}}$, $\mathrm{(DRMV)_{APLI}}$, $\mathrm{(DRMV)_{ACI}}$, and $\mathrm{(DRMV)_{AEI}}$\footnote{Throughout the empirical analysis, portfolio results corresponding to clustering-based asset selection are represented using the notation $\mathrm{(DRMV)}_{\cdot}$.}. To evaluate the effectiveness of the proposed clustering-based portfolio framework, these eight strategies are compared against the following benchmark portfolio strategies:

\begin{itemize}

\item \textbf{$\mathrm{(DRMV)_A}$: Full-universe (DRMV) model.}  
This portfolio is obtained by solving the (DRMV) model over the entire asset universe without applying clustering or asset selection.

\item \textbf{Na\"ive.}  
An equally weighted portfolio in which capital is allocated uniformly across the selected assets.

\item \textbf{Index.}  
The underlying market index is used as a passive investment benchmark.

\item \textbf{$\mathrm{B\&H_S}$ and $\mathrm{B\&H_I}$: Buy-and-hold strategies.}  
These portfolios represent passive investment strategies and are used as benchmarks to compare the proposed actively managed framework. They are constructed by solving the (DRMV) model once at the beginning of the investment horizon and holding the resulting portfolio without rebalancing throughout the out-of-sample period\footnote{For a 3-month in-sample period, the investment horizon is 9 months, while for a 4-month in-sample period, it is 8 months.}. Here, $\mathrm{B\&H_S}$ and $\mathrm{B\&H_I}$ denote the buy-and-hold portfolios under the indicator--sentiment and indicator-only settings, respectively. In both cases, asset selection is based on the AWD distance measure\footnote{The AWD distance measure is adopted due to its superior out-of-sample performance.}.
\end{itemize}

Before analyzing the out-of-sample performance, we first examine the structural characteristics of the portfolios generated under the eight distance specifications reported in Table \ref{tab: clustering results}. The average number of traded stocks across all clustering-based strategies ranges from \(8\) to \(9\) for both \(\alpha=95\%\) and \(\alpha=99\%\), indicating that the optimization model further refines the selected asset universe into a sparse portfolio. Across all strategies, the average number of long positions varies between \(5\) and \(7\), while the average number of short positions ranges from \(2\) to \(3\), demonstrating that the proposed framework consistently constructs sparse long--short portfolios irrespective of the distance measure or confidence level.

Figure \ref{fig: no. of stocks} illustrates the numbers of long and short positions across rolling windows, where the upper boundary of the shaded region represents the total number of traded stocks in each window. The results show that portfolio composition evolves dynamically over time while remaining sparse throughout the investment horizon. In particular, the number of long positions remains relatively stable, typically between 5 and 7 stocks, whereas the number of short positions is higher in the early rolling windows and gradually declines in later periods, reaching 1 or even 0 in several windows. Consequently, the total number of traded assets exhibits greater variability during the early part of the sample and becomes more stable over time. Similar patterns are observed across the remaining distance specifications; therefore, for brevity, only their average values are reported in Table \ref{tab: clustering results}.

\textbf{Rolling window framework with 3-month in-sample and 3-day rebalancing.}

Table \ref{Tab: 3m-3d}\footnote{In all tables, the best-performing indicator values among all models are highlighted in bold, while the second-best values are highlighted in bold italics.} represents the average out-of-sample performance of the proposed filtering-based (DRMV) model under eight different distance-measure specifications for a 3-month in-sample period and 3-day rebalancing frequency. In addition, the last column, $\mathrm{(DRMV)_A}$ represents the baseline (DRMV) model without any filtering of assets. The Table \ref{tab: benchmark_3m_3d} reports the performance of four benchmark portfolios, namely $\mathrm{{B\&H}_S}$, $\mathrm{{B\&H}_I}$, Na\"ive, and Index models.

We first compare the out-of-sample performance of the first four columns corresponding to the indicator-sentiment based filtering strategies in Table \ref{Tab: 3m-3d}. Under both confidence levels \(\alpha=0.95\) and \(\alpha=0.99\), \(\mathrm{(DRMV)_{APLS}}\) consistently emerges as the best-performing portfolio among the four. In particular, it attains the highest mean return, SR, STARR, Sterling, and cumulative return across both confidence levels. It also achieves the highest maximum return and the least negative minimum return. 

Among the remaining sentiment-based strategies, \(\mathrm{(DRMV)_{AWDS}}\) performs next best in terms of return and risk-adjusted measures, which further reinforces the effectiveness of the proposed TDA-based filtering framework. Further, $\mathrm{(DRMV)_{ACS}}$ performs relatively well in terms of volatility, VaR, and CVaR for $\alpha = 0.95$, whereas $\mathrm{(DRMV)_{AWDS}}$ provides more favorable downside risk measures under $\alpha = 0.99$. In particular, for $\alpha = 0.99$, $\mathrm{(DRMV)_{AWDS}}$ demonstrates strong overall performance, ranking second in terms of mean return and risk-adjusted ratios, while achieving the best results across key risk measures such as standard deviation, VaR, and CVaR. On the other hand, $\mathrm{(DRMV)_{AES}}$ yields the smallest MDD and ADD across both confidence levels among the indicator-sentiment-based filtering strategies. %, indicating superior drawdown control. 
Since the portfolios are rebalanced at a high frequency, the primary objective is to achieve higher returns by taking a slightly higher level of risk. From this perspective, $\mathrm{(DRMV)_{APLS}}$ offers the most favorable trade-off between return and risk-adjusted measures among all sentiment-based approaches, followed by $\mathrm{(DRMV)_{AWDS}}$ as the second-best performing model.

Next, we compare the four indicator-only filtering strategies, namely $\mathrm{(DRMV)_{AWDI}}$, $\mathrm{(DRMV)_{APLI}}$, $\mathrm{(DRMV)_{ACI}}$, and $\mathrm{(DRMV)_{AEI}}$, as reported in Table \ref{Tab: 3m-3d}. Similar to the sentiment-based case, the TDA-based strategies outperform their non-TDA counterparts. In particular, $\mathrm{(DRMV)_{AWDI}}$ emerges as the strongest performer within this group across both confidence levels in terms of return and risk-adjusted measures. Moreover, for $\alpha = 99\%$, it also attains lower risk measures such as SD, VaR, and CVaR, highlighting its superior overall performance. Among the remaining strategies, $\mathrm{(DRMV)_{APLI}}$ delivers comparatively better return and risk-adjusted performance for $\alpha = 99\%$, whereas $\mathrm{(DRMV)_{AEI}}$ ranks second for $\alpha = 95\%$ in terms of these measures. On the other hand, $\mathrm{(DRMV)_{ACI}}$ performs relatively better from a downside risk perspective, emerging as the best among the four for $\alpha = 95\%$ and the second-best for $\alpha = 99\%$. Overall, these findings suggest that even in the absence of sentiment information, TDA-based filtering models provide a more effective and return-oriented portfolio construction framework than the non-TDA approaches.

Next, we compare the indicator-sentiment based filtering strategies with their corresponding indicator-only counterparts, as reported in Table \ref{Tab: 3m-3d}, i.e., $\mathrm{(DRMV)_{AWDS}}$ vs.\ $\mathrm{(DRMV)_{AWDI}}$, $\mathrm{(DRMV)_{APLS}}$ vs.\ $\mathrm{(DRMV)_{APLI}}$, $\mathrm{(DRMV)_{ACS}}$ vs.\ $\mathrm{(DRMV)_{ACI}}$, and $\mathrm{(DRMV)_{AES}}$ vs.\ $\mathrm{(DRMV)_{AEI}}$ for both $\alpha = 95\%$ and $99\%$. The results show that the indicator--sentiment based strategies consistently outperform the indicator only approaches across almost all distance measures in terms of return and risk-adjusted performance. However, these improvements are generally accompanied by a marginal increase in risk. We further compare the proposed framework, comprising an initial filtering stage followed by portfolio allocation through the (DRMV) model, with $\mathrm{(DRMV)_A}$, as presented in Table \ref{Tab: 3m-3d}. Specifically, the eight filtering-based models, ranging from $\mathrm{(DRMV)_{AWDS}}$ to $\mathrm{(DRMV)_{AEI}}$, are evaluated against $\mathrm{(DRMV)_A}$. The results reveal that all eight filtering-based models outperform the benchmark in terms of return and risk-adjusted performance. In addition, one-sided $t$-tests and Sharpe-tests for both $\alpha = 95\%$ and 99\% on out-of-sample return series confirm that the mean return and SR of $\mathrm{(DRMV)_{APLS}}$ are significantly higher than those of $\mathrm{(DRMV)_{AES}}$, $\mathrm{(DRMV)_{APLI}}$, $\mathrm{(DRMV)_{ACI}}$, $\mathrm{(DRMV)_{AEI}}$, and $\mathrm{(DRMV)_A}$.

\textbf{Rolling window framework with 3-month in-sample and 5-day rebalancing.}

Similar to Table \ref{Tab: 3m-3d}, Table \ref{Tab: 3m-5d} reports the average out-of-sample performance of the proposed filtering-based (DRMV) strategies for the case of a 3-month in-sample period and 5-day rebalancing frequency. The results remain broadly consistent with those obtained under the 3-day rebalancing setting, again highlighting the effectiveness of sentiment-enriched and TDA-based filtering in the portfolio construction process. For the sentiment-indicator strategies, \(\mathrm{(DRMV)_{AWDS}}\) records the highest mean return, SR, STARR, and cumulative return under \(\alpha=0.95\), whereas \(\mathrm{(DRMV)_{APLS}}\) attains the highest values of these performance measures under \(\alpha=0.99\), along with the highest Sterling value. A similar pattern is observed for the indicator-only strategies as in Table \ref{Tab: 3m-3d}. The TDA-based portfolios \(\mathrm{(DRMV)_{AWDI}}\) and \(\mathrm{(DRMV)_{APLI}}\) generally record higher mean returns, stronger SR and STARR values, and larger cumulative returns than the non-TDA strategies \(\mathrm{(DRMV)_{ACI}}\) and \(\mathrm{(DRMV)_{AEI}}\). In particular, \(\mathrm{(DRMV)_{AWDI}}\) remains the most favorable specification within the indicator-only setting, reflecting more efficient return generation relative to risk.

We also compare the indicator-sentiment-based filtering strategies with their corresponding indicator-only counterparts in Table \ref{Tab: 3m-5d}, for both $\alpha = 95\%$ and $99\%$. Overall, the empirical evidence under the 5-day rebalancing setting remains consistent with that of the 3-day case. In most pairwise comparisons, the sentiment-enhanced strategies outperform their indicator-only counterparts in terms of return and risk-adjusted performance. The results are further supported by one-sided $t$-tests and Sharpe-tests. For $\alpha = 95\%$, the $\mathrm{(DRMV)_{AWDS}}$ model exhibits a significantly higher mean return compared to $\mathrm{(DRMV)_{ACS}}$, $\mathrm{(DRMV)_{AES}}$, $\mathrm{(DRMV)_{APLI}}$, $\mathrm{(DRMV)_{ACI}}$, $\mathrm{(DRMV)_{AEI}}$, and $\mathrm{(DRMV)_A}$. In addition, the $\mathrm{(DRMV)_{AWDS}}$ model achieves a significantly higher SR than $\mathrm{(DRMV)_{ACS}}$, $\mathrm{(DRMV)_{AES}}$, $\mathrm{(DRMV)_{AEI}}$, and $\mathrm{(DRMV)_A}$.

\textbf{Rolling window framework with 3-month in-sample and 10-day rebalancing.}

Table \ref{Tab: 3m-10d} reports the average out-of-sample performance of the proposed filtering-based (DRMV) strategies for the case of a 3-month in-sample period and 10-day rebalancing frequency. Similar to the 3 months in-sample and 3-day or 5-day rebalancing settings, the results continue to support the usefulness of sentiment-enriched and TDA-based filtering in improving portfolio performance. For the sentiment-based strategies, \(\mathrm{(DRMV)_{AWDS}}\) records the highest mean return, SR, STARR, and cumulative return under \(\alpha=0.95\), whereas \(\mathrm{(DRMV)_{APLS}}\) attains the highest mean return, SR, Sterling, and cumulative return under \(\alpha=0.99\).

A similar trend is observed in the indicator-only setting. The TDA-based portfolios \(\mathrm{(DRMV)_{AWDI}}\) and \(\mathrm{(DRMV)_{APLI}}\) generally yield higher mean returns, stronger SR and STARR values, and larger cumulative returns than the non-TDA strategies \(\mathrm{(DRMV)_{ACI}}\) and \(\mathrm{(DRMV)_{AEI}}\). In particular, \\ \(\mathrm{(DRMV)_{AWDI}}\) remains the most favorable specification within the indicator-only group under both confidence levels. We also compare the indicator-sentiment based filtering strategies with their corresponding indicator-only counterparts, as presented in Table \ref{Tab: 3m-10d}. The empirical findings under the 10-day rebalancing setting remain broadly consistent with those observed in the 3-day and 5-day cases. In almost all pairwise comparisons, the sentiment-enhanced strategies deliver superior performance relative to their indicator-only counterparts, particularly in terms of return and risk-adjusted measures.
 
Furthermore, all filtering-based (DRMV) strategies continue to outperform the classical full-universe model, $\mathrm{(DRMV)_A}$, in terms of mean return and risk-adjusted measures, indicating that the proposed two-stage framework remains effective even under a relatively lower rebalancing frequency. The superiority of the proposed approach is further validated through a one-sided $t$-test and the Sharpe-test. For $\alpha = 95\%$, the $\mathrm{(DRMV)_{AWDS}}$ portfolio delivers a statistically significant improvement in mean return over $\mathrm{(DRMV)_{ACS}}$, $\mathrm{(DRMV)_{AES}}$, $\mathrm{(DRMV)_{APLI}}$, $\mathrm{(DRMV)_{ACI}}$, $\mathrm{(DRMV)_{AEI}}$, and $\mathrm{(DRMV)_A}$. Moreover, the SR of $\mathrm{(DRMV)_{AWDS}}$ is found to be significantly higher than those of $\mathrm{(DRMV)_{ACS}}$, $\mathrm{(DRMV)_{AEI}}$, and $\mathrm{(DRMV)_A}$.

\begin{figure*}[htbp]
    \centering
    \begin{minipage}{0.30\textwidth}
        \centering
        \fbox{\includegraphics[width=\linewidth, height=3.5cm]{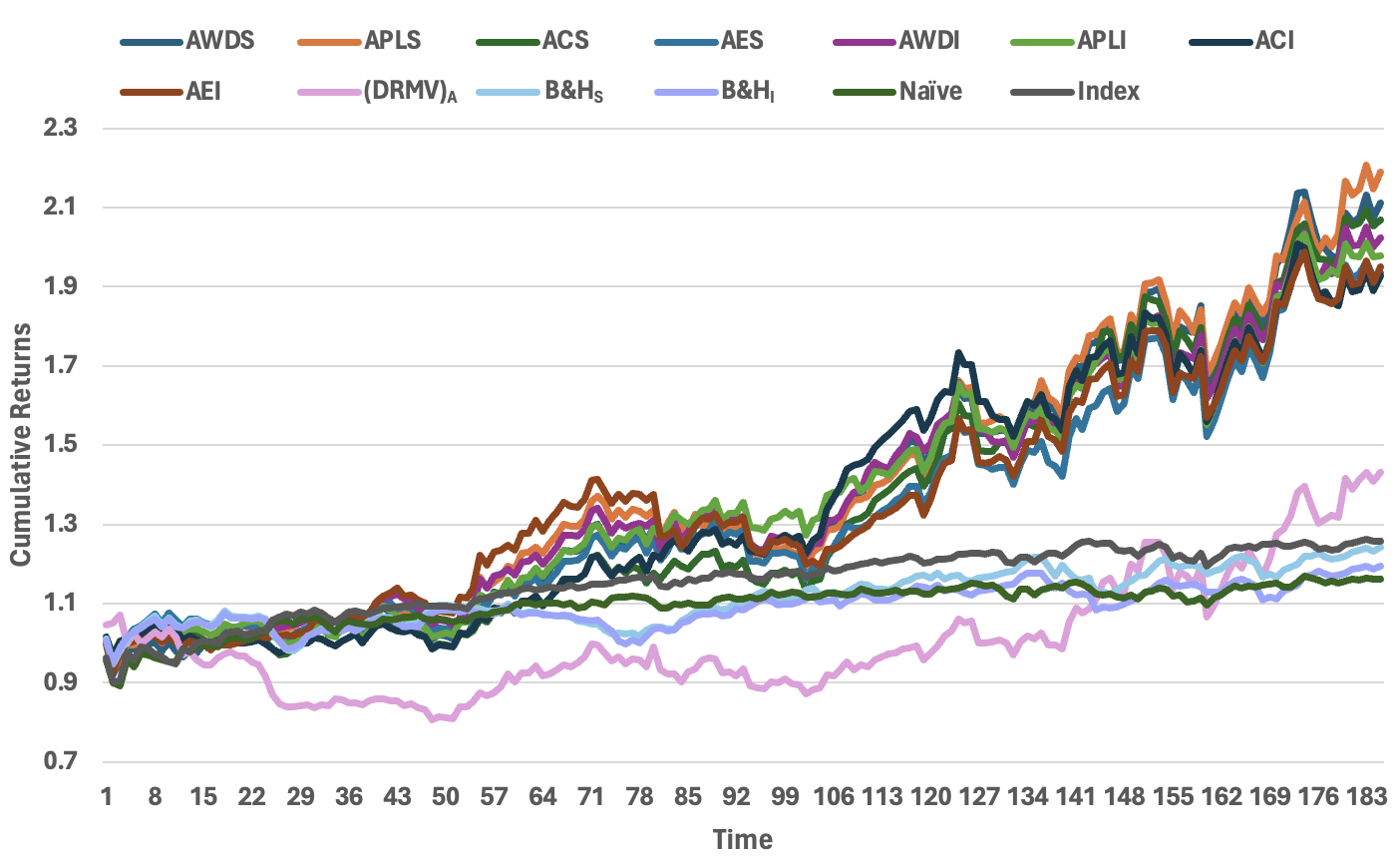}}
    \end{minipage}
    \hfill
    \begin{minipage}{0.30\textwidth}
        \centering
        \fbox{\includegraphics[width=\linewidth, height=3.5cm]{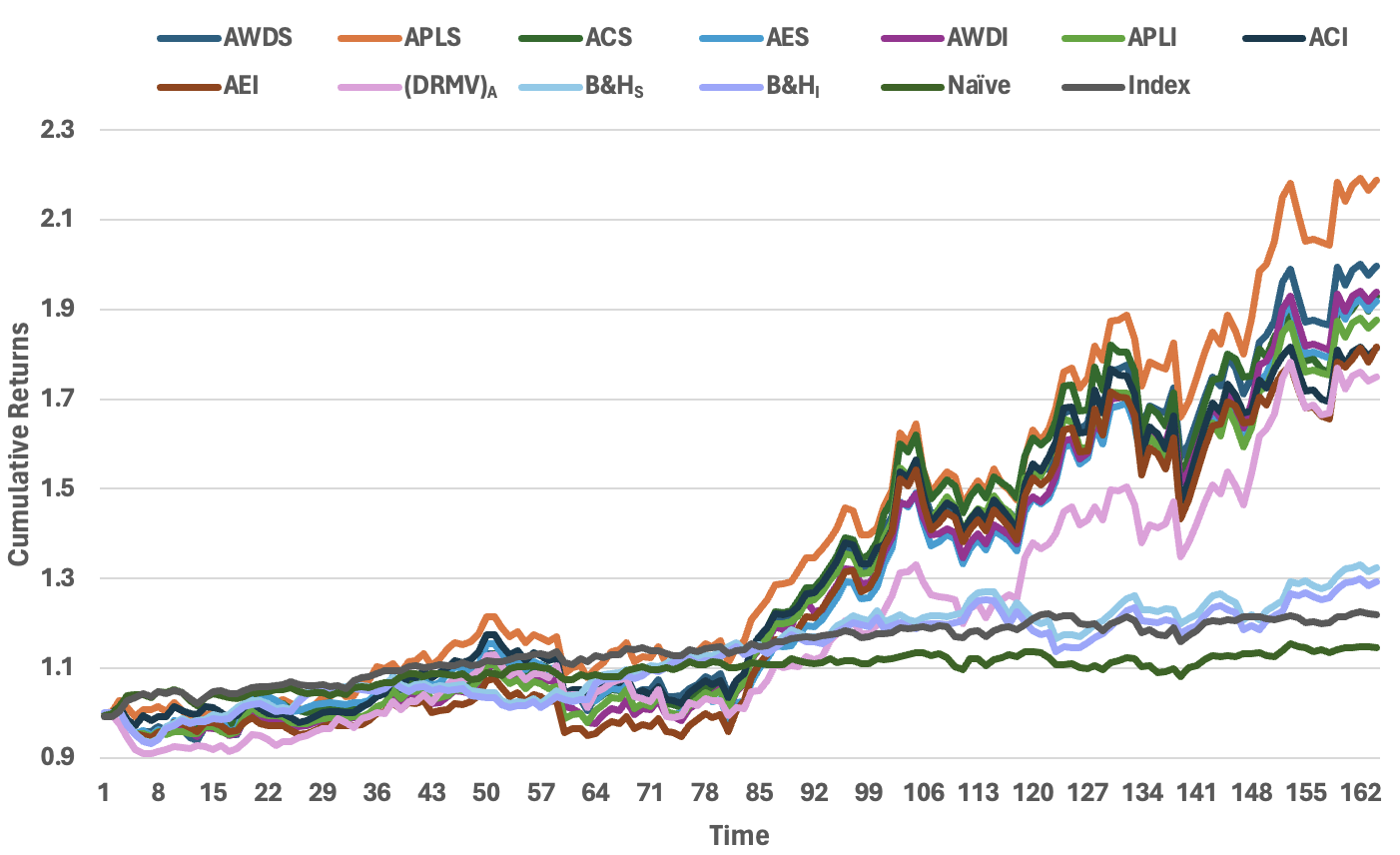}}
    \end{minipage}
    \hfill
    \begin{minipage}{0.30\textwidth}
        \centering
        \fbox{\includegraphics[width=\linewidth, height=3.5cm]{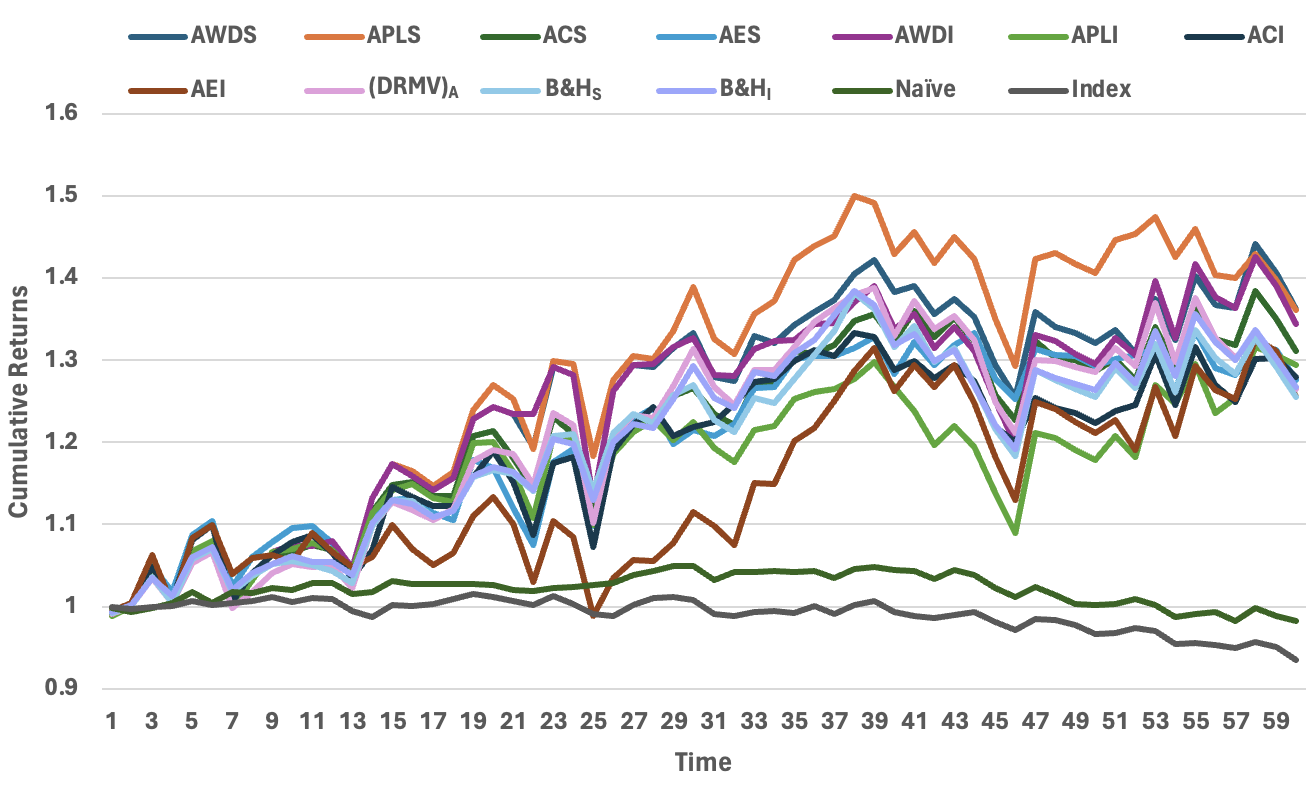}}
    \end{minipage}
    \caption{Out-of-sample cumulative returns for all considered models, including the benchmark models $\mathrm{B\&H_S}$, $\mathrm{B\&H_I}$, Naïve, and Index, under a 3-month in-sample period with 3-day rebalancing (left), a 4-month in-sample period with 3-day rebalancing (center), and a 3-month in-sample period with 3-day rebalancing during the U.S.--Israel--Iran conflict (right).}
    \label{fig: cumulative returns}
\end{figure*}

Finally, we compare the proposed models reported in Tables \ref{Tab: 3m-3d}, \ref{Tab: 3m-5d}, and \ref{Tab: 3m-10d} with the four benchmark portfolios presented in Table \ref{tab: benchmark_3m_3d}, namely $\mathrm{B\&H_S}$, $\mathrm{B\&H_I}$, Na\"ive, and Index. The results clearly show that all nine models in Tables \ref{Tab: 3m-3d}, \ref{Tab: 3m-5d}, and \ref{Tab: 3m-10d} including the full-universe benchmark $\mathrm{(DRMV)_A}$, outperform the benchmark portfolios in Table \ref{tab: benchmark_3m_3d} in terms of return (see Figure \ref{fig: cumulative returns} (left)) and risk-adjusted measures. This superior performance is, to some extent, expected, as the proposed portfolios are rebalanced every 3, 5, or 10 days, enabling them to respond more actively to evolving market conditions compared to the relatively static benchmark portfolios. The results are further reinforced by a one-sided $t$-test and the Sharpe-test. For both $\alpha = 95\%$ and $\alpha = 99\%$, the $\mathrm{(DRMV)_{AWDS}}$ and $\mathrm{(DRMV)_{APLS}}$ models exhibit significantly higher mean returns and SR compared to the $\mathrm{B\&H_{S}}$, $\mathrm{B\&H_{I}}$, Na\"ive, and Index portfolios. However, such frequent rebalancing also leads to higher portfolio turnover, which is reflected in the relatively larger transaction costs reported in Table \ref{tab: tc_values}. Moreover, the higher rebalancing frequency may also result in elevated portfolio risk in certain cases. Nevertheless, despite these additional costs and increases in risk, the proposed filtering-based strategies deliver substantially higher mean returns and superior risk-adjusted performance than all benchmark portfolios, thereby demonstrating the effectiveness of the proposed framework.

\textbf{Comparison across 3, 5, and 10 days rebalancing horizons.}

A comparison across the three rebalancing horizons, namely 3-day, 5-day, and 10-day (see Tables \ref{Tab: 3m-3d}, \ref{Tab: 3m-5d}, \ref{Tab: 3m-10d}), reveals a clear effect of rebalancing frequency on portfolio performance. Across most of the eight filtering specifications, the 3-day rebalancing setting yields higher mean returns, SR, STARR, Sterling, and cumulative returns, followed by the 5-day and then the 10-day rebalancing horizon. For example, the cumulative return of \(\mathrm{(DRMV)_{APLS}}\) decreases from \(2.18968\) under the 3-day rebalancing horizon to \(2.04184\) under the 5-day horizon and further to \(1.92273\) under the 10-day horizon for \(\alpha=0.95\). A similar decline is observed under \(\alpha=0.99\). However, there are some instances in which less frequent rebalancing yields higher returns. For example, under \(\alpha = 0.95\), \(\mathrm{(DRMV)_{AWDS}}\) with 5-day rebalancing achieves higher mean return and cumulative return than its 3-day counterpart. Similarly, under \(\alpha = 0.99\), \(\mathrm{(DRMV)_{AES}}\) and \(\mathrm{(DRMV)_{AWDI}}\) generate higher returns and cumulative performance under the 5-day rebalancing horizon compared to the 3-day setting. In addition, under \(\alpha = 0.99\), \(\mathrm{(DRMV)_{ACS}}\) records higher mean return and cumulative return under the 10-day rebalancing horizon than under the 5-day horizon. Nevertheless, in the majority of configurations, more frequent rebalancing leads to stronger return generation.

At the same time, a reduction in rebalancing frequency is often associated with a moderation in return variability and, in certain cases, lower downside risk measures such as SD, VaR, CVaR, and drawdown-related quantities. This indicates a trade-off between return enhancement and downside containment: while the 3-day rebalancing horizon generally delivers stronger return and risk-adjusted performance, the 10-day horizon tends to provide relatively more stable portfolio behavior in some specifications, albeit at the cost of lower cumulative wealth. An important practical consideration in this comparison is transaction cost, reported in Table \ref{tab: tc_values}. As expected, more frequent rebalancing leads to higher transaction costs, with the 3-day horizon incurring the highest costs, followed by the 5-day and 10-day horizons. In contrast, passive buy-and-hold strategies involve significantly lower trading costs. Overall, the empirical evidence suggests that more frequent rebalancing enhances the ability of the proposed framework to generate higher returns and higher risk-adjusted ratios, although this comes at the expense of increased trading costs and higher risk. From an investment perspective, such actively managed strategies are suitable for relatively aggressive investors who are willing to tolerate higher trading intensity and risk in pursuit of superior returns.

\begin{table*}[H]
\centering
\caption{Performance comparison of eight different distance based clustering portfolios under $\alpha=95\%$ and $\alpha=99\%$, including the $\mathrm{(DRMV)_A}$ model for in-sample period $T_1 = 63$ and out-of-sample period $T_2 = 3$.}
\label{Tab: 3m-3d}
\resizebox{0.70\textwidth}{!}{
\begin{tabular}{lccccccccc}
\toprule
$\alpha = 0.95$ 
& $\mathrm{(DRMV)_{AWDS}}$ & $\mathrm{(DRMV)_{APLS}}$ & $\mathrm{(DRMV)_{ACS}}$ & $\mathrm{(DRMV)_{AES}}$
& $\mathrm{(DRMV)_{AWDI}}$ & $\mathrm{(DRMV)_{APLI}}$ & $\mathrm{(DRMV)_{ACI}}$ & $\mathrm{(DRMV)_{AEI}}$ & $\mathrm{(DRMV)_A}$  \\
\cmidrule(lr){1-5} \cmidrule(lr){6-9} \cmidrule(lr){10-10}
Mean   & 0.00435 & \textbf{0.00454} & 0.00417 & 0.00386 & \textbf{\textit{0.00408}} & 0.00391 & 0.00381 & 0.00399 & 0.00225 \\
SD     & 0.02469 & 0.02378 & \textbf{0.02243} & 0.02459 & 0.02386 & 0.02501 & \textbf{\textit{0.02365}} & 0.02446 & 0.02542 \\
VaR    & 0.03030 & 0.03279 & \textbf{0.02901} & 0.03223 & 0.03243 & 0.03569 & \textbf{\textit{0.02980}} & 0.03125 & 0.03572 \\
CVaR   & 0.04965 & 0.05015 & \textbf{0.04642} & 0.05289 & 0.05105 & 0.05735 & \textbf{\textit{0.04794}} & 0.05333 & 0.05693 \\
Min    & -0.11358 & \textbf{-0.08882} & -0.09986 & -0.10067 & \textbf{\textit{-0.08882}} & -0.08887 & -0.10281 & -0.10825 & -0.10784 \\
Max    & 0.07718 & \textbf{0.08405} & 0.07422 & 0.07121 & 0.07453 & \textbf{\textit{0.08401}} & 0.07216 & 0.07444 & 0.07546 \\
MDD    & -0.54043 & -0.54815 & -0.53853 & -0.52384 & -0.53821 & -0.53115 & \textbf{\textit{-0.51835}} & -0.53384 & \textbf{-0.43730} \\
ADD    & 0.35492 & 0.36648 & 0.35899 & 0.33446 & 0.32796 & 0.32554 & 0.32467 & \textbf{\textit{0.31780}} & \textbf{0.29552} \\
SR     & 0.17618 & \textbf{0.19092} & 0.18582 & 0.15693 & \textbf{\textit{0.17095}} & 0.15626 & 0.16124 & 0.16305 & 0.08851 \\
STARR  & 0.08761 & \textbf{0.09053} & 0.08980 & 0.07296 & 0.07989 & 0.06814 & \textbf{\textit{0.07956}} & 0.07479 & 0.03953 \\
Sterling & 0.01226 & \textbf{\textit{0.01238}} & 0.01161 & 0.01154 & \textbf{0.01243} & 0.01220 & 0.01175 & 0.01230 & 0.00761 \\
Cumulative Return & 2.11184 & \textbf{2.18968} & 2.06939 & 1.93566 & \textbf{\textit{2.02296}} & 1.94965 & 1.92764 & 1.97637 & 1.43056 \\

\cmidrule(lr){1-5} \cmidrule(lr){6-9} \cmidrule(lr){10-10}
$\alpha = 0.99$ & \\
\cmidrule(lr){1-5} \cmidrule(lr){6-9} \cmidrule(lr){10-10}
Mean   & 0.00435 & \textbf{0.00503} & 0.00415 & 0.00383 & \textbf{\textit{0.00387}} & 0.00381 & 0.00369 & 0.00358 & 0.00225 \\
SD     & \textbf{0.02255} & 0.02540 & 0.02571 & 0.02473 & \textbf{\textit{0.02385}} & 0.02500 & 0.02451 & 0.02505 & 0.02542 \\
VaR    & \textbf{0.02902} & 0.03415 & 0.03297 & 0.03423 & \textbf{\textit{0.03180}} & 0.03661 & 0.03417 & 0.03254 & 0.03572 \\
CVaR   & \textbf{0.04829} & 0.05335 & 0.05680 & 0.05346 & \textbf{\textit{0.05033}} & 0.05510 & 0.05291 & 0.05512 & 0.05693 \\
Min    & -0.09967 & \textbf{\textit{-0.08894}} & -0.11340 & -0.09005 & \textbf{-0.08861} & -0.08897 & -0.10071 & -0.10583 & -0.10784 \\
Max    & 0.07388 & \textbf{0.08405} & 0.07718 & 0.07720 & 0.07453 & \textbf{\textit{0.08405}} & 0.07124 & 0.07106 & 0.07546 \\
MDD    & -0.55272 & -0.56657 & -0.55606 & -0.52701 & -0.52294 & -0.52542 & \textbf{\textit{-0.50874}} & -0.62760 & \textbf{-0.43730} \\
ADD    & 0.34228 & 0.35602 & 0.33871 & 0.32860 & 0.31875 & \textbf{\textit{0.30694}} & 0.34489 & 0.36700 & \textbf{0.29552} \\
SR     & 0.19281 & \textbf{0.19819} & 0.16143 & 0.15483 & \textbf{\textit{0.16231}} & 0.15241 & 0.15060 & 0.14295 & 0.08851 \\
STARR  & 0.09005 & \textbf{0.09434} & 0.07307 & 0.07164 & \textbf{\textit{0.07692}} & 0.06916 & 0.06975 & 0.06496 & 0.03953 \\
Sterling & 0.01270 & \textbf{0.01413} & 0.01225 & 0.01165 & \textbf{\textit{0.01242}} & 0.01242 & 0.01070 & 0.00975 & 0.00761 \\
Cumulative Return & 2.13837 & \textbf{2.39762} & 2.03221 & 1.92413 & \textbf{\textit{1.94702}} & 1.91503 & 1.87692 & 1.83432 & 1.43056 \\
\bottomrule
\end{tabular}}
\end{table*}

\begin{table*}[htbp]
\centering
\caption{Performance comparison of eight different distance based clustering portfolios under $\alpha=95\%$ and $\alpha=99\%$, including the $\mathrm{(DRMV)_A}$ model for in-sample period $T_1 = 63$ and out-of-sample period $T_2 = 5$.}
\label{Tab: 3m-5d}
\resizebox{0.70\textwidth}{!}{
\begin{tabular}{lccccccccc}
\toprule
$\alpha = 0.95$ & $\mathrm{(DRMV)_{AWDS}}$& $\mathrm{(DRMV)_{APLS}}$& $\mathrm{(DRMV)_{ACS}}$& $\mathrm{(DRMV)_{AES}}$& $\mathrm{(DRMV)_{AWDI}}$& $\mathrm{(DRMV)_{APLI}}$& $\mathrm{(DRMV)_{ACI}}$& $\mathrm{(DRMV)_{AEI}}$ & $\mathrm{(DRMV)_A}$\\
\cmidrule(lr){2-5} \cmidrule(lr){6-9}\cmidrule(lr){10-10}

Mean  
& \textbf{0.00452} & 0.00420 & 0.00318 & 0.00342 
& \textbf{\textit{0.00396}} & 0.00373 & 0.00352 & 0.00321 & 0.00124 \\

SD    
& 0.02625 & 0.02603 & 0.02653 & 0.02541 
& 0.02579 & 0.02659 & 0.02503 & \textbf{0.02458} & \textbf{\textit{0.02599}} \\

VaR   
& 0.03368 & \textbf{\textit{0.03300}} & 0.03305 & 0.03687 
& 0.03391 & 0.03398 & \textbf{0.03204} & 0.03429 & 0.03885 \\

CVaR  
& 0.05533 & \textbf{\textit{0.05477}} & 0.05882 & 0.05797 
& 0.05588 & 0.05703 & \textbf{0.05250} & 0.05412 & 0.06056 \\

Min   
& -0.10939 & \textbf{-0.08919} & -0.10931 & -0.10363 
& -0.10933 & -0.10930 & -0.10852 & \textbf{\textit{-0.10154}} & -0.10934 \\

Max   
& 0.08405 & \textbf{0.08626} & 0.08405 & 0.07688 
& 0.07622 & \textbf{\textit{0.08625}} & 0.08405 & 0.07118 & 0.08167 \\

MDD      
& -0.58347 & -0.55536 & \textbf{\textit{-0.44982}} & -0.48709  & -0.52719 & -0.51434 & -0.48782 & {-0.46205} & \textbf{-0.33517} \\

ADD      
& 0.36082  & 0.32394  & \textbf{\textit{0.29288}}  & 0.32157   
& 0.33625  & 0.32347  & 0.32283  & {0.30955} & \textbf{0.23237} \\

SR
& \textbf{0.17211} & 0.16146 & 0.11978 & 0.13452 
& \textbf{\textit{0.15354}} & 0.14031 & 0.14073 & 0.13048 & 0.04788 \\

STARR  
& \textbf{0.08166} & 0.07673 & 0.05403 & 0.05897 
& \textbf{\textit{0.07087}} & 0.06542 & 0.06709 & 0.05925 & 0.02055 \\

Sterling 
& 0.01252  & \textbf{0.01297}  & 0.01085  & 0.01063   
& \textbf{\textit{0.01178}}  & 0.01154  & 0.01091  & 0.01009 & 0.00536 \\

Cumulative Return 
& \textbf{2.16128} & 2.04184 & 1.68567 & 1.77148 
& \textbf{\textit{1.95434}} & 1.86618 & 1.80901 & 1.71527 & 1.18244 \\

\cmidrule(lr){2-5} \cmidrule(lr){6-9}\cmidrule(lr){10-10}
$\alpha = 0.99$ & \\
\cmidrule(lr){2-5} \cmidrule(lr){6-9}\cmidrule(lr){10-10}

Mean  
& 0.00403 & \textbf{0.00440} & 0.00359 & 0.00388 
& \textbf{\textit{0.00396}} & 0.00372 & 0.00368 & 0.00354 & 0.00124 \\

SD    
& 0.02599 & 0.02625 & 0.02577 & \textbf{\textit{0.02570}} 
& 0.02582 & 0.02661 & 0.02661 & \textbf{0.02532} & 0.02599 \\

VaR   
& 0.03288 & \textbf{0.03198} & 0.03278 & 0.03224 
& \textbf{\textit{0.03386}} & 0.03446 & 0.03386 & 0.03518 & 0.03885 \\

CVaR  
& \textbf{0.05373} & 0.05576 & 0.05541 & 0.05505 
& \textbf{\textit{0.05558}} & 0.05816 & 0.05798 & 0.05685 & 0.06056 \\

Min   
& -0.08922 & -0.08918 & \textbf{-0.08886} & -0.10937 
& -0.10930 & -0.10927 & -0.10927 & \textbf{\textit{-0.10600}} & -0.10934 \\

Max   
& 0.08405 & 0.08405 & \textbf{0.08812} & 0.08031 
& 0.07619 & \textbf{\textit{0.08405}} & 0.08405 & 0.07720 & 0.08167 \\

MDD      
& -0.54256 & -0.57201 & -0.49208 & -0.52805   
& -0.52723 & -0.51427 & \textbf{\textit{-0.51077}} & -0.49684 & \textbf{-0.33517} \\

ADD      
& 0.32412  & 0.34011  & 0.31638  & 0.31702    
& 0.32206  & 0.31672  & \textbf{\textit{0.31583}}  & 0.31689 & \textbf{0.23237} \\

SR
& 0.15494 & \textbf{0.16742} & 0.13933 & 0.15103 
& \textbf{\textit{0.15342}} & 0.13980 & 0.13834 & 0.13983 & 0.04788 \\

STARR  
& 0.07495 & \textbf{0.07883} & 0.06480 & 0.07051 
& \textbf{\textit{0.07128}} & 0.06396 & 0.06348 & 0.06229 & 0.02055 \\

Sterling 
& 0.01242  & \textbf{0.01292}  & 0.01135  & 0.01224   
& \textbf{\textit{0.01230}}  & 0.01174  & 0.01165  & 0.01117 & 0.00536 \\

Cumulative Return 
& 1.97723 & \textbf{2.11324} & 1.82600 & 1.92686 
& \textbf{\textit{1.95462}} & 1.86212 & 1.84881 & 1.81270 & 1.18244 \\

\bottomrule
\end{tabular}}
\end{table*}

\begin{table*}[htbp]
\centering
\caption{Performance comparison of eight different distance based clustering portfolios under $\alpha=95\%$ and $\alpha=99\%$, including the $\mathrm{(DRMV)_A}$ model for in-sample period $T_1 = 63$ and out-of-sample period $T_2 = 10$.}
\label{Tab: 3m-10d}
\resizebox{0.70\textwidth}{!}{
\begin{tabular}{lccccccccc}
\toprule
$\alpha = 0.95$ & $\mathrm{(DRMV)_{AWDS}}$& $\mathrm{(DRMV)_{APLS}}$& $\mathrm{(DRMV)_{ACS}}$& $\mathrm{(DRMV)_{AES}}$
 & $\mathrm{(DRMV)_{AWDI}}$& $\mathrm{(DRMV)_{APLI}}$& $\mathrm{(DRMV)_{ACI}}$& $\mathrm{(DRMV)_{AEI}}$ & $\mathrm{(DRMV)_A}$\\
\cmidrule(lr){2-5} \cmidrule(lr){6-9}\cmidrule(lr){10-10}

Mean   
& \textbf{0.00405} & 0.00395 & 0.00315 & 0.00322 
& \textbf{\textit{0.00381}} & 0.00343 & 0.00309 & 0.00322 & 0.00104 \\

SD     
& 0.02541 & 0.02512 & 0.02652 & 0.02393 
& 0.02290 & 0.02528 & 0.02267 & 0.02517 & \textbf{0.02259} \\

VaR    
& 0.03286 & 0.03342 & 0.03238 & 0.03219 
& 0.02969 & 0.03365 & \textbf{\textit{0.02773}} & 0.03245 & \textbf{0.03643} \\

CVaR   
& 0.05445 & 0.05659 & 0.06017 & 0.05383 
& 0.04675 & 0.05607 & \textbf{\textit{0.04527}} & 0.05383 & \textbf{0.05558} \\

Min    
& -0.08995 & \textbf{-0.08939} & -0.10937 & -0.09118 
& \textbf{\textit{-0.08998}} & -0.09014 & -0.09008 & -0.10945 & -0.09303 \\

Max    
& 0.08405 & 0.07720 & \textbf{0.08498} & 0.07531 
& \textbf{\textit{0.07720}} & 0.07720 & 0.07474 & 0.07412 & 0.05980 \\

MDD      
& -0.55573 & -0.54975 & \textbf{\textit{-0.45231}} & -0.49105 
& -0.53167 & -0.50177 & -0.50799 & -0.52291 & \textbf{-0.34580} \\

ADD      
& 0.35065  & 0.35120  & \textbf{\textit{0.27975}}  & 0.31640   
& 0.36975  & 0.33891  & {0.33163}  & 0.35076  & \textbf{0.23193} \\

SR
& 0.15940 & 0.15738 & 0.11869 & 0.13458 
& \textbf{0.16618} & 0.13585 & 0.13627 & 0.12804 & 0.04617 \\

STARR  
& 0.07439 & 0.06985 & 0.05231 & 0.05983 
& \textbf{0.08140} & 0.06125 & 0.06822 & 0.05987 & 0.01876 \\

Sterling 
& \textbf{0.01155}  & 0.01126  & 0.01125  & 0.01018   
& \textbf{\textit{0.01029}}  & 0.01013  & 0.00931  & 0.00919  & 0.00450 \\

Cumulative Return 
& \textbf{1.95437} & 1.92273 & 1.65297 & 1.69466 
& \textbf{\textit{1.89067}} & 1.75076 & 1.66410 & 1.68625 & 1.15207 \\

\cmidrule(lr){2-5} \cmidrule(lr){6-9}\cmidrule(lr){10-10}
$\alpha = 0.99$ & \\
\cmidrule(lr){2-5} \cmidrule(lr){6-9}\cmidrule(lr){10-10}

Mean   
& 0.00394 & \textbf{0.00407} & 0.00400 & 0.00375 
& \textbf{\textit{0.00372}} & 0.00346 & 0.00341 & 0.00328 & 0.00104 \\

SD     
& 0.02492 & 0.02551 & 0.02754 & 0.02528 
& \textbf{0.02378} & 0.02622 & 0.02626 & 0.02456 & \textbf{0.02259} \\

VaR    
& \textbf{0.03074} & 0.03249 & 0.03877 & 0.03290 
& \textbf{\textit{0.03102}} & 0.03413 & 0.03720 & 0.03222 & 0.03643 \\

CVaR   
& 0.05336 & 0.05826 & 0.05684 & 0.05397 
& \textbf{0.05164} & 0.06292 & 0.05680 & 0.05325 & 0.05558 \\

Min    
& -0.08955 & \textbf{-0.08940} & -0.10941 & -0.09016 
& -0.10939 & -0.10937 & \textbf{\textit{-0.08996}} & -0.10855 & -0.09303 \\

Max    
& 0.08405 & 0.08405 & \textbf{0.11887} & 0.08405 
& 0.07557 & 0.08405 & \textbf{\textit{0.09945}} & 0.07473 & 0.05980 \\

MDD      
& -0.53984 & -0.54713 & -0.51892 & -0.53151 
& -0.51591 & -0.49352 & -0.48534 & \textbf{\textit{-0.47427}} & \textbf{-0.34580} \\

ADD      
& 0.33592  & 0.33176  & 0.32700  & 0.33543   
& 0.31956  & 0.30777  & \textbf{\textit{0.30264}}  & 0.32015  & \textbf{0.23193} \\

SR
& 0.15815 & \textbf{0.15947} & 0.14519 & 0.14839 
& \textbf{\textit{0.15623}} & 0.13213 & 0.12987 & 0.13366 & 0.04617 \\

STARR  
& \textbf{0.07387} & 0.06983 & 0.07034 & 0.06952 
& \textbf{\textit{0.07195}} & 0.05505 & 0.06004 & 0.06165 & 0.01876 \\

Sterling 
& 0.01173  & \textbf{0.01226}  & 0.01223  & 0.01118   
& \textbf{\textit{0.01163}}  & 0.01125  & 0.01127  & 0.01025  & 0.00450 \\

Cumulative Return 
& 1.92069 & \textbf{1.95954} & 1.91722 & 1.85335 
& \textbf{\textit{1.85305}} & 1.75185 & 1.73563 & 1.70896 & 1.15207 \\

\bottomrule
\end{tabular}}
\end{table*}

\begin{table*}[htbp]
\centering
\caption{Performance of four benchmark models for an in-sample period $T_1 = 63$ (3 months) and out-of-sample period $T_2 = 187$ (9 months), i.e., without rebalancing.}
\label{tab: benchmark_3m_3d}
\resizebox{0.70\textwidth}{!}{
\begin{tabular}{lcccccc}
\toprule
 & \multicolumn{2}{c}{$\mathrm{B\&H_S}$} & \multicolumn{2}{c}{$\mathrm{B\&H_I}$} & \multirow{2}{*}{Na\"ive} & \multirow{2}{*}{Index} \\
\cmidrule(lr){2-3} \cmidrule(lr){4-5}
 & $\alpha = 0.95$ & $\alpha = 0.99$ & $\alpha = 0.95$ & $\alpha = 0.99$ &  &  \\
\midrule
Mean   & 0.00124 & 0.00141 & 0.00102 & 0.00095 & 0.00088 & 0.00130 \\
SD     & 0.01181 & 0.01341 & 0.01077 & 0.01165 & 0.01248 & 0.01250 \\
VaR    & 0.01576 & 0.01517 & 0.01479 & 0.01436 & 0.01270 & 0.01366 \\
CVaR   & 0.02920 & 0.02992 & 0.02729 & 0.02719 & 0.02707 & 0.02793 \\
Min    & -0.06406 & -0.07143 & -0.05473 & -0.05564 & -0.05887 & -0.06393 \\
Max    & 0.06541 & 0.07533 & 0.06549 & 0.06450 & 0.07463 & 0.07812 \\
MDD    & -0.23731 & -0.27295 & -0.20484 & -0.19474 & -0.23721 & -0.28579 \\
ADD    & 0.09369 & 0.09159 & 0.08626 & 0.08122 & 0.06490 & 0.08812 \\
SR     & 0.10499 & 0.10532 & 0.09471 & 0.08152 & 0.07071 & 0.10442 \\
STARR  & 0.04242 & 0.04713 & 0.03746 & 0.03492 & 0.03259 & 0.04672 \\
Sterling & 0.01323 & 0.01539 & 0.01185 & 0.01169 & 0.01359 & 0.01481 \\
Cumulative Return & 1.24088 & 1.27874 & 1.19395 & 1.17821 & 1.16144 & 1.25628 \\
\bottomrule
\end{tabular}}
\end{table*}

% \begin{frame}{Cumulative Returns}
%     \begin{columns}[T]
%         \begin{column}{0.5\textwidth}
%             \begin{figure}
%                 \centering
%                 \fbox{\includegraphics[width=0.95\linewidth, height=3.5cm]{Graphs/3M3D.png}}
%                 \caption{Caption 1}
%                 \label{fig:fig1}
%             \end{figure}
%         \end{column}
%         \begin{column}{0.5\textwidth}
%             \begin{figure}
%                 \centering
%                 \fbox{\includegraphics[width=0.95\linewidth, height=3.5cm]{Graphs/4M3D.png}}
%                 \caption{Caption 2}
%                 \label{fig:fig2}
%             \end{figure}
%         \end{column}
%     \end{columns}
% \end{frame}

% \begin{figure}
%     \centering
%     \fbox{\includegraphics[width=0.8\linewidth]{3M3D.png}}
%     \caption{Out-of-sample cumulative returns obtained under a 3-month in-sample period with portfolio rebalancing every 3 days for all considered models, including the benchmark models $\mathrm{B\&H_S}$, $\mathrm{B\&H_I}$, Naïve, and Index.}
%     \label{fig: 3M3D}
% \end{figure}

% \begin{figure}
%     \centering
%     \fbox{\includegraphics[width=0.8\linewidth]{4M3D.png}}
%     \caption{Out-of-sample cumulative returns obtained under a 4-month in-sample period with portfolio rebalancing every 3 days for all considered models, including the benchmark models $\mathrm{B\&H_S}$, $\mathrm{B\&H_I}$, Naïve, and Index.}
%     \label{fig: 4M3D}
% \end{figure}

% \begin{figure}
%     \centering
%     \fbox{\includegraphics[width=0.8\linewidth]{3M3D_War.png}}
%     \caption{Out-of-sample cumulative returns obtained under a 3-month in-sample period with portfolio rebalancing every 3 days for all considered models, including the benchmark models $\mathrm{B\&H_S}$, $\mathrm{B\&H_I}$, Naïve, and Index during the U.S.--Israel--Iran conflict.}
%     \label{fig: 3M3D_war}
% \end{figure}

\begin{table*}[ht]
\centering
\caption{Transaction cost (\%) for eight clustering based portfolios under $\alpha=95\%$ including $\mathrm{(DRMV)_A}$ model, for in-sample period $T_1 = 63$ and out-of-sample period $T_2 = 3, 5, \text{ and, } 10$. The table reports results along with $\mathrm{B\&H_S}$ and $\mathrm{B\&H_I}$ models, which are implemented without rebalancing.}
\label{tab: tc_values}
\renewcommand{\arraystretch}{1.2}
\resizebox{0.70\textwidth}{!}{
\begin{tabular}{lccc lccc}
\hline
TC & 3D & 5D & 10D & TC & 3D & 5D & 10D \\
\hline
$\mathrm{(DRMV)_{AWDS}}$ & 3.68174 & 3.35026 & 2.86638 & $\mathrm{(DRMV)_{AWDI}}$ & 3.34919 & 3.10257 & 2.70163 \\
$\mathrm{(DRMV)_{APLS}}$ & 3.74900 & 3.30013 & 2.95862   & $\mathrm{(DRMV)_{APLI}}$ & 3.14923 & 2.85087 & 2.79962 \\
$\mathrm{(DRMV)_{ACS}}$  & 3.55926 & 2.98252 & 2.33242  & $\mathrm{(DRMV)_{ACI}}$  & 2.92264 & 2.65826 & 2.37898 \\
$\mathrm{(DRMV)_{AES}}$  & 3.21652 & 3.06161 & 2.64509  & $\mathrm{(DRMV)_{AEI}}$  & 3.36298 & 2.76815 & 2.69718 \\ 
$\mathrm{B\&H_S}$  & 0.34988 & 0.34988 & 0.34988 & $\mathrm{B\&H_I}$  & 0.32487 & 0.32487 & 0.32487 \\
$\mathrm{(DRMV)_A}$ & 3.91892 & 3.32951 & 2.93689 & \\
\hline
\end{tabular}}
\end{table*}

\subsection{Sensitivity check to in-sample window size}
To examine the robustness of the proposed strategy, we extend the in-sample period from 63 trading days (approximately 3 months) to 84 trading days (approximately 4 months). Tables \ref{tab: 4m_3d}, \ref{Tab: 4m-5d}, and \ref{Tab: 4m-10d} present the average out-of-sample performance when the portfolio is rebalanced every 3 days, 5 days, and 10 days, respectively.

In particular, the TDA-based sentiment-indicator based filtering and allocation framework, followed by the (DRMV) model, continues to exhibit relatively stronger mean returns (see Figure \ref{fig: cumulative returns} (middle)) and superior risk-adjusted performance across most configurations. Among the proposed strategies, \(\mathrm{(DRMV)_{APLS}}\) remains one of the best-performing specifications followed by \(\mathrm{(DRMV)_{AWDS}}\) in terms of return generation and risk-adjusted measures, while also maintaining competitive performance across all three rebalancing horizons, namely 3-day, 5-day, and 10-day (see Tables \ref{tab: 4m_3d}, \ref{Tab: 4m-5d}, and \ref{Tab: 4m-10d}). In addition, we compare Tables \ref{tab: 4m_3d}, \ref{Tab: 4m-5d}, and \ref{Tab: 4m-10d} across the three rebalancing frequencies of 3, 5, and 10 days. In general, the results indicate that the 3-day rebalancing strategy yields higher mean returns and stronger risk-adjusted measures in most cases, highlighting the ability of the proposed framework to effectively capture short-term market movements through more frequent portfolio revision.

These filtering-based models are also compared with the one-stage $\mathrm{(DRMV)_A}$ model. Across all configurations, the proposed two-stage framework consistently yields more favorable mean returns, cumulative returns, and risk-adjusted measures than $\mathrm{(DRMV)_A}$ benchmark model across all eight filtering-based specifications, including both TDA and non-TDA based approaches under sentiment-indicator and indicator-only settings. Models in Tables \ref{tab: 4m_3d}, \ref{Tab: 4m-5d}, and \ref{Tab: 4m-10d} outperform all benchmark models in Table \ref{tab: benchmark_4m_3d} in terms of returns and risk-adjusted ratios. This indicates that the preliminary filtering stage and dynamic rebalancing play an important role in removing less informative assets and enabling the optimization model to focus on a more relevant and investable subset of stocks.

Overall, the consistency of results across both 63 days and 84 days in-sample specifications indicates that the proposed two-stage framework is robust to changes in the estimation window. The empirical findings confirm that incorporating sentiment information and employing TDA-based distance measures consistently improve portfolio performance in terms of returns across different sample configurations and rebalancing horizons.

\begin{table*}[htbp]
\centering
\caption{Performance comparison of eight different distance based clustering portfolios under $\alpha=95\%$ and $\alpha=99\%$, including the $\mathrm{(DRMV)_A}$ model for in-sample period $T_1 = 84$ and out-of-sample period $T_2 = 3$.}
\label{tab: 4m_3d}
\resizebox{0.70\textwidth}{!}{
\begin{tabular}{lccccccccc}
\toprule
$\alpha = 0.95$ 
& $\mathrm{(DRMV)_{AWDS}}$ & $\mathrm{(DRMV)_{APLS}}$ & $\mathrm{(DRMV)_{ACS}}$ & $\mathrm{(DRMV)_{AES}}$
& $\mathrm{(DRMV)_{AWDI}}$ & $\mathrm{(DRMV)_{APLI}}$ & $\mathrm{(DRMV)_{ACI}}$ & $\mathrm{(DRMV)_{AEI}}$ & $\mathrm{(DRMV)_A}$ \\
\cmidrule(lr){2-5} \cmidrule(lr){6-9}\cmidrule(lr){10-10}
Mean   & 0.00448 & \textbf{0.00506} & 0.00431 & 0.00429 & \textbf{\textit{0.00428}} & 0.00414 & 0.00394 & 0.00395 & 0.00365 \\
SD     & 0.02352 & 0.02489 & 0.02500 & 0.02567 & \textbf{\textit{0.02278}} & 0.02537 & 0.02556 & 0.02545 & \textbf{0.02257} \\
VaR    & 0.03401 & 0.03448 & 0.03167 & 0.03640 & \textbf{\textit{0.03160}} & 0.03445 & 0.03342 & 0.03439 & \textbf{0.03116} \\
CVaR   & 0.04882 & 0.05200 & 0.05449 & 0.05384 & \textbf{{0.04603}} & 0.05471 & 0.05593 & 0.05509 & \textbf{\textit{0.04852}} \\
Min    & -0.09028 & -0.09011 & -0.11033 & -0.11734 & \textbf{\textit{-0.08997}} & -0.10944 & -0.11602 & -0.11141 & \textbf{-0.08452} \\
Max    & 0.06905 & 0.08495 & \textbf{\textit{0.08573}} & 0.07514 & 0.06870 & 0.08369 & \textbf{0.09192} & 0.08510 & 0.07377 \\
MDD    & -0.53068 & -0.55730 & -0.51065 & -0.50703 & -0.51349 & -0.49640 & \textbf{-0.46464} & -0.47997 & \textbf{\textit{-0.48904}} \\
ADD    & 0.34751 & 0.36860 & \textbf{\textit{0.32127}} & 0.33914 & 0.34654 & 0.32341 & \textbf{0.31291} & 0.31461 & 0.33032 \\
SR     & 0.19033 & \textbf{0.20340} & 0.17227 & 0.16696 & \textbf{\textit{0.18768}} & 0.16338 & 0.15410 & 0.15511 & 0.16179 \\
STARR  & 0.09168 & \textbf{0.09736} & 0.07902 & 0.07959 & \textbf{\textit{0.09290}} & 0.07575 & 0.07043 & 0.07166 & 0.07526 \\
Sterling & 0.01288 & \textbf{0.01374} & 0.01340 & 0.01264 & 0.01234 & \textbf{\textit{0.01281}} & 0.01259 & 0.01255 & 0.01106 \\
Cumulative Return & 1.99693 & \textbf{2.18728} & 1.93032 & 1.91830 & \textbf{\textit{1.93792}} & 1.87684 & 1.81306 & 1.81644 & 1.75018 \\
\cmidrule(lr){2-5} \cmidrule(lr){6-9}\cmidrule(lr){10-10}
$\alpha = 0.99$ 
&  \\
\cmidrule(lr){2-5} \cmidrule(lr){6-9}\cmidrule(lr){10-10}
Mean   & \textbf{0.00451} & 0.00418 & 0.00412 & 0.00434 & \textbf{\textit{0.00446}} & 0.00401 & 0.00394 & 0.00403 & 0.00365 \\
SD     & 0.02493 & 0.02504 & 0.02586 & 0.02426 & \textbf{\textit{0.02288}} & 0.02508 & 0.02556 & 0.02526 & \textbf{0.02257} \\
VaR    & 0.03733 & 0.03154 & 0.03494 & 0.03271 & \textbf{\textit{0.03152}} & 0.03330 & 0.03342 & 0.03609 & \textbf{0.03116} \\
CVaR   & 0.05509 & 0.05408 & 0.05567 & 0.05057 & \textbf{\textit{0.04726}} & 0.05230 & 0.05593 & 0.05356 & 0.04852 \\
Min    & -0.11734 & -0.10930 & -0.11613 & -0.09402 & \textbf{\textit{-0.08993}} & -0.11403 & -0.11602 & -0.11331 & \textbf{-0.08452} \\
Max    & 0.08356 & \textbf{\textit{0.08499}} & 0.08459 & 0.08154 & 0.06867 & 0.09133 & \textbf{0.09192} & 0.08494 & 0.07377 \\
MDD    & -0.51032 & \textbf{\textit{-0.47310}} & -0.49245 & -0.51063 & -0.48810 & -0.46787 & \textbf{-0.46464} & -0.48699 & {-0.48904} \\
ADD    & 0.33340 & \textbf{0.29197} & 0.31933 & 0.34620 & 0.33588 & 0.30275 & \textbf{\textit{0.29896}} & 0.32138 & {0.33032} \\
SR     & 0.18079 & 0.16683 & 0.15936 & 0.17873 & \textbf{0.19511} & 0.15974 & 0.15410 & 0.15940 & 0.16179 \\
STARR  & 0.08181 & 0.07724 & 0.07402 & \textbf{\textit{0.08573}} & \textbf{0.09446} & 0.07659 & 0.07043 & 0.07517 & 0.07526 \\
Sterling & 0.01352 & \textbf{0.01431} & 0.01290 & 0.01252 & 0.01328 & \textbf{\textit{0.01323}} & 0.01318 & 0.01253 & 0.01106 \\
Cumulative Return & \textbf{\textit{1.99549}} & 1.88947 & 1.86569 & 1.94577 & \textbf{1.99791} & 1.83693 & 1.81306 & 1.84145 & 1.75018 \\
\bottomrule
\end{tabular}}
\end{table*}

\begin{table*}[htbp]
\centering
\caption{Performance comparison of eight different distance based clustering portfolios under $\alpha=95\%$ and $\alpha=99\%$, including the $\mathrm{(DRMV)_A}$ model for in-sample period $T_1 = 84$ and out-of-sample period $T_2 = 5$.}
\label{Tab: 4m-5d}
\resizebox{0.70\textwidth}{!}{
\begin{tabular}{lccccccccc}
\toprule
$\alpha = 0.95$ & $\mathrm{(DRMV)_{AWDS}}$& $\mathrm{(DRMV)_{APLS}}$& $\mathrm{(DRMV)_{ACS}}$& $\mathrm{(DRMV)_{AES}}$& $\mathrm{(DRMV)_{AWDI}}$& $\mathrm{(DRMV)_{APLI}}$& $\mathrm{(DRMV)_{ACI}}$& $\mathrm{(DRMV)_{AEI}}$ & $\mathrm{(DRMV)_A}$\\
\cmidrule(lr){2-5} \cmidrule(lr){6-9}\cmidrule(lr){10-10} 

Mean  
& 0.00449 & \textbf{0.00469} & 0.00431 & 0.00409 
& \textbf{\textit{0.00416}} & 0.00410 & 0.00385 & 0.00392 & 0.00370 \\

SD    
& \textbf{0.02300} & 0.02518 & 0.02368 & 0.02561 
& 0.02472 & 0.02409 & 0.02547 & \textbf{\textit{0.02334}} & 0.02325 \\

VaR   
& 0.03210 & 0.03322 & \textbf{\textit{0.03205}} & 0.03466 
& 0.03342 & 0.03657 & 0.03646 & \textbf{0.02926} & 0.03219 \\

CVaR  
& \textbf{0.04723} & 0.05059 & 0.05009 & 0.05354 
& 0.05182 & 0.05116 & 0.05574 & \textbf{\textit{0.04730}} & 0.04919 \\

Min   
& \textbf{\textit{-0.07638}} & -0.09213 & -0.09368 & -0.11752 
& -0.09506 & -0.09138 & -0.11282 & \textbf{-0.09033} & -0.08953 \\

Max   
& 0.07662 & \textbf{\textit{0.09017}} & 0.08457 & 0.08456 
& \textbf{0.09021} & 0.07627 & 0.08457 & 0.06881 & 0.06089 \\

MDD      
& -0.52899 & -0.53015 & -0.51385 & -0.49949 
& -0.50400 & -0.50347 & \textbf{-0.48575} & -0.48992 & \textbf{\textit{-0.49259}} \\

ADD      
& 0.35275  & 0.34869  & 0.33724  & 0.32855   
& 0.32584  & 0.33810  & \textbf{0.30439}  & 0.33622  & \textbf{\textit{0.33720}} \\

SR
& \textbf{0.19533} & 0.18613 & 0.18211 & 0.15960 
& \textbf{\textit{0.17016}} & 0.16835 & 0.15132 & 0.16784 & 0.15931 \\

STARR  
& \textbf{0.09513} & 0.09264 & 0.08611 & 0.07634 
& 0.08030 & 0.08014 & 0.06915 & \textbf{\textit{0.08281}} & 0.07530 \\

Sterling 
& 0.01274  & \textbf{0.01344}  & 0.01279  & 0.01244   
& \textbf{\textit{0.01277}}  & 0.01213  & 0.01266  & 0.01165  & 0.01099 \\

Cumulative Return 
& 2.00672 & \textbf{2.05403} & 1.94315 & 1.85727 
& \textbf{\textit{1.88747}} & 1.87309 & 1.78859 & 1.82304 & 1.76077 \\

\cmidrule(lr){2-5} \cmidrule(lr){6-9}\cmidrule(lr){10-10}

$\alpha = 0.99$ &\\
\cmidrule(lr){2-5} \cmidrule(lr){6-9}\cmidrule(lr){10-10}

Mean  
& 0.00448 & \textbf{0.00479} & 0.00411 & 0.00410 
& \textbf{\textit{0.00437}} & 0.00398 & 0.00366 & 0.00378 & 0.00370 \\

SD    
& 0.02584 & 0.02539 & 0.02565 & 0.02575 
& 0.02400 & 0.02360 & 0.02547 & \textbf{0.02319} & \textbf{\textit{0.02325}} \\

VaR   
& 0.03950 & 0.03400 & 0.03408 & 0.03577 
& 0.03430 & 0.03379 & 0.03478 & \textbf{0.03019} & \textbf{\textit{0.03219}} \\

CVaR  
& 0.05489 & 0.05348 & 0.05529 & 0.05548 
& 0.05048 & 0.04839 & 0.05539 & \textbf{0.04702} & \textbf{\textit{0.04919}} \\

Min   
& -0.11728 & -0.10937 & -0.11584 & -0.11909 
& -0.09512 & -0.08992 & -0.11280 & \textbf{\textit{-0.09002}} & \textbf{-0.08953} \\

Max   
& 0.08458 & \textbf{\textit{0.08459}} & \textbf{\textit{0.08459}} & 0.07661 
& 0.07628 & \textbf{0.10406} & 0.08572 & 0.06878 & 0.06089 \\

MDD      
& -0.50678 & -0.53013 & -0.49625 & -0.49815 
& -0.51756 & -0.46730 & \textbf{-0.45128} & -0.48579 & \textbf{\textit{-0.49259}} \\

ADD      
& \textbf{\textit{0.30873}}  & 0.33176  & 0.30878  & 0.31248   
& 0.34601  & 0.31608  & \textbf{0.26715}  & 0.33202  & 0.33720 \\

SR
& 0.17334 & \textbf{0.18860} & 0.16031 & 0.15918 
& \textbf{\textit{0.18228}} & 0.16858 & 0.14366 & 0.16295 & 0.15931 \\

STARR  
& 0.08161 & \textbf{0.08953} & 0.07437 & 0.07388 
& \textbf{\textit{0.08666}} & 0.08222 & 0.06606 & 0.08037 & 0.07530 \\

Sterling 
& \textbf{0.01451}  & 0.01443  & 0.01332  & 0.01312   
& 0.01263  & 0.01259  & \textbf{\textit{0.01370}}  & 0.01138  & 0.01099 \\

Cumulative Return 
& 1.97893 & \textbf{2.08622} & 1.86454 & 1.85970 
& \textbf{\textit{1.96020}} & 1.84020 & 1.73204 & 1.78304 & 1.76077 \\

\bottomrule
\end{tabular}}
\end{table*}

\begin{table*}[htbp]
\centering
\caption{Performance comparison of eight different distance based clustering portfolios under $\alpha=95\%$ and $\alpha=99\%$, including the $\mathrm{(DRMV)_A}$ model for in-sample period $T_1 = 84$ and out-of-sample period $T_2 = 10$.}
\label{Tab: 4m-10d}
\resizebox{0.70\textwidth}{!}{
\begin{tabular}{lccccccccc}
\toprule
$\alpha = 0.95$ & $\mathrm{(DRMV)_{AWDS}}$& $\mathrm{(DRMV)_{APLS}}$& $\mathrm{(DRMV)_{ACS}}$& $\mathrm{(DRMV)_{AES}}$& $\mathrm{(DRMV)_{AWDI}}$& $\mathrm{(DRMV)_{APLI}}$& $\mathrm{(DRMV)_{ACI}}$& $\mathrm{(DRMV)_{AEI}}$ & $\mathrm{(DRMV)_A}$\\
\cmidrule(lr){2-5} \cmidrule(lr){6-9}\cmidrule(lr){10-10}

Mean  
& 0.00419 & \textbf{0.00472} & 0.00402 & 0.00423 
& 0.00394 & \textbf{\textit{0.00402}} & 0.00381 & 0.00392 & 0.00336 \\

SD    
& \textbf{0.02313} & 0.02552 & 0.02425 & 0.02445 
& \textbf{\textit{0.02345}} & 0.02484 & 0.02663 & 0.02347 & 0.02417 \\

VaR   
& \textbf{\textit{0.03187}} & 0.03354 & 0.03278 & 0.03601 
& \textbf{0.02933} & 0.03773 & 0.03968 & 0.03594 & 0.03380 \\

CVaR  
& \textbf{\textit{0.05047}} & 0.05217 & 0.05291 & 0.05304 
& \textbf{0.04664} & 0.05440 & 0.06034 & 0.05140 & 0.05792 \\

Min   
& -0.09029 & -0.09218 & -0.09022 & -0.09095 
& -0.08697 & -0.09186 & -0.11563 & \textbf{-0.08114} & \textbf{\textit{-0.08844}} \\

Max   
& 0.06902 & \textbf{0.09019} & 0.08456 & 0.08350 
& 0.06883 & \textbf{\textit{0.08837}} & 0.08456 & 0.08456 & 0.06344 \\

MDD      
& -0.50019 & -0.54002 & -0.48460 & -0.50224 
& -0.50142 & -0.48691 & \textbf{\textit{-0.46737}} & -0.48192 & \textbf{-0.44757} \\

ADD      
& 0.34833  & 0.38091  & 0.32764  & 0.35051  
& 0.35810  & 0.33053  & \textbf{\textit{0.31407}}  & 0.33456 & \textbf{0.29737} \\

SR
& 0.18110 & \textbf{0.18500} & 0.16569 & 0.17285 
& \textbf{\textit{0.16812}} & 0.16192 & 0.14317 & 0.16717 & 0.13901 \\

STARR  
& 0.08301 & \textbf{0.09050} & 0.07594 & 0.07969 
& \textbf{\textit{0.08452}} & 0.07393 & 0.06318 & 0.07631 & 0.05801 \\

Sterling 
& 0.01203  & \textbf{0.01239}  & 0.01226  & 0.01206  
& 0.01101  & \textbf{\textit{0.01216}}  & 0.01214  & 0.01172 & 0.01130 \\

Cumulative Return 
& 1.87079 & \textbf{2.01812} & 1.81298 & 1.87287 
& 1.79675 & \textbf{\textit{1.80996}} & 1.73727 & 1.79093 & 1.63236 \\

\cmidrule(lr){2-5} \cmidrule(lr){6-9}\cmidrule(lr){10-10}

$\alpha = 0.99$ &\\
\cmidrule(lr){2-5} \cmidrule(lr){6-9}\cmidrule(lr){10-10}

Mean  
& 0.00423 & \textbf{0.00440} & 0.00418 & 0.00420 
& 0.00405 & \textbf{\textit{0.00407}} & 0.00369 & 0.00383 & 0.00336 \\

SD    
& \textbf{0.02336} & 0.02479 & 0.02609 & 0.02445 
& \textbf{\textit{0.02365}} & 0.02440 & 0.02622 & 0.02485 & 0.02417 \\

VaR   
& \textbf{0.02977} & 0.03793 & 0.03288 & 0.03601 
& \textbf{\textit{0.03186}} & 0.03292 & 0.03966 & 0.03774 & 0.03380 \\

CVaR  
& \textbf{\textit{0.05090}} & 0.05218 & 0.05861 & 0.05304 
& \textbf{0.04950} & 0.05113 & 0.06020 & 0.05363 & 0.05792 \\

Min   
& -0.09025 & -0.09008 & -0.11729 & -0.09091 
& \textbf{-0.08590} & {-0.08598} & -0.11493 & -0.08769 & \textbf{\textit{-0.08844}} \\

Max   
& 0.06897 & 0.08348 & \textbf{\textit{0.08457}} & 0.08350 
& 0.06909 & \textbf{0.09196} & 0.08456 & 0.08457 & 0.06344 \\

MDD      
& -0.49894 & -0.49412 & -0.49323 & -0.49977 
& -0.50961 & -0.49011 & \textbf{\textit{-0.45814}} & -0.47146 & \textbf{-0.44757} \\

ADD      
& 0.35278  & 0.33704  & 0.32077  & 0.35024  
& 0.36049  & 0.33915  & \textbf{\textit{0.30825}}  & 0.32391 & \textbf{0.29737} \\

SR
& \textbf{0.18096} & 0.17744 & 0.16007 & 0.17162 
& \textbf{\textit{0.17132}} & 0.16700 & 0.14091 & 0.15398 & 0.13901 \\

STARR  
& \textbf{0.08304} & 0.08429 & 0.07126 & 0.07911 
& \textbf{\textit{0.08185}} & 0.07969 & 0.06138 & 0.07134 & 0.05801 \\

Sterling 
& 0.01198  & \textbf{0.01305} & 0.01302  & 0.01198  
& 0.01124  & \textbf{\textit{0.01200}} & 0.01199 & 0.01181 & 0.01130 \\

Cumulative Return 
& 1.88052 & \textbf{1.92239} & 1.84511 & 1.86358 
& 1.82700 & \textbf{\textit{1.82864}} & 1.70777 & 1.75422 & 1.63236 \\

\bottomrule
\end{tabular}}
\end{table*}

\begin{table*}[htbp]
\centering
\caption{Performance of four benchmark models for an in-sample period $T_1 = 84$ (4 months) and out-of-sample period $T_2 = 166$ (8 months), i.e., without rebalancing.}
\label{tab: benchmark_4m_3d}
% \resizebox{0.70\textwidth}{!}{
\footnotesize
\begin{tabular}{lcccccc}
\toprule
 & \multicolumn{2}{c}{$\mathrm{B\&H_S}$} & \multicolumn{2}{c}{$\mathrm{B\&H_I}$} & \multirow{2}{*}{Na\"ive} & \multirow{2}{*}{Index} \\
\cmidrule(lr){2-3} \cmidrule(lr){4-5}
 & $\alpha = 0.95$ & $\alpha = 0.99$ & $\alpha = 0.95$ & $\alpha = 0.99$ &  &  \\
\midrule
Mean   & 0.00164 & 0.00176 & 0.00162 & 0.00143 & 0.00086 & 0.00123 \\
SD     & 0.00929 & 0.01011 & 0.00908 & 0.01077 & 0.00730 & 0.00731 \\
VaR    & 0.01802 & 0.01512 & 0.01890 & 0.01429 & 0.01215 & 0.01122 \\
CVaR   & 0.02033 & 0.02051 & 0.02131 & 0.01989 & 0.01445 & 0.01694 \\
Min    & -0.03416 & -0.03384 & -0.03493 & -0.03609 & -0.01744 & -0.02704 \\
Max    & 0.02962 & 0.03353 & 0.03234 & 0.03055 & 0.02373 & 0.02246 \\
MDD    & -0.22619 & -0.22776 & -0.218286 & -0.21132 & -0.13830 & -0.19177 \\
ADD    & 0.09151 & 0.09953 & 0.93796 & 0.09439 & 0.05124 & 0.07052 \\
SR     & 0.17653 & 0.17409 & 0.17841 & 0.13233 & 0.11721 & 0.16872 \\
STARR  & 0.08067 & 0.08564 & 0.07602 & 0.07189 & 0.05920 & 0.07279 \\
Sterling & 0.01792 & 0.01768 & 0.01727 & 0.01515 & 0.01669 & 0.01749 \\
Cumulative Return & 1.29697 & 1.32331 & 1.29276 & 1.25298 & 1.14650 & 1.22017 \\
\bottomrule
\end{tabular}%}
\end{table*}

\section{Robustness analysis during the U.S.--Israel--Iran conflict}

We next evaluate the portfolio performance over a turbulent market period from 1 October 2025 to 31 March 2026, comprising approximately 125 trading days (around six months), which overlaps with the U.S.--Israel--Iran conflict. This period is considered to test the robustness of the proposed portfolio strategies under adverse geopolitical conditions and elevated market volatility. For consistency with the above empirical analysis, we keep the in-sample window fixed at 3 months and consider out-of-sample rebalancing horizons of 3, 5, and 10 days. Based on the rolling window framework, these horizons yield 20, 12, and 6 out-of-sample windows, respectively. 

Table \ref{Tab: 3m-3d_robust} reports the average out-of-sample performance of the proposed framework over the conflict period for a 3-month in-sample and 3-day rebalancing horizon. We first compare the four indicator-sentiment based strategies (first four columns). Under \(\alpha = 0.95\), \(\mathrm{(DRMV)_{AWDS}}\) achieves the highest mean return, maximum return, and cumulative return (see Figure \ref{fig: cumulative returns} (right)), while \(\mathrm{(DRMV)_{APLS}}\) delivers the strongest risk-adjusted performance with the highest Sharpe and STARR ratios. Under \(\alpha = 0.99\), \(\mathrm{(DRMV)_{APLS}}\) emerges as the strongest performer among the sentiment-based models, attaining the highest mean return, SR, STARR, and cumulative return, while \(\mathrm{(DRMV)_{AWDS}}\) ranks closely behind in terms of return performance. Overall, among the sentiment-based strategies, \(\mathrm{(DRMV)_{AWDS}}\) and \(\mathrm{(DRMV)_{APLS}}\) dominate in terms of return and reward-to-risk performance during the conflict period.

We next compare the four indicator-only filtering-based strategies. Under \(\alpha = 0.95\), \(\mathrm{(DRMV)_{AWDI}}\) provides the strongest return performance, attaining the highest mean return, maximum return, and cumulative return among the four, whereas \(\mathrm{(DRMV)_{APLI}}\) performs best in terms of downside risk and risk-adjusted performance, achieving the lowest VaR, minimum return, MDD, and the highest STARR and Sterling ratio. For \(\alpha = 0.99\), \(\mathrm{(DRMV)_{AWDI}}\) again remains the strongest performer in terms of return-based measures. Overall, similar to the normal market period, the TDA-based filtering strategies continue to outperform the non-TDA alternatives even during the conflict period, especially in terms of return. We further compare the indicator-sentiment based models with their corresponding indicator-only counterparts. The results indicate that the inclusion of sentiment information generally improves return performance and risk-adjusted measures. Except for $\alpha= 99\%$, \(\mathrm{(DRMV)_{AWDI}}\) performing better than \(\mathrm{(DRMV)_{AWDS}}\) in terms of returns and risk-adjusted ratios.

Further, we compare the eight filtering-based (DRMV) models with the full asset universe model \(\mathrm{(DRMV)_A}\). The results show that almost all filtering-based strategies, except \(\mathrm{(DRMV)_{AES}}\) and \(\mathrm{(DRMV)_{AEI}}\) in some cases, outperform \(\mathrm{(DRMV)_A}\) in terms of return and risk-adjusted measures under both confidence levels. In particular, the best-performing filtered portfolios achieve notably higher mean returns, SR, STARR values, and cumulative returns than the \(\mathrm{(DRMV)_A}\) model. Moreover, several filtering-based strategies also exhibit improved downside risk characteristics and lower drawdowns relative to \(\mathrm{(DRMV)_A}\).

Similar patterns are observed in Tables \ref{Tab: 3m-5d_robust} and \ref{Tab: 3m-10d_robust}, corresponding to the 5-day and 10-day rebalancing horizons, respectively. The main findings remain qualitatively unchanged during the conflict period. In particular, the proposed TDA-based filtering strategies continue to outperform the benchmark distance-based specifications in terms of return and risk-adjusted performance across both confidence levels. Among them, \(\mathrm{(DRMV)_{AWDS}}\) remains one of the strongest performers across most return-oriented and reward-to-risk measures. Overall, these results further support the robustness of the proposed two-stage filtering framework under heightened geopolitical uncertainty and across lower rebalancing frequencies.

We also compare Tables \ref{Tab: 3m-3d_robust}, \ref{Tab: 3m-5d_robust}, and \ref{Tab: 3m-10d_robust} to assess the impact of rebalancing frequency during the conflict period. Overall, the results indicate that more frequent rebalancing is generally advantageous in terms of return and reward-to-risk performance. In particular, the 3-day rebalancing horizon often delivers higher mean returns, SR, STARR values, and cumulative returns for several of the best-performing models, highlighting the benefit of more responsive portfolio adjustment during periods of elevated market stress. Interestingly, when comparing the lower-frequency rebalancing settings, the 10-day horizon often performs better than the 5-day horizon in terms of return and risk-adjusted measures. This suggests that, beyond a certain point, increasing the rebalancing frequency does not necessarily lead to monotonic improvements in performance. Further, the 5-day and 10-day settings occasionally exhibit relatively better downside risk and drawdown characteristics for some specifications, which is also expected, since more frequent rebalancing may enhance return generation but can simultaneously expose the portfolio to higher risk and trading frictions.

Finally, we compare the proposed filtering-based (DRMV) models reported in Tables \ref{Tab: 3m-3d_robust}, \ref{Tab: 3m-5d_robust}, and \ref{Tab: 3m-10d_robust} with the benchmark portfolios presented in Table \ref{tab: benchmark_new}, namely \(\mathrm{B\&H_S}\), \(\mathrm{B\&H_I}\), Na\"ive, and Index. The results clearly demonstrate the superiority of the proposed two-stage framework during the conflict period. In particular, the Na\"ive and Index portfolios generate negative average returns over the period from January 2026 to March 2026, whereas all proposed filtering-based strategies continue to deliver positive returns. Even the benchmark portfolios, \(\mathrm{B\&H_S}\) and \(\mathrm{B\&H_I}\), perform relatively better than Na\"ive and Index and are able to generate positive returns over the same period. Although Na\"ive and Index exhibit relatively lower risk, such portfolios remain unattractive from an investment perspective due to their negative return performance. Moreover, the 3-day rebalanced filtering-based models substantially outperform \(\mathrm{B\&H_S}\) and \(\mathrm{B\&H_I}\) in terms of return and reward-to-risk measures. For the 5-day and 10-day rebalancing settings, the strongest filtering strategies, particularly \(\mathrm{(DRMV)_{AWDS}}\), \(\mathrm{(DRMV)_{AWDI}}\), \(\mathrm{(DRMV)_{APLS}}\), and \(\mathrm{(DRMV)_{APLI}}\), continue to outperform all benchmark portfolios. In contrast, some of the correlation and Euclidean distance-based filtering strategies, while still superior to Na\"ive and Index, do not consistently dominate \(\mathrm{B\&H_S}\) and \(\mathrm{B\&H_I}\) across all performance measures. Overall, these findings indicate that the proposed two-stage dynamic rebalancing framework remains highly effective even during a period of elevated geopolitical uncertainty and market stress. Moreover, the inclusion of sentiment as an additional feature in measuring asset similarity, together with the proposed TDA-based framework, appears to further strengthen the filtering stage and improve the overall portfolio allocation process by delivering higher returns and superior risk-adjusted performance even during stressed market conditions.

\begin{table*}[H]
\centering
\caption{Performance comparison of eight different distance based clustering portfolios under $\alpha=95\%$ and $\alpha=99\%$, including the $\mathrm{(DRMV)_A}$ model for in-sample period $T_1 = 63$ and out-of-sample period $T_2 = 3$, during the period of heightened U.S.--Israel--Iran conflict.}
\label{Tab: 3m-3d_robust}
% \resizebox{\textwidth}{!}{
\resizebox{0.70\textwidth}{!}{
\begin{tabular}{lccccccccc}
\toprule
$\alpha = 0.95$ 
& $\mathrm{(DRMV)_{AWDS}}$ & $\mathrm{(DRMV)_{APLS}}$ & $\mathrm{(DRMV)_{ACS}}$ & $\mathrm{(DRMV)_{AES}}$
& $\mathrm{(DRMV)_{AWDI}}$ & $\mathrm{(DRMV)_{APLI}}$ & $\mathrm{(DRMV)_{ACI}}$ & $\mathrm{(DRMV)_{AEI}}$ & $\mathrm{(DRMV)_A}$  \\
\cmidrule(lr){1-5} \cmidrule(lr){6-9} \cmidrule(lr){10-10}

Mean   & \textbf{0.00594} & 0.00586 & 0.00515 & 0.00468 & \textbf{\textit{0.00579}} & 0.00503 & 0.00477 & 0.00457 & 0.00458 \\
SD     & 0.03979 & 0.03814 & 0.03585 & \textbf{\textit{0.03545}} & 0.04188 & 0.03840 & \textbf{0.03659} & 0.03994 & 0.03667 \\
VaR    & 0.04090 & 0.04913 & \textbf{0.03916} & 0.04213 & 0.04666 & \textbf{\textit{0.04661}} & 0.04734 & 0.05443 & 0.05245 \\
CVaR   & 0.07910 & 0.07107 & 0.07029 & \textbf{0.06438} & 0.08102 & 0.07031 & 0.07780 & \textbf{\textit{0.06993}} & 0.07316 \\
Min    & -0.11865 & -0.08678 & -0.08511 & \textbf{-0.07901} & -0.11865 & \textbf{\textit{-0.08678}} & -0.09229 & -0.08850 & -0.09836 \\
Max    & \textbf{0.11710} & 0.10114 & 0.07841 & 0.09456 & \textbf{\textit{0.11710}} & 0.11108 & 0.10921 & 0.10570 & 0.09781 \\
MDD    & -0.31450 & -0.34106 & -0.28441 & \textbf{\textit{-0.25952}} & -0.30676 & \textbf{-0.24889} & -0.25741 & -0.25431 & -0.28656 \\
ADD    & 0.13334 & 0.14346 & 0.12054 & \textbf{0.09920} & 0.12993 & 0.10925 & \textbf{\textit{0.10701}} & 0.13448 & {0.12523} \\
Sharpe & 0.14937 & \textbf{0.15358} & 0.14360 & 0.13215 & \textbf{\textit{0.13821}} & 0.13097 & 0.13045 & 0.11453 & 0.12487 \\
STARR  & 0.07513 & \textbf{0.08243} & 0.07324 & 0.07276 & 0.07145 & \textbf{\textit{0.07154}} & 0.06135 & 0.06540 & 0.06259 \\
Sterling & 0.04457 & 0.04083 & 0.04271 & \textbf{0.04722} & 0.04455 & \textbf{\textit{0.04604}} & 0.04460 & 0.03401 & 0.03657 \\
Cumulative Returns & \textbf{1.36254} & 1.36084 & 1.31079 & 1.27618 & \textbf{\textit{1.34357}} & 1.29468 & 1.27959 & 1.25525 & 1.26463 \\

\cmidrule(lr){1-5} \cmidrule(lr){6-9} \cmidrule(lr){10-10}
$\alpha = 0.99$ & \\
\cmidrule(lr){1-5} \cmidrule(lr){6-9} \cmidrule(lr){10-10}

Mean   & 0.00514 & \textbf{0.00572} & 0.00518 & 0.00435 & \textbf{\textit{0.00534}} & 0.00522 & 0.00490 & 0.00401 & 0.00458 \\
SD     & 0.04046 & 0.03872 & \textbf{\textit{0.03749}} & 0.04109 & 0.03801 & 0.03943 & 0.03836 & 0.04187 & \textbf{0.03667} \\
VaR    & 0.04825 & 0.04925 & \textbf{0.04375} & 0.05650 & \textbf{\textit{0.04901}} & 0.05872 & 0.05146 & 0.06293 & 0.05245 \\
CVaR   & 0.08194 & \textbf{\textit{0.06640}} & 0.07361 & 0.07814 & 0.07385 & 0.07709 & \textbf{0.06228} & 0.07652 & \textbf{\textit{0.06640}} \\
Min    & -0.11865 & \textbf{-0.07519} & -0.08678 & -0.08869 & -0.08678 & -0.08707 & \textbf{\textit{-0.07519}} & -0.09801 & -0.09836 \\
Max    & \textbf{0.11710} & 0.09718 & 0.09028 & 0.09246 & 0.09737 & \textbf{\textit{0.10792}} & 0.10326 & 0.10157 & 0.09781 \\
MDD    & -0.33542 & -0.34155 & -0.28482 & \textbf{\textit{-0.26448}} & -0.30292 & -0.31404 & -0.31668 & \textbf{-0.25255} & -0.28656 \\
ADD    & 0.15235 & 0.14687 & 0.12900 & 0.13811 & \textbf{\textit{0.12557}} & 0.13935 & 0.14106 & 0.13472 & \textbf{0.12523} \\
Sharpe & 0.12711 & \textbf{0.14762} & 0.13804 & 0.10577 & \textbf{\textit{0.14045}} & 0.13228 & 0.12772 & 0.09581 & 0.12487 \\
STARR  & 0.06276 & \textbf{0.08609} & 0.07031 & 0.05562 & 0.07228 & 0.06766 & \textbf{\textit{0.07867}} & 0.05243 & 0.06259 \\
Sterling & 0.03376 & 0.03892 & \textbf{\textit{0.04012}} & 0.03147 & \textbf{0.04251} & 0.03743 & 0.03473 & 0.02978 & \textbf{\textit{0.04012}} \\
Cumulative Returns & 1.29689 & \textbf{1.34766} & 1.30838 & 1.23468 & \textbf{\textit{1.31963}} & 1.30581 & 1.28468 & 1.20804 & 1.26463 \\
\bottomrule
\end{tabular}}
\end{table*}

\begin{table*}[H]
\centering
\caption{Performance comparison of eight different distance based clustering portfolios under $\alpha=95\%$ and $\alpha=99\%$, including the $\mathrm{(DRMV)_A}$ model for in-sample period $T_1 = 63$ and out-of-sample period $T_2 = 5$, during the period of heightened U.S.--Israel--Iran conflict.}
\label{Tab: 3m-5d_robust}
\resizebox{0.70\textwidth}{!}{
\begin{tabular}{lccccccccc}
\toprule
$\alpha = 0.95$ 
& $\mathrm{(DRMV)_{AWDS}}$ & $\mathrm{(DRMV)_{APLS}}$ & $\mathrm{(DRMV)_{ACS}}$ & $\mathrm{(DRMV)_{AES}}$
& $\mathrm{(DRMV)_{AWDI}}$ & $\mathrm{(DRMV)_{APLI}}$ & $\mathrm{(DRMV)_{ACI}}$ & $\mathrm{(DRMV)_{AEI}}$ & $\mathrm{(DRMV)_A}$  \\
\cmidrule(lr){1-5} \cmidrule(lr){6-9} \cmidrule(lr){10-10}

Mean   & \textbf{0.00528} & 0.00436 & 0.00440 & 0.00441 & \textbf{\textit{0.00522}} & 0.00433 & 0.00407 & 0.00394 & 0.00406 \\
SD     & 0.04005 & 0.03916 & 0.03394 & 0.03434 & 0.04059 & 0.03469 & 0.03760 & \textbf{\textit{0.03356}} & \textbf{0.03255} \\
VaR    & \textbf{\textit{0.04195}} & 0.05230 & 0.04987 & 0.04953 & 0.05919 & 0.04594 & 0.06381 & 0.04222 & \textbf{0.04048} \\
CVaR   & 0.08453 & 0.07774 & 0.07100 & 0.07409 & 0.09084 & 0.06850 & 0.08105 & \textbf{0.06181} & \textbf{\textit{0.06865}} \\
Min    & -0.11865 & -0.08333 & \textbf{\textit{-0.07519}} & -0.08670 & -0.12761 & -0.07519 & -0.08511 & \textbf{-0.07194} & -0.08629 \\
Max    & 0.09464 & \textbf{\textit{0.10114}} & 0.08249 & 0.09031 & \textbf{0.11116} & 0.08249 & 0.09222 & 0.08433 & 0.07666 \\
MDD    & -0.30134 & -0.28106 & -0.28998 & \textbf{\textit{-0.25683}} & -0.28065 & -0.25911 & -0.24873 & \textbf{{-0.23109}} & -0.26248 \\
ADD    & 0.12780 & 0.11248 & 0.12182 & \textbf{\textit{0.10815}} & 0.11659 & 0.11499 & 0.09256 & \textbf{{0.08197}} & 0.12011 \\
Sharpe & \textbf{0.13184} & 0.11136 & 0.12976 & 0.12842 & \textbf{\textit{0.12873}} & 0.12492 & 0.10838 & 0.11733 & 0.12479 \\
STARR  & \textbf{\textit{0.06247}} & 0.05609 & 0.06202 & 0.05953 & 0.05752 & 0.06327 & 0.05028 & \textbf{0.06371} & 0.05917 \\
Sterling & \textbf{\textit{0.04132}} & 0.03877 & 0.03615 & 0.04078 & 0.04481 & 0.03769 & 0.04402 & \textbf{0.04804} & 0.03382 \\
Cumulative Returns & \textbf{1.30849} & 1.24146 & 1.25852 & 1.25787 & \textbf{\textit{1.30225}} & 1.25152 & 1.22443 & 1.22511 & 1.23634 \\

\cmidrule(lr){1-5} \cmidrule(lr){6-9} \cmidrule(lr){10-10}
$\alpha = 0.99$ & \\
\cmidrule(lr){1-5} \cmidrule(lr){6-9} \cmidrule(lr){10-10}

Mean   & \textbf{0.00512} & 0.00445 & 0.00410 & 0.00418 & \textbf{\textit{0.00452}} & 0.00443 & 0.00441 & 0.00385 & 0.00406 \\
SD     & 0.03654 & 0.03671 & 0.03725 & 0.04104 & 0.03592 & 0.03434 & \textbf{\textit{0.03434}} & 0.04184 & \textbf{0.03255} \\
VaR    & 0.04929 & \textbf{\textit{0.04913}} & 0.05858 & 0.05650 & 0.04959 & 0.05122 & 0.04953 & 0.06293 & \textbf{0.04048} \\
CVaR   & 0.07302 & 0.07103 & 0.07593 & 0.07814 & \textbf{{0.06571}} & 0.07413 & 0.07409 & 0.07652 & \textbf{\textit{0.06865}} \\
Min    & -0.08666 & -0.08678 & \textbf{-0.08207} & -0.08869 & -0.08673 & -0.08921 & -0.08670 & -0.09801 & \textbf{\textit{-0.08629}} \\
Max    & 0.08902 & \textbf{\textit{0.10114}} & 0.08907 & 0.09246 & 0.09463 & 0.10077 & 0.09031 & \textbf{0.10157} & 0.07666 \\
MDD    & -0.29601 & -0.31026 & \textbf{-0.23225} & -0.26448 & -0.26184 & \textbf{\textit{-0.23409}} & -0.25683 & -0.25255 & -0.26248 \\
ADD    & 0.12690 & 0.12807 & \textbf{\textit{0.09947}} & 0.13798 & 0.10344 & \textbf{{0.09096}} & 0.10815 & 0.13460 & 0.12011 \\
Sharpe & \textbf{0.14014} & 0.12133 & 0.11011 & 0.10182 & 0.12596 & \textbf{\textit{0.12885}} & 0.12842 & 0.09192 & 0.12479 \\
STARR  & \textbf{0.07012} & 0.06270 & 0.05402 & 0.05348 & \textbf{\textit{0.06885}} & 0.05970 & 0.05953 & 0.05026 & 0.05917 \\
Sterling & 0.04035 & 0.03478 & \textbf{\textit{0.04124}} & 0.03029 & 0.04374 & \textbf{0.04865} & 0.04078 & 0.02857 & \textbf{\textit{0.04124}} \\
Cumulative Returns & \textbf{1.30667} & 1.25530 & 1.22757 & 1.22256 & \textbf{\textit{1.26268}} & 1.25908 & 1.25787 & 1.19618 & 1.23634 \\
\bottomrule
\end{tabular}}
\end{table*}

\begin{table*}[H]
\centering
\caption{Performance comparison of eight different distance based clustering portfolios under $\alpha=95\%$ and $\alpha=99\%$, including the $\mathrm{(DRMV)_A}$ model for in-sample period $T_1 = 63$ and out-of-sample period $T_2 = 10$, during the period of heightened U.S.--Israel--Iran conflict.}
\label{Tab: 3m-10d_robust}
\resizebox{0.70\textwidth}{!}{
\begin{tabular}{lccccccccc}
\toprule
$\alpha = 0.95$ 
& $\mathrm{(DRMV)_{AWDS}}$ & $\mathrm{(DRMV)_{APLS}}$ & $\mathrm{(DRMV)_{ACS}}$ & $\mathrm{(DRMV)_{AES}}$
& $\mathrm{(DRMV)_{AWDI}}$ & $\mathrm{(DRMV)_{APLI}}$ & $\mathrm{(DRMV)_{ACI}}$ & $\mathrm{(DRMV)_{AEI}}$ & $\mathrm{(DRMV)_A}$  \\
\cmidrule(lr){1-5} \cmidrule(lr){6-9} \cmidrule(lr){10-10}

Mean   & \textbf{0.00535} & 0.00429 & 0.00423 & 0.00383 & \textbf{\textit{0.00508}} & 0.00431 & 0.00418 & 0.00349 & 0.00393 \\
SD     & 0.03933 & 0.04044 & \textbf{\textit{0.02867}} & 0.03384 & 0.03954 & 0.03627 & 0.03887 & 0.03114 & \textbf{0.02855} \\
VaR    & 0.04912 & 0.05233 & 0.04652 & \textbf{\textit{0.04542}} & 0.04896 & 0.04580 & 0.06044 & 0.05438 & \textbf{0.03878} \\
CVaR   & 0.07571 & 0.08761 & \textbf{\textit{0.05942}} & 0.06719 & 0.07077 & 0.07200 & 0.08092 & 0.06542 & \textbf{0.04908} \\
Min    & -0.08706 & -0.12476 & -0.07211 & -0.07687 & -0.08666 & -0.08395 & -0.08473 & \textbf{\textit{-0.07157}} & \textbf{-0.05463} \\
Max    & 0.09628 & \textbf{\textit{0.10100}} & 0.08952 & 0.09205 & \textbf{0.11157} & 0.09084 & 0.09222 & 0.06838 & 0.08316 \\
MDD    & -0.34467 & -0.27947 & -0.29450 & -0.27071 & -0.31882 & -0.27907 & -0.25850 & \textbf{-0.21191} & \textbf{\textit{-0.24977}} \\
ADD    & 0.16709 & 0.12739 & 0.12919 & 0.12256 & 0.14061 & 0.13055 & 0.09802 & \textbf{-0.08672} & \textbf{\textit{0.12003}} \\
Sharpe & 0.13615 & 0.10604 & \textbf{0.14743} & 0.11321 & 0.12860 & 0.11871 & 0.11220 & 0.10746 & \textbf{\textit{0.13780}} \\
STARR  & 0.07073 & 0.04895 & 0.07113 & 0.05702 & \textbf{\textit{0.07185}} & 0.05980 & 0.05162 & 0.05341 & \textbf{0.08015} \\
Sterling & 0.03205 & \textbf{\textit{0.03366}} & 0.03272 & 0.03126 & 0.03616 & 0.03298 & \textbf{0.04262} & 0.04029 & 0.03277 \\
Cumulative Returns & \textbf{1.31697} & 1.23178 & 1.25749 & 1.21660 & \textbf{\textit{1.29555}} & 1.24524 & 1.22855 & 1.19800 & 1.23604 \\

\cmidrule(lr){1-5} \cmidrule(lr){6-9} \cmidrule(lr){10-10}
$\alpha = 0.99$ & \\
\cmidrule(lr){1-5} \cmidrule(lr){6-9} \cmidrule(lr){10-10}

Mean   & \textbf{0.00529} & 0.00512 & 0.00438 & 0.00401 & 0.00457 & \textbf{\textit{0.00494}} & 0.00450 & 0.00400 & 0.00393 \\
SD     & 0.03708 & 0.03549 & 0.03904 & 0.03566 & 0.03701 & 0.04020 & \textbf{\textit{0.03251}} & 0.04030 & \textbf{0.02855} \\
VaR    & 0.04909 & \textbf{\textit{0.04740}} & 0.05589 & 0.05189 & 0.05144 & 0.05207 & 0.05866 & 0.06576 & \textbf{0.03878} \\
CVaR   & 0.07160 & 0.06997 & 0.07394 & \textbf{\textit{0.06826}} & 0.07145 & 0.08366 & 0.07294 & 0.09147 & \textbf{0.04908} \\
Min    & -0.08666 & -0.08395 & -0.08473 & -0.08867 & -0.08473 & -0.11865 & \textbf{\textit{-0.08323}} & -0.10263 & \textbf{-0.05463} \\
Max    & 0.09669 & 0.08691 & \textbf{\textit{0.11234}} & 0.07996 & 0.10104 & \textbf{0.11710} & 0.10252 & 0.09548 & 0.08316 \\
MDD    & -0.32658 & -0.30958 & -0.30943 & \textbf{-0.24610} & -0.27921 & -0.33249 & -0.28357 & -0.25868 & \textbf{\textit{-0.24977}} \\
ADD    & 0.14342 & 0.14308 & 0.14563 & \textbf{\textit{0.11927}} & \textbf{ 0.10904} & 0.17205 & 0.12341 & 0.12290 & 0.12003 \\
Sharpe & 0.14264 & \textbf{0.14413} & 0.11229 & 0.11252 & 0.12361 & 0.12295 & \textbf{\textit{0.13835}} & 0.09936 & 0.13780 \\
STARR  & \textbf{\textit{0.07386}} & 0.07311 & 0.05928 & 0.05878 & 0.06402 & 0.05909 & 0.06166 & 0.04378 & \textbf{0.08015} \\
Sterling & \textbf{\textit{0.03687}} & 0.03575 & 0.03010 & 0.03364 & \textbf{0.04195} & 0.02873 & \textbf{\textit{0.03687}} & 0.03258 & 0.03277 \\
Cumulative Returns & \textbf{1.31845} & 1.30914 & 1.24372 & 1.22512 & 1.26356 & \textbf{\textit{1.28212}} & 1.26908 & 1.21142 & 1.23604 \\
\bottomrule
\end{tabular}}
\end{table*}

\begin{table*}[htbp]
\centering
\caption{Performance of four benchmark models for an in-sample period $T_1 = 63$ (3 months) and out-of-sample period $T_2 = 62$ (3 months), i.e., without rebalancing, during the period of heightened U.S.--Israel--Iran conflict.}
\label{tab: benchmark_new}
\footnotesize
\begin{tabular}{lccccccc}
\toprule
 & \multicolumn{2}{c}{$\mathrm{B\&H_S}$} & \multicolumn{2}{c}{$\mathrm{B\&H_I}$} & \multirow{2}{*}{Na\"ive} & \multirow{2}{*}{Index} \\
\cmidrule(lr){2-3} \cmidrule(lr){4-5}
 & $\alpha = 0.95$ & $\alpha = 0.99$ & $\alpha = 0.95$ & $\alpha = 0.99$ &  &  \\
\midrule
Mean   & 0.00428 & 0.00438 & 0.00439 & 0.00431 & -0.00027 & -0.00108 \\
SD     & 0.03151 & 0.03069 & 0.02991 & 0.02964 & 0.00798 & 0.00774 \\
VaR    & 0.04268 & 0.03899 & 0.03788 & 0.03788 & 0.01343 & 0.01378 \\
CVaR   & 0.05020 & 0.04997 & 0.04885 & 0.04885 & 0.01557 & 0.01677 \\
Min    & -0.05753 & -0.05964 & -0.05702 & -0.05702 & -0.01700 & -0.01772 \\
Max    & 0.08752 & 0.07184 & 0.08108 & 0.08108 & 0.01665 & 0.01523 \\
MDD    & -0.28210 & -0.28830 & -0.28263 & -0.27692 & -0.06442 & -0.07828 \\
ADD    & 0.13193 & 0.13842 & 0.12920 & 0.13010 & 0.02732 & 0.02355 \\
Sharpe & 0.13584 & 0.14275 & 0.14690 & 0.14541 & - & - \\
STARR  & 0.08528 & 0.08766 & 0.08994 & 0.08822 & - & - \\
Sterling & 0.03245 & 0.03165 & 0.03401 & 0.03313 & - & - \\
Cumulative Returns & 1.25540 & 1.26471 & 1.26749 & 1.26172 & 0.98211 & 0.93529 \\
\bottomrule
\end{tabular}
\end{table*}

\section{Conclusion}\label{sec: Conclusion}
In this work, we introduced distance measures based on topological features derived via PH on multi-dimensional data, integrating technical indicators and news-based sentiment, to quantify similarity among assets. Notably, these measures capture complex, non-linear dependencies and incorporate behavioral market signals that are often overlooked by traditional covariance and Euclidean distance-based approaches. These distance measures are subsequently used as inputs to an agglomerative clustering framework for asset selection, followed by the development of a dynamic rebalancing mean–variance model that incorporates practical constraints such as transaction costs, short-selling, and lower and upper bounds on portfolio weights. Specifically, we propose a novel, fully data-driven framework grounded in TDA and contextual sentiment analysis for constructing sparse portfolios within a practically implementable, actively managed PO setting.

Our empirical findings demonstrate that sparse portfolios selected using TDA on multi-dimensional data, including sentiment, are not only comparable to existing methods but often outperform them. The results remain robust under volatile market conditions and across multiple performance measures, including returns and risk-adjusted ratios, albeit with a marginal increase in risk. Furthermore, the inclusion of sentiment enhances the model’s ability to capture market regime shifts and investor behavior, thereby improving adaptability and robustness. Thus, this study contributes to the literature on sparse portfolio selection and advances the application of TDA and sentiment-informed modeling in finance.

\textbf{Acknowledgements} 

The author sincerely thanks Prof. Aparna Mehra, Department of Mathematics, IIT Delhi, for her valuable guidance, insightful suggestions, and careful review of the manuscript. Puneet Pasricha, Department of Mathematics, IIT Ropar, for his insightful comments. The author sincerely acknowledges the financial support provided through the Prime Minister Research Fellowship under award number PMRF ID-11CS142003.

\textbf{Declaration of Generative AI and AI-assisted technologies in the manuscript preparation process} 

During the preparation of this work, the author used GPT-Go (ChatGPT) to improve the language, readability, and clarity of the manuscript. In addition, Google Gemini was used to assist in the creation of explanatory images included in the manuscript. Following the use of these tools, the author carefully reviewed, edited, and verified all content and images and takes full responsibility for the accuracy and integrity of the published work.

% \textbf{Acknowledgements} 

% The author sincerely thanks Prof. Aparna Mehra, Department of Mathematics, IIT Delhi, for her valuable feedback and suggestions on the manuscript, and Prof. Puneet Pasricha, Department of Mathematics, IIT Ropar, for his insightful comments. The author sincerely acknowledges the financial support provided through the Prime Minister Research Fellowship by the Ministry of Education, Government of India, under award number PMRF ID-11CS142003.

\appendix
\section*{Appendix A. Technical indicators}
\textbf{Relative Strength Index (RSI):}
RSI is a momentum-based technical indicator that measures the speed and magnitude of recent price changes to assess overbought or oversold conditions in a stock. It is computed using the ratio of average gains to average losses over a specified look-back period, typically 14 trading days, and is scaled to lie between 0 and 100. Higher RSI values indicate strong upward momentum and potential overbought conditions, whereas lower values suggest downward momentum and potential oversold conditions. %In the proposed framework, RSI is employed to capture short-term price momentum and reversal signals, thereby providing valuable insights into the strength and sustainability of recent price movements.

The average gain \( AG_t \) and average loss \( AL_t \) over a look-back window of length \( L \) (typically \( L = 10 \)) are computed as
\[
AG_t = \frac{1}{L} \sum_{i=1}^{L} \max(\Delta C_{t-i+1}, 0), \]
\[
AL_t = \frac{1}{L} \sum_{i=1}^{L} \max(-\Delta C
_{t-i+1}, 0), \quad \text{where} \quad \Delta C_t = C_t - C_{t-1}.
\]

\[ 
RSI_t = 100 - \frac{100}{1 + RS_t}\,, \quad \text{where} \quad RS_t = \frac{AG_t}{AL_t}.
\]

where \( RSI_t \in [0, 100] \). \( RSI_t \geq 70 \) indicate stronger upward momentum and is overbought, while \( RSI_t \leq 30 \) suggest stronger downward momentum and is oversold.

\textbf{Stochastic Oscillator (SO):}  
SO is a momentum-based technical indicator that evaluates the position of a stock’s closing price relative to its recent trading range. It is designed to capture overbought and oversold conditions by comparing the current closing price with the highest and lowest prices over a specified look-back period. Let \( H_t \), \( L_t \), and \( C_t \) denote the highest, lowest, and closing prices of a stock at time \( t \), respectively. It is defined as
\[
SO_t = \frac{C_t - L_{t}}{H_{t} - L_{t}} \times 100.
\]
SO index has a value between 0 and 100. Similar to RSI, a value above 90 is overbought and below 10 is oversold.

\textbf{Moving Average Convergence Divergence (MACD):}
MACD \cite{appel2005technical} is a momentum-based technical indicator widely used to identify changes in trend direction and to generate buy–sell signals. It measures the relationship between two exponential moving averages (EMAs) of a stock’s closing price. A positive MACD value indicates upward momentum, while a negative value suggests downward momentum. In practice, a buy signal is typically generated when the MACD crosses from negative to positive, whereas a sell signal is indicated when it moves from positive to negative. It is computed as
\[
\text{MACD}_t = 2 \left( \text{DIFF}_t - \text{DEA}_t \right),
\]
\begin{align*}
\text{DIFF}_t &= \text{EMA}_t^{(\text{SHORT})} - \text{EMA}_t^{(\text{LONG})}, \\
\text{DEA}_t  &= \frac{2}{M+1} \text{DIFF}_t + \frac{M-1}{M+1} \text{DEA}_{t-1},
\end{align*}

where, $\text{EMA}_t^{(N)}$ denotes the exponential moving average of the closing price with window length $N$, which is computed recursively as
\begin{equation}
\text{EMA}_t^{(N)} = \frac{2}{N+1} C_t + \frac{N-1}{N+1} \text{EMA}_{t-1}^{(N)}.
\end{equation}

Following standard practice in technical analysis, we set the parameters $\text{SHORT}=12$, $\text{LONG}=26$, and $M=9$ in this study. These values are commonly used to balance short-term responsiveness with long-term trend stability and are adopted to ensure comparability with widely used trading benchmarks.

\section*{\textcolor{black}{Appendix B. Correlation and Euclidean distances}}\label{Appendix:B}
\setcounter{definition}{0}
\renewcommand{\thedefinition}{B.\arabic{definition}}
As a benchmark for comparison, we also consider a correlation-based distance between the multivariate time series. Correlation measures the degree of co-movement between two time series and is widely used in financial applications to quantify similarity in asset dynamics. Unlike TDA-based distances, which capture the geometric structure of the underlying point clouds, correlation-based distance reflects the linear dependence between the corresponding time-series components.

\begin{definition}
The correlation-based distance between the two multivariate time series $Z_i$ and $Z_j$ is defined as
\[
d_{\mathrm{AC}}(Z_i,Z_j) =
\sqrt{2\left(1-\rho(Z_i,Z_j)\right)},
\]
\[
\rho(Z_i,Z_j)=\frac{1}{d}\sum_{k=1}^{d}\rho (Z_{ik},Z_{jk}),
\]

where $\rho (Z_{ik},Z_{jk})$ denotes the Pearson correlation coefficient between the $k$-th components of the time series $Z_i$ and $Z_j$.
\end{definition}

% The above distance captures similarity in the co-movement patterns of the underlying variables while remaining invariant to differences in scale. 

\begin{definition}
The Euclidean distance between two multivariate time series $Z_i$ and $Z_j$ is defined as
\[
d_{\mathrm{AE}}(Z_i,Z_j)
=
\left\| Z_i - Z_j \right\|_2
=
\sqrt{\sum_{k=1}^{d} \left\| Z_{ik} - Z_{jk} \right\|_2^2},
\]
where $Z_{ik}$ and $Z_{jk}$ denote the $k$-th component time series of $Z_i$ and $Z_j$, respectively, and $\|\cdot\|_2$ represents the Euclidean norm.
\end{definition}

% \section*{Appendix D. Algorithms}
\begin{algorithm}[h]\footnotesize
\caption{Algorithm to compute the TDA-based distance between assets}
\label{alg:pointcloud}

1: For asset $i$, represent the multivariate time series in the in-sample period as
\[
Z_i=\{z_{it}\in\mathbb{R}^d\}_{t=1}^{T_1}, \qquad d=3 \text{ or }4,
\]
where $z_{it}=(z_{it}^{(1)},z_{it}^{(2)},z_{it}^{(3)},z_{it}^{(4)})$ denotes the four-dimensional observation at time $t$, and $T_1$ is the length of the in-sample period.

2: The corresponding point cloud representation for $d=4$ is
\[
Z_i=
\begin{bmatrix}
z_{i1}\\
z_{i2}\\
\vdots\\
z_{iT_1}
\end{bmatrix}
=
\begin{bmatrix}
z_{i1}^{(1)} & z_{i1}^{(2)} & z_{i1}^{(3)} & z_{i1}^{(4)}\\
z_{i2}^{(1)} & z_{i2}^{(2)} & z_{i2}^{(3)} & z_{i2}^{(4)}\\
\vdots & \vdots & \vdots & \vdots\\
z_{iT_1}^{(1)} & z_{iT_1}^{(2)} & z_{iT_1}^{(3)} & z_{iT_1}^{(4)}
\end{bmatrix}.
\]

3: Form the first point cloud using a time-window of size $L$, resulting in a matrix of order $L\times d$ with $L=21$,
\[
Z_{i1}=
\begin{bmatrix}
z_{i1}^{(1)} & z_{i1}^{(2)} & z_{i1}^{(3)} & z_{i1}^{(4)}\\
z_{i2}^{(1)} & z_{i2}^{(2)} & z_{i2}^{(3)} & z_{i2}^{(4)}\\
\vdots & \vdots & \vdots & \vdots\\
z_{iL}^{(1)} & z_{iL}^{(2)} & z_{iL}^{(3)} & z_{iL}^{(4)}
\end{bmatrix}.
\]

4: Apply Algorithm~\ref{alg:rips} to construct the Rips filtration 
$\{R(Z_{i1},\epsilon_j)\}_{j=1}^{N}$ and Algorithm~\ref{alg:persistence} to compute the persistence diagram $\mathcal{D}_{Z_{i1}}$ and the persistence landscape $\zeta_{Z_{i1}}$.

5: Using a sliding step of one day, construct the next point cloud
\[
Z_{i2}=
\begin{bmatrix}
z_{i2}^{(1)} & z_{i2}^{(2)} & z_{i2}^{(3)} & z_{i2}^{(4)}\\
z_{i3}^{(1)} & z_{i3}^{(2)} & z_{i3}^{(3)} & z_{i3}^{(4)}\\
\vdots & \vdots & \vdots & \vdots\\
z_{i(L+1)}^{(1)} & z_{i(L+1)}^{(2)} & z_{i(L+1)}^{(3)} & z_{i(L+1)}^{(4)}
\end{bmatrix}.
\]

6: For each subsequent window $k=2,\dots,K$, repeat the above procedure to construct the point cloud $Z_{ik}$ and compute its persistence diagram $\mathcal{D}_{Z_{ik}}$ and persistence landscape $\zeta_{Z_{ik}}$ using Algorithms~\ref{alg:rips} and \ref{alg:persistence}.

7: This process yields $K=T_1-L+1$ point clouds $\{Z_{ik}\}_{k=1}^{K}$ and their corresponding persistence diagrams and persistence landscapes for each asset $i=1,2,\dots,n$.

8: For two assets $i$ and $j$, compute the average Wasserstein distance (AWD) between the persistence diagrams
\[
d_{\mathrm{AWD}}(Z_i,Z_j)
=
\sum_{k=1}^{K}
w_k\,W_p(\mathcal{D}_{Z_{ik}},\mathcal{D}_{Z_{jk}}),
\]
as defined in (6). Similarly, the average landscape distance (ALD) is computed using (7) in the manuscript.

9: By computing these distances for all pairs of assets, obtain the symmetric distance matrix $D \in \mathbb{R}^{n \times n}$, where $n$ denotes the number of assets considered for portfolio construction.

\end{algorithm}

\begin{algorithm}[h]\footnotesize
\caption{Construction of Vietoris--Rips filtration from a point cloud}
\label{alg:rips}

\textbf{Data:} A point cloud 
\[
Z_{ik}^T=\{z_{ik},z_{i(k+1)},\ldots,z_{i(k+L-1)}\}, \qquad z_{it}\in\mathbb{R}^{d},
\]
and a sequence of scales $0 < \epsilon_1 < \epsilon_2 < \cdots < \epsilon_N$.

\textbf{Result:} Vietoris--Rips filtration $\{R(Z_{ik},\epsilon_j)\}_{j=1}^{N}$.

1: \textbf{for} $j=1,2,\ldots,N$ \textbf{do}

2: \qquad Draw a ball of radius $\epsilon_j$ centered at each point of the point cloud 
$\{z_{ik},z_{i(k+1)},\ldots,z_{i(k+L-1)}\}$.

3: \qquad Connect each pair of points in $Z_{ik}$ with an edge if
\[
d(z_{ia},z_{ib}) < 2\epsilon_j .
\]

4: \qquad Connect each triplet of points in $Z_{ik}$ to form a triangle if
\[
d(z_{ia},z_{ib}),\; d(z_{ia},z_{ic}),\; d(z_{ib},z_{ic}) < 2\epsilon_j .
\]

5: \qquad More generally, connect any set of $p$ points to form a $(p-1)$-simplex if all pairwise distances between them are less than $2\epsilon_j$.

6: \textbf{return} the filtration $\{R(Z_{ik},\epsilon_j)\}_{j=1}^{N}$.

\end{algorithm}

\begin{algorithm}[h]\footnotesize
\caption{Algorithm to obtain persistence diagram and persistence landscape from Rips filtration}
\label{alg:persistence}

\textbf{Data:} Rips filtration $\{R(Z_{ik},\epsilon_j)\}_{j=1}^{N}$.

\textbf{Result:} Persistence diagram $\mathcal{D}_{Z_{ik}}$ and persistence landscape $\zeta_{Z_{ik}}$.

1: Compute the $r$-dimensional homology groups
\[
H_r(R(Z_{ik},\epsilon_j)), \qquad r=0,1.
\]

2: Record the birth and death scales of each topological feature
\[
\mathcal{D}_{Z_{ik}}=\{(b_\ell,d_\ell)\}_{\ell=1}^{m}.
\]

3: For each pair $(b_\ell,d_\ell)$ construct the triangular function
\[
\Lambda_{(b_\ell,d_\ell)}(t).
\]

4: Construct the persistence landscape
\[
\zeta_h(t)=
h\text{-th largest value of }
\{\Lambda_{(b_\ell,d_\ell)}(t)\}_{\ell=1}^{m}.
\]

\qquad For $h > m$, set $\zeta_h(t)=0$.

5: \textbf{return} $\mathcal{D}_{Z_{ik}}$ and $\zeta_{Z_{ik}}$.

\end{algorithm}

\section*{\textcolor{black}{Appendix C. Agglomerative clustering algorithm}}\label{Appendix:C}
% Agglomerative clustering is a hierarchical unsupervised learning method that groups observations based on pairwise similarity. Unlike partition-based approaches such as k-means, agglomerative clustering does not require the number of clusters to be specified in advance and naturally accommodates the detection of outliers.

% The algorithm follows a bottom-up strategy. Initially, each stock is treated as an individual cluster. At each iteration, the two closest clusters are merged according to a chosen linkage criterion. In this study, we employ average linkage, where the distance between two clusters is computed as the average of all pairwise distances between observations belonging to the two clusters.

% The clustering process generates a hierarchical structure represented through a dendrogram. Figure 2 illustrates the mechanism using five points $p,q,r,s,t$. Initially, each point forms an individual cluster. The closest points are merged first, resulting in clusters such as $(p,q)$ and $(s,t)$. Subsequent iterations merge nearby clusters until the inter-cluster distance exceeds the threshold $\epsilon$. The procedure then terminates, yielding the final cluster partition.

% The stopping criterion is determined through a linkage threshold $\epsilon$, computed as the $\alpha$-th percentile of the nearest-neighbor distances among all stocks. An important advantage of this framework is its ability to identify outliers. Stocks that remain as singleton clusters at the termination stage are treated as outliers and removed from further analysis.

Clustering methods operate in an unsupervised manner and aim to uncover inherent group structures in data without relying on labeled observations. In this study, we employ agglomerative clustering to group stocks based on pairwise distance measures derived from their multivariate time-series representations. Prior to clustering, pairwise distances between all stocks are computed using the measures introduced in Section 4.1, namely the topological distances $d_{\mathrm{AWD}}$ and $d_{\mathrm{ALD}}$, along with the benchmark distances $d_{\mathrm{AC}}$ and $d_{\mathrm{AE}}$. These distances are used to construct a symmetric distance matrix, which serves as the input to the clustering algorithm.

Compared to widely used clustering methods such as k-means, agglomerative clustering does not require the number of clusters to be specified in advance and naturally accommodates the identification of outliers. In contrast, k-means assigns every observation to a cluster, even if certain stocks exhibit behavior that is significantly different from others. Agglomerative clustering follows a bottom-up strategy, where each stock is initially treated as an individual cluster. At each iteration, the two closest clusters are merged according to a chosen linkage criterion. In this study, we adopt the average linkage method, which defines the distance between two clusters as the mean of all pairwise distances between their constituent elements. Specifically, let $A$ and $B$ denote two clusters containing $N_A$ and $N_B$ observations, respectively. The inter-cluster distance is computed as
\begin{equation}
d_{\text{average}}(A,B)
=
\frac{1}{N_A N_B}
\sum_{Z_i \in A} \sum_{Z_j \in B}
d(Z_i,Z_j),
\end{equation}
where $d(Z_i,Z_j)$ represents the pairwise distance between observations $Z_i$ and $Z_j$, corresponding to one of the four distance measures considered in this study, namely $d_{\mathrm{AWD}}$, $d_{\mathrm{ALD}}$, $d_{\mathrm{AC}}$, or $d_{\mathrm{AE}}$.

The iterative merging process produces a hierarchical structure, typically represented by a dendrogram (see Figure 2 in the manuscript), which provides insight into the nested relationships among clusters. A key component of the clustering procedure is the stopping criterion, which determines when the merging process terminates. Instead of specifying the number of clusters in advance, we employ a distance-based threshold $\epsilon$, referred to as the linkage distance. Clusters are merged only if their inter-cluster distance does not exceed $\epsilon$, allowing the number of clusters to emerge endogenously from the data.

\textbf{Hyperparameter $\epsilon$ selection:}
To determine an appropriate value of $\epsilon$, we adopt an approach based on nearest-neighbor distances. For each stock, the minimum distance to its closest neighbor is computed, and $\epsilon$ is defined as the $\alpha$-th percentile of these distances. This percentile-based rule controls the granularity of clustering: smaller values of $\alpha$ lead to a larger number of smaller clusters, while larger values produce fewer and more compact clusters. In our empirical analysis, we consider $\alpha = 0.95$ and $\alpha = 0.99$. Figure 2 in the manuscript illustrates the clustering mechanism using five data points $p, q, r, s, t$.

Overall, agglomerative clustering (see Algorithm 4) provides a flexible and interpretable framework for grouping assets based on their similarity structure. By applying the clustering procedure separately for each distance measure, we are able to examine how different notions of similarity influence the resulting cluster configurations and, consequently, the performance of the proposed (DRMV) portfolio strategy.

\begin{algorithm}[H]\footnotesize
\caption{Agglomerative clustering}
\begin{algorithmic}[1]\label{Algo: clustering}
\Require $D \in \mathbb{R}^{n\times n}$: pairwise distance matrix computed using one of the distance measures ($d_{\mathrm{AWD}}$, $d_{\mathrm{ALD}}$, $d_{\mathrm{AC}}$, or $d_{\mathrm{AE}}$); $\alpha$: percentile parameter
\State Compute $\epsilon$ as the $\alpha$-th percentile of nearest-neighbor distances
\State Initialize each stock as an individual cluster
\While{minimum inter-cluster distance $< \epsilon$}
    \State Compute inter-cluster distances using average linkage
    \State Identify the closest pair of clusters
    \State Merge the selected pair
    \State Remove singleton clusters and treat them as outliers
\EndWhile
\State \Return Final clusters
\end{algorithmic}
\end{algorithm}

\section*{\textcolor{black}{Appendix D. Statistical significance test}}\label{Appendix:D}

{\color{black} We conduct a one-sided paired t-test to examine whether the mean return of the one strategy $s_1$ is significantly greater than that of the other strategy $s_2$. The hypotheses are $H_0:\mu_{s_1}\leq \mu_{s_2} \text{ and }
H_1:\mu_{{s_1}}>\mu_{s_2},$ where $\mu_{{s_1}}$ and $\mu_{s_2}$ denote the out-of-sample mean returns of the strategies $s_1$ and $s_2$, respectively. The statistics used in the t-test are $t= \frac{\bar{d}}{s_d/\sqrt{n}},$ where $\bar{d}$ and $s_d$ denote the sample mean and sample SD of the paired return differences, respectively. 

Finally, we employ a one-sided bootstrap-based Sharpe ratio test following the methodology proposed in \cite{ledoit2008robust} to examine whether the $s_1$ strategy provides significantly superior risk-adjusted performance relative to the $s_2$ strategy. The hypotheses are $H_0: SR_{s_1}\leq SR_{s_2} \text{ and } H_1: SR_{s_1}>SR_{s_2},$ where $SR_{s_1}$ and $SR_{s_2}$ denote the out-of-sample Sharpe ratios of the ${s_1}$ and ${s_2}$ strategies, respectively.
}

\bibliographystyle{plain}\scriptsize
\bibliography{reference}

@article{Markowitz1952PortfolioSelection,
    title = {{Portfolio selection}},
    year = {1952},
    journal = {Journal of Finance},
    author = {Markowitz, Harry M},
    number = {1},
    pages = {71--91},
    volume = {7}
}

@article{rockafellar2002deviation,
  title={Deviation measures in risk analysis and optimization},
  author={Rockafellar, R Tyrrell and Uryasev, Stanislav P and Zabarankin, Michael},
  journal={University of Florida, Department of Industrial \& Systems Engineering Working Paper},
  number={7},
  year={2002}
}

@article{linsmeier1996risk,
  title={Risk measurement: An introduction to value at risk},
  author={Linsmeier, Thomas J and Pearson, Neil D},
  year={1996}
}

@article{yu2017incorporating,
  title={Incorporating transaction costs, weighting management, and floating required return in robust portfolios},
  author={Yu, Jing-Rung and Chiou, Wan-Jiun Paul and Liu, Ren-Ting},
  journal={Computers \& Industrial Engineering},
  volume={109},
  pages={48--58},
  year={2017},
  publisher={Elsevier}
}

@article{dhingra2024comprehensive,
  title={A comprehensive evaluation of constrained mean-expectile portfolios with short selling},
  author={Dhingra, Vrinda and Sharma, Amita and Gupta, Shiv Kumar},
  journal={Annals of Operations Research},
  pages={1--39},
  year={2024},
  publisher={Springer}
}

@article{sharpe1998sharpe,
  title={The {S}harpe ratio},
  author={Sharpe, William F},
  journal={Streetwise--the Best of the Journal of Portfolio Management},
  pages={169--185},
  year={1998},
  publisher={Princeton University Press NJ}
}

@article{han2023pairs,
  title={Pairs trading via unsupervised learning},
  author={Han, Chulwoo and He, Zhaodong and Toh, Alenson Jun Wei},
  journal={European Journal of Operational Research},
  volume={307},
  number={2},
  pages={929--947},
  year={2023},
  publisher={Elsevier}
}

@article{leon2017clustering,
  title={Clustering algorithms for risk-adjusted portfolio construction},
  author={Le{\'o}n, Diego and Arag{\'o}n, Arbey and Sandoval, Javier and Hern{\'a}ndez, Germ{\'a}n and Ar{\'e}valo, Andr{\'e}s and Ni{\~n}o, Jaime},
  journal={Procedia Computer Science},
  volume={108},
  pages={1334--1343},
  year={2017},
  publisher={Elsevier}
}

@article{cohen2010lipschitz,
  title={Lipschitz functions have L p-stable persistence},
  author={Cohen-Steiner, David and Edelsbrunner, Herbert and Harer, John and Mileyko, Yuriy},
  journal={Foundations of Computational Mathematics},
  volume={10},
  number={2},
  pages={127--139},
  year={2010},
  publisher={Springer}
}

@article{bubenik2015statistical,
  title={Statistical topological data analysis using persistence landscapes.},
  author={Bubenik, Peter and others},
  journal={ The Journal of Machine Learning Research},
  volume={16},
  number={1},
  pages={77--102},
  year={2015}
}

@article{gidea2018topological,
  title={Topological data analysis of financial time series: Landscapes of crashes},
  author={Gidea, Marian and Katz, Yuri},
  journal={Physica A: Statistical Mechanics and its Applications},
  volume={491},
  pages={820--834},
  year={2018},
  publisher={Elsevier}
}

@book{appel2005technical,
  title={Technical analysis: power tools for active investors},
  author={Appel, Gerald},
  year={2005},
  publisher={FT Press}
}

@article{basak2019predicting,
  title={Predicting the direction of stock market prices using tree-based classifiers},
  author={Basak, Suryoday and Kar, Saibal and Saha, Snehanshu and Khaidem, Luckyson and Dey, Sudeepa Roy},
  journal={The North American Journal of Economics and Finance},
  volume={47},
  pages={552--567},
  year={2019},
  publisher={Elsevier}
}

@article{weng2017stock,
  title={Stock market one-day ahead movement prediction using disparate data sources},
  author={Weng, Bin and Ahmed, Mohamed A and Megahed, Fadel M},
  journal={Expert Systems with Applications},
  volume={79},
  pages={153--163},
  year={2017},
  publisher={Elsevier}
}

@article{de1990noise,
  title={Noise trader risk in financial markets},
  author={De Long, J Bradford and Shleifer, Andrei and Summers, Lawrence H and Waldmann, Robert J},
  journal={Journal of Political Economy},
  volume={98},
  number={4},
  pages={703--738},
  year={1990},
  publisher={The University of Chicago Press}
}

@article{baker2007investor,
  title={Investor sentiment in the stock market},
  author={Baker, Malcolm and Wurgler, Jeffrey},
  journal={Journal of Economic Perspectives},
  volume={21},
  number={2},
  pages={129--151},
  year={2007},
  publisher={American Economic Association}
}

@article{checkley2017hasty,
  title={The hasty wisdom of the mob: How market sentiment predicts stock market behavior},
  author={Checkley, MS and Hig{\'o}n, D A{\~n}{\'o}n and Alles, H},
  journal={Expert Systems with Applications},
  volume={77},
  pages={256--263},
  year={2017},
  publisher={Elsevier}
}

@article{nardo2016walking,
  title={Walking down wall street with a tablet: A survey of stock market predictions using the web},
  author={Nardo, Michela and Petracco-Giudici, Marco and Naltsidis, Min{\'a}s},
  journal={Journal of Economic Surveys},
  volume={30},
  number={2},
  pages={356--369},
  year={2016},
  publisher={Wiley Online Library}
}

@article{zhou2021big,
  title={Big data and portfolio optimization: A novel approach integrating DEA with multiple data sources},
  author={Zhou, Zhongbao and Gao, Meng and Xiao, Helu and Wang, Rui and Liu, Wenbin},
  journal={Omega},
  volume={104},
  pages={102479},
  year={2021},
  publisher={Elsevier}
}

@article{loughran2011liability,
  title={When is a liability not a liability? Textual analysis, dictionaries, and 10-Ks},
  author={Loughran, Tim and McDonald, Bill},
  journal={The Journal of Finance},
  volume={66},
  number={1},
  pages={35--65},
  year={2011},
  publisher={Wiley Online Library}
}

@inproceedings{devlin2019bert,
  title={Bert: Pre-training of deep bidirectional transformers for language understanding},
  author={Devlin, Jacob and Chang, Ming-Wei and Lee, Kenton and Toutanova, Kristina},
  booktitle={Proceedings of the 2019 Conference of the North American Chapter of the Association for Computational Linguistics: Human Language Technologies, volume 1 (long and short papers)},
  pages={4171--4186},
  year={2019}
}

@article{araci2019finbert,
  title={Finbert: Financial sentiment analysis with pre-trained language models},
  author={Araci, Dogu},
  journal={arXiv preprint arXiv:1908.10063},
  year={2019}
}

@article{konstantinidis2024finllama,
  title={Finllama: Financial sentiment classification for algorithmic trading applications},
  author={Konstantinidis, Thanos and Iacovides, Giorgos and Xu, Mingxue and Constantinides, Tony G and Mandic, Danilo},
  journal={arXiv preprint arXiv:2403.12285},
  year={2024}
}

@inproceedings{sidogi2021stock,
  title={Stock price prediction using sentiment analysis},
  author={Sidogi, Thendo and Mbuvha, Rendani and Marwala, Tshilidzi},
  booktitle={2021 IEEE International Conference on Systems, Man, and Cybernetics (SMC)},
  pages={46--51},
  year={2021},
  organization={IEEE}
}

@inproceedings{jun2024predicting,
  title={Predicting stock prices with finbert-lstm: Integrating news sentiment analysis},
  author={jun Gu, Wen and hao Zhong, Yi and zun Li, Shi and song Wei, Chang and ting Dong, Li and yue Wang, Zhuo and Yan, Chao},
  booktitle={Proceedings of the 2024 8th International Conference on Cloud and Big Data Computing},
  pages={67--72},
  year={2024}
}

@article{lee2025large,
  title={Large language models in finance (finllms)},
  author={Lee, Jean and Stevens, Nicholas and Han, Soyeon Caren},
  journal={Neural Computing and Applications},
  volume={37},
  number={30},
  pages={24853--24867},
  year={2025},
  publisher={Springer}
}

@article{mantshimuli2025sentiment,
  title={Sentiment-Aware Portfolio Optimization: CVaR-Based Diversification with Deep Reinforcement Learning},
  author={Mantshimuli, Lamukanyani A},
  journal={IEEE Access},
  year={2025},
  publisher={IEEE}
}

@article{touvron2023llama,
  title={Llama: Open and efficient foundation language models},
  author={Touvron, Hugo and Lavril, Thibaut and Izacard, Gautier and Martinet, Xavier and Lachaux, Marie-Anne and Lacroix, Timoth{\'e}e and Rozi{\`e}re, Baptiste and Goyal, Naman and Hambro, Eric and Azhar, Faisal and others},
  journal={arXiv preprint arXiv:2302.13971},
  year={2023}
}

@inproceedings{mahendran2025comparative,
  title={Comparative advances in financial sentiment analysis: A review of BERT, FinBert, and large language models},
  author={Mahendran, Manish Barath and Gokul, Aswin Kumar and Lakshmi, Poornima and Pavithra, S},
  booktitle={2025 3rd International Conference on Intelligent Data Communication Technologies and Internet of Things (IDCIoT)},
  pages={39--45},
  year={2025},
  organization={IEEE}
}

@inproceedings{takens2006detecting,
  title={Detecting strange attractors in turbulence},
  author={Takens, Floris},
  booktitle={Dynamical Systems and Turbulence, Warwick 1980: proceedings of a symposium held at the University of Warwick 1979/80},
  pages={366--381},
  year={2006},
  organization={Springer}
}

@article{goel2020topological,
  title={Topological data analysis in investment decisions},
  author={Goel, Anubha and Pasricha, Puneet and Mehra, Aparna},
  journal={Expert Systems with Applications},
  volume={147},
  pages={113222},
  year={2020},
  publisher={Elsevier}
}

@article{goel2025risk,
  title={Risk reduced sparse index tracking portfolio: A topological data analysis approach},
  author={Goel, Anubha and Pasricha, Puneet and Kanniainen, Juho},
  journal={Omega},
  pages={103432},
  year={2025},
  publisher={Elsevier}
}

@article{ghrist2008barcodes,
  title={Barcodes: the persistent topology of data},
  author={Ghrist, Robert},
  journal={Bulletin of the American Mathematical Society},
  volume={45},
  number={1},
  pages={61--75},
  year={2008}
}

@article{yu2011portfolio,
  title={Portfolio rebalancing model using multiple criteria},
  author={Yu, Jing-Rung and Lee, Wen-Yi},
  journal={European Journal of Operational Research},
  volume={209},
  number={2},
  pages={166--175},
  year={2011},
  publisher={Elsevier}
}

@article{yu2022dynamic,
  title={Dynamic rebalancing portfolio models with analyses of investor sentiment},
  author={Yu, Jing-Rung and Chiou, W Paul and Hung, Cing-Hung and Dong, Wen-Kuei and Chang, Yi-Hsuan},
  journal={International Review of Economics \& Finance},
  volume={77},
  pages={1--13},
  year={2022},
  publisher={Elsevier}
}

@article{keynes1983keynes,
  title={Keynes as an investor},
  author={Keynes, John Maynard and Johnson, E and Moggridge, D},
  journal={The Collected Works of John Maynard Keynes},
  volume={12},
  pages={1--113},
  year={1983}
}

@article{brodie2009sparse,
  title={Sparse and stable Markowitz portfolios},
  author={Brodie, Joshua and Daubechies, Ingrid and De Mol, Christine and Giannone, Domenico and Loris, Ignace},
  journal={Proceedings of the National Academy of Sciences},
  volume={106},
  number={30},
  pages={12267--12272},
  year={2009},
  publisher={National Academy of Sciences}
}

@article{dai2018some,
  title={Some improved sparse and stable portfolio optimization problems},
  author={Dai, Zhifeng and Wen, Fenghua},
  journal={Finance Research Letters},
  volume={27},
  pages={46--52},
  year={2018},
  publisher={Elsevier}
}

@article{kremer2022sparse,
  title={Sparse index clones via the sorted $\ell_1$-norm},
  author={Kremer, Philipp J and Brzyski, Damian and Bogdan, Ma{\l}gorzata and Paterlini, Sandra},
  journal={Quantitative Finance},
  volume={22},
  number={2},
  pages={349--366},
  year={2022},
  publisher={Taylor \& Francis}
}

@article{goel2025sparse,
  title={Sparse portfolio selection via topological data analysis based clustering},
  author={Goel, Anubha and Filipovi{\'c}, Damir and Pasricha, Puneet},
  journal={Quantitative Finance},
  volume={25},
  number={8},
  pages={1261--1291},
  year={2025},
  publisher={Taylor \& Francis}
}

@article{xu2024efficient,
  title={An efficient global optimal method for cardinality constrained portfolio optimization},
  author={Xu, Wei and Tang, Jie and Yiu, Ka Fai Cedric and Peng, Jian Wen},
  journal={INFORMS Journal on Computing},
  volume={36},
  number={2},
  pages={690--704},
  year={2024},
  publisher={INFORMS}
}

@article{kalayci2020efficient,
  title={An efficient hybrid metaheuristic algorithm for cardinality constrained portfolio optimization},
  author={Kalayci, Can B and Polat, Olcay and Akbay, Mehmet A},
  journal={Swarm and Evolutionary Computation},
  volume={54},
  pages={100662},
  year={2020},
  publisher={Elsevier}
}

@article{wei2017informativeness,
  title={Informativeness of the market news sentiment in the Taiwan stock market},
  author={Wei, Yu-Chen and Lu, Yang-Cheng and Chen, Jen-Nan and Hsu, Yen-Ju},
  journal={The North American Journal of Economics and Finance},
  volume={39},
  pages={158--181},
  year={2017},
  publisher={Elsevier}
}

@article{siganos2017divergence,
  title={Divergence of sentiment and stock market trading},
  author={Siganos, Antonios and Vagenas-Nanos, Evangelos and Verwijmeren, Patrick},
  journal={Journal of Banking \& Finance},
  volume={78},
  pages={130--141},
  year={2017},
  publisher={Elsevier}
}

@article{deng2017adapting,
  title={Adapting sentiment lexicons to domain-specific social media texts},
  author={Deng, Shuyuan and Sinha, Atish P and Zhao, Huimin},
  journal={Decision Support Systems},
  volume={94},
  pages={65--76},
  year={2017},
  publisher={Elsevier}
}

@article{renault2017intraday,
  title={Intraday online investor sentiment and return patterns in the US stock market},
  author={Renault, Thomas},
  journal={Journal of Banking \& Finance},
  volume={84},
  pages={25--40},
  year={2017},
  publisher={Elsevier}
}

@article{aromi2021topological,
  title={Topological features of multivariate distributions: Dependency on the covariance matrix},
  author={Aromi, Lloyd L and Katz, Yuri A and Vives, Josep},
  journal={Communications in Nonlinear Science and Numerical Simulation},
  volume={103},
  pages={105996},
  year={2021},
  publisher={Elsevier}
}

@article{majumdar2023pairs,
  title={Pairs trading with topological data analysis},
  author={Majumdar, Sourav and Laha, Arnab Kumar},
  journal={International Journal of Theoretical and Applied Finance},
  volume={26},
  number={08},
  pages={2450002},
  year={2023},
  publisher={World Scientific}
}

@article{karan2021time,
  title={Time series classification via topological data analysis},
  author={Karan, Alperen and Kaygun, Atabey},
  journal={Expert Systems with Applications},
  volume={183},
  pages={115326},
  year={2021},
  publisher={Elsevier}
}

@article{carlsson2014topological,
  title={Topological pattern recognition for point cloud data},
  author={Carlsson, Gunnar},
  journal={Acta Numerica},
  volume={23},
  pages={289--368},
  year={2014},
  publisher={Cambridge University Press}
}

@article{lum2013extracting,
  title={Extracting insights from the shape of complex data using topology},
  author={Lum, Pek Y and Singh, Gurjeet and Lehman, Alan and Ishkanov, Tigran and Vejdemo-Johansson, Mikael and Alagappan, Muthu and Carlsson, John and Carlsson, Gunnar},
  journal={Scientific Reports},
  volume={3},
  number={1},
  pages={1236},
  year={2013},
  publisher={Nature Publishing Group UK London}
}

@article{papamarkou2024position,
  title={Position: Topological deep learning is the new frontier for relational learning},
  author={Papamarkou, Theodore and Birdal, Tolga and Bronstein, Michael and Carlsson, Gunnar and Curry, Justin and Gao, Yue and Hajij, Mustafa and Kwitt, Roland and Lio, Pietro and Di Lorenzo, Paolo and others},
  journal={Proceedings of Machine Learning Research},
  volume={235},
  pages={39529},
  year={2024}
}

@inproceedings{rivera2019topological,
  title={Topological data analysis for portfolio management of cryptocurrencies},
  author={Rivera-Castro, Rodrigo and Pilyugina, Polina and Burnaev, Evgeny},
  booktitle={2019 International Conference on Data Mining Workshops (ICDMW)},
  pages={238--243},
  year={2019},
  organization={IEEE}
}

@inproceedings{sokerin2024portfolio,
  title={Portfolio selection via topological data analysis},
  author={Sokerin, Petr and Kuznetsov, Kristian and Makhneva, Elizaveta and Zaytsev, Alexey},
  booktitle={Sixteenth International Conference on Machine Vision (ICMV 2023)},
  volume={13072},
  pages={371--379},
  year={2024},
  organization={SPIE}
}

@article{millington2021construction,
  title={Construction of minimum spanning trees from financial returns using rank correlation},
  author={Millington, Tristan and Niranjan, Mahesan},
  journal={Physica A: Statistical Mechanics and its Applications},
  volume={566},
  pages={125605},
  year={2021},
  publisher={Elsevier}
}

@article{michis2022multiscale,
  title={Multiscale partial correlation clustering of stock market returns},
  author={Michis, Antonis A},
  journal={Journal of Risk and Financial Management},
  volume={15},
  number={1},
  pages={24},
  year={2022},
  publisher={MDPI}
}

@article{jung2016clustering,
  title={Clustering stocks using partial correlation coefficients},
  author={Jung, Sean S and Chang, Woojin},
  journal={Physica A: Statistical Mechanics and its Applications},
  volume={462},
  pages={410--420},
  year={2016},
  publisher={Elsevier}
}

@article{puerto2020clustering,
  title={Clustering and portfolio selection problems: A unified framework},
  author={Puerto, Justo and Rodr{\'\i}guez-Madrena, Mois{\'e}s and Scozzari, Andrea},
  journal={Computers \& Operations Research},
  volume={117},
  pages={104891},
  year={2020},
  publisher={Elsevier}
}

@article{mattera2025time,
  title={Time series clustering for high-dimensional portfolio selection: a comparative study: R. Mattera et al.},
  author={Mattera, Raffaele and Scepi, Germana and Kaur, Parmjit},
  journal={Soft Computing},
  volume={29},
  number={8},
  pages={4219--4231},
  year={2025},
  publisher={Springer}
}

@article{rousseeuw1987silhouettes,
  title={Silhouettes: a graphical aid to the interpretation and validation of cluster analysis},
  author={Rousseeuw, Peter J},
  journal={Journal of Computational and Applied Mathematics},
  volume={20},
  pages={53--65},
  year={1987},
  publisher={Elsevier}
}

@article{cen2022financial,
  title={Financial market correlation analysis and stock selection application based on TCN-deep clustering},
  author={Cen, Yuefeng and Luo, Mingxing and Cen, Gang and Zhao, Cheng and Cheng, Zhigang},
  journal={Future Internet},
  volume={14},
  number={11},
  pages={331},
  year={2022},
  publisher={MDPI}
}

@article{embrechts2003using,
  title={Using copulae to bound the value-at-risk for functions of dependent risks},
  author={Embrechts, Paul and H{\"o}ing, Andrea and Juri, Alessandro},
  journal={Finance and Stochastics},
  volume={7},
  number={2},
  pages={145--167},
  year={2003},
  publisher={Springer}
}

@article{al2018outperformance,
  title={Outperformance and tracking: Dynamic asset allocation for active and passive portfolio management},
  author={Al-Aradi, Ali and Jaimungal, Sebastian},
  journal={Applied Mathematical Finance},
  volume={25},
  number={3},
  pages={268--294},
  year={2018},
  publisher={Taylor \& Francis}
}

@article{garg2025enhanced,
  title={Enhanced indexing using cumulative prospect theory utility function with expectile risk},
  author={Garg, Divyanee and Khan, Ahmad Zaman and Mehra, Aparna},
  journal={Omega},
  pages={103444},
  year={2025},
  publisher={Elsevier}
}

@article{wu2024sparse,
  title={Sparse portfolio optimization via $\ell_1$ over $\ell_2$ regularization},
  author={Wu, Zhongming and Sun, Kexin and Ge, Zhili and Allen-Zhao, Zhihua and Zeng, Tieyong},
  journal={European Journal of Operational Research},
  volume={319},
  number={3},
  pages={820--833},
  year={2024},
  publisher={Elsevier}
}

@article{wang2026exact,
  title={An exact algorithm for a cardinality-constrained index tracking model considering investment preferences in portfolio optimization},
  author={Wang, Chun and Wu, Zhongming and Xu, Wei and Yuan, Yu},
  journal={Journal of Industrial and Management Optimization},
  volume={22},
  number={1},
  pages={612--641},
  year={2026}
}

@article{ran2023comprehensive,
  title={Comprehensive survey on hierarchical clustering algorithms and the recent developments},
  author={Ran, Xingcheng and Xi, Yue and Lu, Yonggang and Wang, Xiangwen and Lu, Zhenyu},
  journal={Artificial Intelligence Review},
  volume={56},
  number={8},
  pages={8219--8264},
  year={2023},
  publisher={Springer}
}

@article{ledoit2008robust,
  title={Robust performance hypothesis testing with the Sharpe ratio},
  author={Ledoit, Oliver and Wolf, Michael},
  journal={Journal of Empirical Finance},
  volume={15},
  number={5},
  pages={850--859},
  year={2008},
  publisher={Elsevier}
}

@article{zhang2012generalized,
  title={Generalized adjusted rand indices for cluster ensembles},
  author={Zhang, Shaohong and Wong, Hau-San and Shen, Ying},
  journal={Pattern Recognition},
  volume={45},
  number={6},
  pages={2214--2226},
  year={2012},
  publisher={Elsevier}
}

@article{yu2024dynamic,
  title={Dynamic portfolio optimization with the MARCOS approach under uncertainty},
  author={Yu, Pengrui and Ge, Zhipeng and Gong, Xiaomin and Cao, Xiao},
  journal={International Review of Financial Analysis},
  volume={96},
  pages={103565},
  year={2024},
  publisher={Elsevier}
}

@article{migliavacca2023bibliometric,
  title={A bibliometric review of portfolio diversification literature},
  author={Migliavacca, Milena and Goodell, John W and Paltrinieri, Andrea},
  journal={International Review of Financial Analysis},
  volume={90},
  pages={102836},
  year={2023},
  publisher={Elsevier}
}
\end{document}